\documentclass[useAMS,usenatbib]{mn2e}

\usepackage{graphicx,color}
\usepackage{amssymb,amsmath}
\usepackage{amsmath} %Alinhamento das equações
\usepackage{multicol}
\usepackage{latexsym}
\usepackage{amsfonts}
\usepackage{amssymb}
\usepackage{wasysym}
\usepackage{textcomp}
\usepackage{longtable,lscape}
\usepackage{mathtools}

\usepackage{aas_macros}

\usepackage{natbib}
\usepackage{amssymb,amsmath}
\title[Open clusters in the Sagittarius spiral arm]{Characterizing dynamical stages of open clusters located in the Sagittarius spiral arm}
\author[M. S. Angelo et al.]{M. S. Angelo$^{1}$\thanks{E-mail:
mateusangelo@cefet.mg.br}, A.E. Piatti$^{2,3}$, W.S. Dias$^{4}$ and F.F.S. Maia$^{5}$ \\ %\newauthor and  \\ \\ 
\noindent
$^1$Centro Federal de Educa\c{c}\~ao Tecnol\'ogica de Minas Gerais, Av. Monsenhor Luiz de Gonzaga, 103, 37250-000 Nepomuceno, MG, Brazil\\
$^2$Consejo Nacional de Investigaciones Cient\'ificas y T\'ecnicas, Av. Rivadavia 1917, C1033AAJ, Buenos Aires, Argentina \\
$^3$Observatorio Astron\'omico, Universidad Nacional de C\'ordoba, Laprida 854, 5000, C\'ordoba, Argentina\\
$^4$Instituto de F\'isica e Qu\'imica, Universidade Federal de Itajub\'a, Av. BPS 1303 Pinheirinho, 37500-903 Itajub\'a, MG, Brazil \\
$^5$Instituto de F\'isica, Universidade Federal do Rio de Janeiro, 21941-972, Rio de Janeiro, Brazil
}

\begin{document}

\date{Accepted XXX. Received XXX; in original form XXX}

\pagerange{\pageref{firstpage}--\pageref{lastpage}} \pubyear{XXXX}

\maketitle

\label{firstpage}

\begin{abstract}

%(manter máx de 250 palavras)

The study of dynamical properties of Galactic open clusters is a fundamental prerequisite for the comprehension of their dissolution processes. In this work, we characterized 12 open clusters, namely: Collinder\,258, NGC\,6756, Czernik\,37, NGC\,5381, Ruprecht\,111, Ruprecht\,102, NGC\,6249, Basel\,5, Ruprecht\,97, Trumpler\,25, ESO\,129-SC32 and BH\,150, projected against dense stellar fields. In
order to do that, we employed Washington $CT_{1}$ photometry and GAIA DR2 astrometry, combined
with a decontamination algorithm applied to the three-dimensional astrometric space of proper motions and parallaxes. From the derived membership likelihoods, we built decontaminated colour-magnitude diagrams, while structural parameters were obtained from King profiles fitting. 
Our analysis revealed that they are relatively young open clusters  (log($t$\,yr$^{-1}$) $\sim7.3-8.6$), 
placed along the Sagittarius spiral arm, and at different internal dynamical stages. We found that
the half-light radius to Jacobi radius ratio, the concentration parameter and the age to relaxation time
ratio describe satisfactorily their different stages of dynamical evolution.
Those relative more dynamically evolved open clusters have apparently 
experienced more important low-mass star loss.

\end{abstract}


\section{Introduction}

It is known that the majority of stars are born embedded within giant molecular clouds \citep{Lada:2003} and form stellar aggregates named associations or open clusters (OCs). Whilst the first are loose and gravitationally unbound groups (typical dissolution times between $\sim10-100\,$Myr; \citeauthor{Moraux:2016}\,\,\citeyear{Moraux:2016}), the latter are long-lived stellar structures and their diversity in terms of age, stellar content and metallicity makes them ideal tracers of the Galaxy structure, providing information regarding its kinematical evolution and chemical enrichment.  

The initial evolutionary stages of the OCs have critical impact on their subsequent dynamical evolution, since the early gas-expulsion process (caused by, e.g., supernova and stellar winds during the first $\sim3\,$Myr; \citeauthor{Portegies-Zwart:2010}\,\,\citeyear{Portegies-Zwart:2010}) cause less than 10\% of embedded OCs to survive emergence from molecular clouds \citep{Lada:2003}. Those that are massive enough to survive this initial phase enter subsequent phases, when dynamical timescales become continuously smaller than mass loss timescales due to stellar evolution. The investigation of these structures helps to constrain the initial conditions assumed by, e.g., $N-$body simulations aimed at reproducing observable quantities of OCs at later evolutionary stages.

During the long-term evolution, the interplay between several destructive effects (e.g., tidal stripping, collisions with molecular clouds, evaporation of stars due to internal relaxation) causes the OC's stellar content to be gradually depleted until its dissolution amongst the general Galactic field. How stellar OCs dissolve is a debated topic (\citeauthor{Pavani:2007}\,\,\citeyear{Pavani:2007}; \citeauthor{Bica:2001}\,\,\citeyear{Bica:2001}; \citeauthor{Piatti:2017a}\,\,\citeyear{Piatti:2017a}) and uniform characterizations of evolved OCs, possibly covering different parameters and positions within the Galaxy, are important to constrain evolutionary models and thus to clarify this subject.  
%\citeauthor{Angelo:2018}\,\,\citeyear{Angelo:2018}

In this context, the second release of the GAIA catalogue \citep{Gaia-Collaboration:2018} inaugurated a new era in astronomy. The availability of astrometric information with high-precision (typically $\lesssim$\,0.1\,mas and $\lesssim$\,0.1\,mas\,yr$^{-1}$ for parallaxes and proper motion components, respectively) allowed the discovery of new OCs (\citeauthor{Cantat-Gaudin:2018a}\,\,\citeyear{Cantat-Gaudin:2018a}; \citeauthor{Borissova:2018}\,\,\citeyear{Borissova:2018}; \citeauthor{Ferreira:2019}\,\,\citeyear{Ferreira:2019}) and a more precise characterization of already catalogued ones (e.g., \citeauthor{Kos:2018}\,\,\citeyear{Kos:2018}; \citeauthor{Dias:2018a}\,\,\citeyear{Dias:2018a}). 

In the present paper, we took full advantage of GAIA DR2 data combined with mostly unpublished Washington $CT_1$ photometry to select member stars and analyse the dynamical stage of a set of 11 OCs (namely Collinder\,258, NGC\,6756, Czernik\,37, NGC\,5381, Ruprecht\,111, Ruprecht\,102, NGC\,6249, Basel\,5, Ruprecht\,97, Trumpler\,25 and ESO\,129-SC32). We also discuss the case of BH\,150, which existence as a physical system could not be confirmed in previous studies \citep{Carraro:2005a} and it was classified as a dubious object by inspection of photographic data (\citeauthor{Kharchenko:2013}\,\,\citeyear{Kharchenko:2013} and references therein). It is currently classified as an asterism in the \cite{Dias:2002} catalogue. We have included this OC in our sample in order to establish conclusively its physical nature. Since these 12 OCs are projected against dense stellar population from the Galactic disc, their colour-magnitude diagrams (CMDs) are severely contaminated, which makes the disentanglement of OC and field stars and the search for evolutionary sequences non-trivial tasks. We have combined astrometric and photometric data together with a decontamination algorithm which searches for overdensities across the 3-dimensional astrometric space ($\mu_{\alpha}, \mu_{\delta}, \varpi$) and check their significance by statistical comparisons with the dispersion of data from field stars. To the best of our knowledge, this is the first work that employs GAIA DR2 data in the analysis of the present sample.

Interestingly, 11 of the investigated OCs share similar ages and Galactocentric distances ($R_G$, uncertainties considered), which allows us to explore the relationships between parameters associated to their dynamical evolution. Besides, their concentration parameters are relatively low ($c$=log$(r_t/r_c)\lesssim0.75$), which places them in the lower regime of $c$ values for OCs with similar ages (Section \ref{discussion}). This characteristics makes this sample an unique one. Although they are not in advanced dynamical stages, according to their $c$ parameters, our OCs present signals of dynamical evolution, as evidenced by their age/$t_{\textrm{rh}}$ ratios, where $t_{\textrm{rh}}$ is the half-light relaxation time, and signals of low-mass star depletion in their mass functions (Section \ref{mass_functions}).

This paper is organized as follows: in Section \ref{data_collection_reduction}, we present our sample and give some details about the data reduction steps. In Section \ref{method}, we present the analysis procedure (structural parameters and membership determination). The results are presented in Section \ref{results} and discussed in Section \ref{discussion}. In Section \ref{conclusions}, we summarize our conclusions.

\section{Data collection and reduction}
\label{data_collection_reduction}

\begin{table*}
 \small
 %\begin{minipage}{300mm}
  \caption{Observations log of the studied OCs.}
  \label{log_observations}
 \begin{tabular}{lccccccc}
 
  \hline

Cluster & $\rmn{RA}$     & $\rmn{DEC}$     & $\ell$     & $b$     & Filter     & Exposure     & Airmass  \\   
            &  ($\rmn{h}$:$\rmn{m}$:$\rmn{s}$) & ($\degr$:$\arcmin$:$\arcsec$) & ($^{\circ}$) & ($^{\circ}$) &   & (s)  &    \\
\hline

Collinder\,258   & 12:27:16   & -60:46:42  &  299.9843     &  01.9550    & $C$ & 15,150            & 1.2,1.3               \\
                 &            &            &               &             & $R$ & 2,20              & 1.3,1.3               \\

NGC\,6756        & 19:08:45   & 04:43:01   &   39.1046     &  -01.6865   & $C$ & 30,90,90,900      & 1.2,1.2,1.2,1.2       \\
                 &            &            &               &             & $R$ & 20,20,150,150     & 1.2,1.2,1.2,1.2       \\
                   
Czernik\,37      & 17:53:12   & -27:22:00  &   2.2053      &  -0.6255    & $C$ & 30,450            & 1.0,1.0               \\
                 &            &            &               &             & $R$ & 5,45              & 1.0,1.0               \\
                                                                                   
NGC\,5381        & 14:00:40   & -59:36:18  &   311.5940    &   02.0975   & $C$ & 90,120,600,600    & 1.1,1.1,1.2,1.2       \\
                 &            &            &               &             & $R$ & 30,30,120,120     & 1.2,1.2,1.2,1.2       \\
                                                                                   
Trumpler\,25     & 17:24:24   & -39:01:01  &   349.1460    &  -01.7594   & $C$ & 30,300            & 1.0,1.0               \\
                 &            &            &               &             & $R$ & 5,300             & 1.0,1.0               \\
                                                                                   
BH\,150          & 13:38:04   & -63:20:42  &   308.1334    &  -0.9451    & $C$ & 60,60,600,600     & 1.2,1.2,1.2,1.2       \\
                 &            &            &               &             & $R$ & 5,15,15,90,90     & 1.2,1.2,1.2,1.2,1.2   \\
                                                                                   
Ruprecht\,111    & 14:36:00   & -59:58:48  &   315.6658    &  0.2811     & $C$ & 30,45,450         & 1.2,1.2,1.2           \\
                 &            &            &               &             & $R$ & 15,15             & 1.2,1.2               \\
                                                                                   
Ruprecht\,102    & 12:13:34   & -62:43:48  &  298.6088     & -0.1766     & $C$ & 45,450            & 1.2,1.2               \\
                 &            &            &               &             & $R$ & 7,45              & 1.2,1.2               \\
                                                                                   
NGC\,6249        & 16:57:36   & -44:49:00  &  341.5242     & -01.1772    & $C$ & 30,30             & 1.0,1.0               \\
                 &            &            &               &             & $R$ & 5,5               & 1.0,1.0               \\
                                                                                   
Basel\,5         & 17:52:24   & -30:06:00  & 359.7616      & -01.8636    & $C$ & 30,30,600,600     & 1.0,1.0,1.0,1.0       \\
                 &            &            &               &             & $R$ &15,15,120,120      & 1.0,1.0,1.0,1.0       \\
                                                                                   
Ruprecht\,97     & 11:57:28   & -62:43:00  & 296.7920      & -0.4901     & $C$ & 60,90,900         & 1.3,1.3,1.3           \\
                 &            &            &               &             & $R$ & 60,180,180        & 1.3,1.3,1.3           \\
                                                                                   
ESO\,129-SC32    & 11:44:11   & -61:03:29  & 294.8851      & 0.7587      & $C$ & 60,60,450,450     & 1.3,1.3,1.3,1.3       \\
                 &            &            &               &             & $R$ & 10,10,60,120      & 1.3,1.3,1.3,1.3       \\

\hline
\end{tabular}
%\end{minipage}
\end{table*}

%A pipeline do Wilton toma o menor tempo de exposição e vai adicionando as estrelas não coincidentes encontradas nas imagens de texp mais longo.

\begin{figure*}
\begin{center}
%\parbox[c]{1.0\textwidth}
%{

    \includegraphics[width=0.325\textwidth]{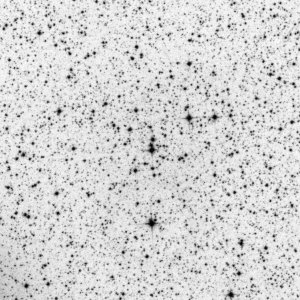}    
    \includegraphics[width=0.325\textwidth]{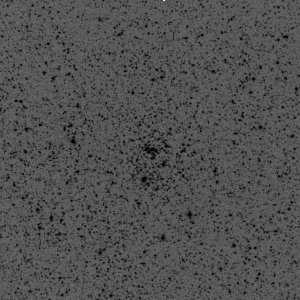}   
    \includegraphics[width=0.325\textwidth]{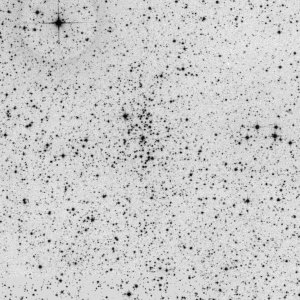}  
    \includegraphics[width=0.325\textwidth]{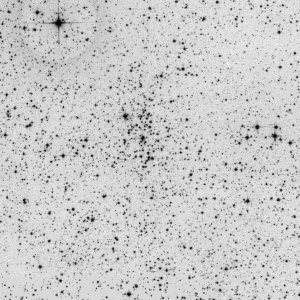}    
    \includegraphics[width=0.325\textwidth]{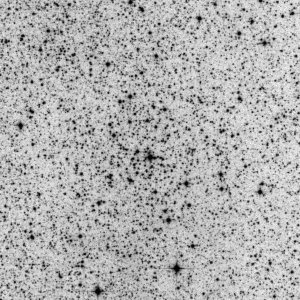}   
    \includegraphics[width=0.325\textwidth]{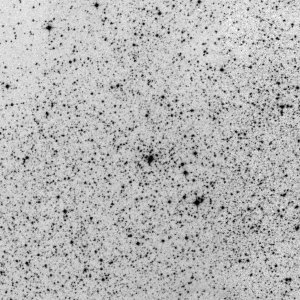}
    \includegraphics[width=0.325\textwidth]{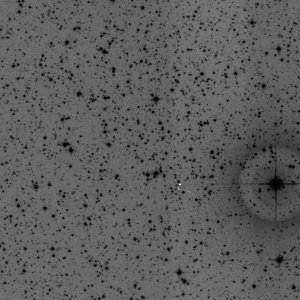}    
    \includegraphics[width=0.325\textwidth]{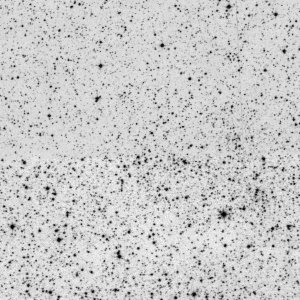}   
    \includegraphics[width=0.325\textwidth]{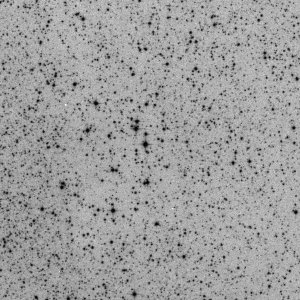}  
    \includegraphics[width=0.325\textwidth]{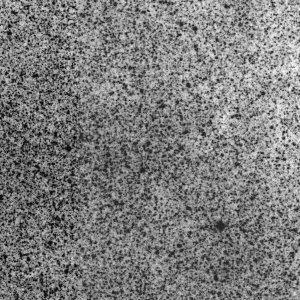}    
    \includegraphics[width=0.325\textwidth]{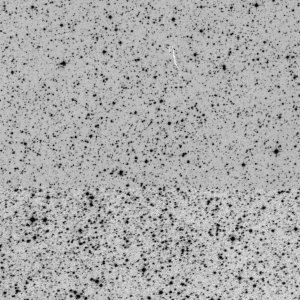}   
    \includegraphics[width=0.325\textwidth]{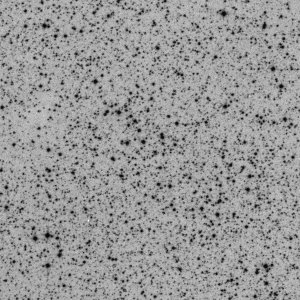} 
%}  
\caption{ DSS2 near-IR images of the OCs (from top-left to bottom-right) Collinder\,258, NGC\,6756, Czernik\,37, NGC\,5381, Trumpler\,25, BH\,150, Ruprecht\,111, Ruprecht\,102, NGC\,6249, Basel\,5, Ruprecht\,97 and ESO\,129-SC32. Image sizes are $14^{\arcmin}\times14^{\arcmin}$. North is up and East to the left. }

\label{images_clusters_parte1}
\end{center}
\end{figure*}

Images taken in $R$ (Kron-Kousins) and $C$ (Washington) filters were downloaded from the National Optical Astronomy Observatory (NOAO) public archive\footnote[1]{http://www.noao.edu/sdm/archives.php}. Observations were carried out during the nights 2008\,May\,08 to 2008\,May\,12 with the Tek2K CCD imager (scale of 0.4 arcsec\,pixel$^{-1}$, which provides a field of view of 13.6 arcmin$^2$) attached to the 0.9-m telescope at the Cerro Tololo Inter-American Observatory (CTIO, Chile; programme no. 2008A-0001, PI: Clari\'a). The observations log is showed in Table \ref{log_observations}. To illustrate the reader, images for the 12 selected OCs are showed in Figure \ref{images_clusters_parte1}. %, while those for the remaining OCs are shown in the Appendix.
% We adopt the same procedure throughout for other figures.

Calibration (bias, dome and sky flats in both $C$ and $R$ filters) and standard star field images (SA\,101, SA\,107, SA\,110; \citeauthor{Landolt:1992}\,\,\citeyear{Landolt:1992}; \citeauthor{Geisler:1996}\,\,\citeyear{Geisler:1996}) were also downloaded together with the images for the investigated OCs. The data reduction steps followed the standard procedures employed for optical CCD photometry: overscan and bias subtraction and division by normalized flat fields. All procedures were implented with the use of {\fontfamily{ptm}\selectfont QUADRED} package in {\fontfamily{ptm}\selectfont IRAF}\footnote[2]{{\fontfamily{ptm}\selectfont 
IRAF} is distributed by the National Optical Astronomy Observatories, which is operated by the Association of Universities for Research in Astronomy, Inc., under contract with the National Science Foundation.}. Images were taken with four amplifiers and were properly mosaiced in a single image extension. Bad pixels masks were also built and defective regions were corrected via linear interpolations performed on the images.

%(see also section 2 of \citeauthor{Angelo:2018}\,\,\citeyear{Angelo:2018})

The photometry was performed on the reduced images using a point spread function (PSF)-fitting algorithm. We used a modified version of the STARFINDER code \citep{Diolaiti:2000}, which draws empirical PSFs from pre-selected stars on the images (with high signal-to-noise and relatively isolated from nearby sources) and cross-correlates them with every point source detected above a defined threshold. The main modification consisted in automatising the code, minimising the user intervention during the choice of proper sources for PSF modelling. This allowed us to deal with a relatively large number of images taken in crowded fields. In this step, we only kept in the photometric catalogues those stars for which the correlation coefficients between the measured profile and the modelled PSF resulted greater than 0.7. This criterion minimized the introduction of spurious detections and, at the same time, allowed the detection of faint stars contaminated by the background noise. 

Astrometric solutions were computed for the whole set of images by mapping the positions of stars in each CCD frame with the corresponding coordinates as given in the GAIA DR2 catalogue for the observed region. A set of linear equations were calibrated, which allowed the transformation between the CCD reference system and the equatorial system (see \citeauthor{Caetano:2015}\,\,\citeyear{Caetano:2015} and references therein for further details) with an astrometric precision better than $\sim0.1\,$mas. Finally, our pipeline builds a final master table (containing the instrumental $c$ and $r$ magnitudes) by registering the fainter stars detected in the longer exposure frames and successively including those brighter sources identified in shorter exposures, thus avoiding the inclusion of saturated objects.            

Nearly 70 magnitudes of standard stars per filter per night were measured in order to calibrate the transformation equations between the instrumental and standard systems. Standard star field SA101 was observed repeatedly in a wide range of airmass ($\sim1.1-2.5$), which allowed the determination of the extinction coefficients. We used the following calibration equations:

%(using the {\fontfamily{ptm}\selectfont APPHOT} package within {\fontfamily{ptm}\selectfont IRAF})

\begin{align}
    c\, & =\,c_{1} + C + c_{2}\times X_{C} + c_{3}\times(C-T_{1}),  \\
    r\, & =\,t_{11} + T_{1} + t_{12}\times X_{T_{1}} + t_{13}\times(C-T_{1}), 
\end{align}
\noindent
       
\noindent
where $c,r$ represent instrumental magnitudes and $C,T_1$ are the standard ones; $c_1$ and $t_{11}$ are the zero-point coefficients, $c_2$ and $t_{12}$ are the extinction coefficients, $c_3$ and $t_{13}$ are the colour terms. The airmass in both filters are represented as $X_C$ and $X_{T_{1}}$. The coefficients were obatined via multiple linear regression as implemented in the {\fontfamily{ptm}\selectfont IRAF} task {\fontfamily{ptm}\selectfont FITPARAMS}. The results are presented in Table \ref{results_FITPARAMS}. Instead of using instrumental $t_1$ magnitudes, \cite{Geisler:1996} proposed that the $R$ filter is an excellent substitute of the $T_1$ filter due to increased transmission at all wavelengths.

\begin{table}
% \small
 \begin{minipage}{85mm}
  \caption{Mean values of the fitted coefficients and residuals for the presently calibrated $CT_1$ photometric data set.}
  \label{results_FITPARAMS}
 \begin{tabular}{lcccc}
 
\hline

Filter   &   Zero                              &   Extinction                    &  Colour             &  Residual              \\
           &   point                              &   coefficient                   &    term               &   (mag)                 \\

\hline

$C$          &  3.884$\pm$0.023      &   0.282$\pm$0.002      &   -0.173$\pm$0.011   &  0.011           \\
$T_{1}$    &  3.306$\pm$0.024      &   0.089$\pm$0.002      &   -0.041$\pm$0.004   &  0.010           \\
      
\hline
\end{tabular}
\end{minipage}
\end{table}

\begin{table}
 \tiny
\begin{minipage}{85mm}
  \caption{Stars in the field of NGC\,5381: Identifiers, coordinates, magnitudes and photometric uncertainties.}
  \label{excerpt_NGC5381}
 \begin{tabular}{ccccc}
 
\hline

Star ID   &   $\rmn{RA}$       &   $\rmn{DEC}$       &  $C$           &  $T_{1}$               \\
              &   ($^{\circ}$)         &   ($^{\circ}$)           &   (mag)       &   (mag)                  \\
\hline

%$-$   &        $-$             &           $-$             &                   $-$           &              $-$               \\ 
  1        &  210.1042480   & -59.5857430        & 13.429$\pm$0.001    & 12.631$\pm$0.001   \\   %ID 4 in the original table 
  2        &  210.1034546   & -59.5751610        & 13.619$\pm$0.001    & 13.024$\pm$0.001   \\   %ID 5 in the original table
  3        &  210.1103973   &  -59.5492859       & 14.035$\pm$0.001    & 12.383$\pm$0.002    \\   %ID 6 in the original table  
  4        &  210.0230560   &  -59.6616936       & 16.398$\pm$0.007     & 14.996$\pm$0.004    \\  %ID 104 in the original table
  5        &  210.3296967   &  -59.5831680       & 17.027$\pm$0.010     & 15.274$\pm$0.006    \\  %ID 175 in the original table
$-$      &        $-$             &        $-$                 &          $-$                    &            $-$                   \\
\hline 
\end{tabular}
\end{minipage}
\end{table}

\begin{figure}
\begin{center}
 \includegraphics[width=8.5cm]{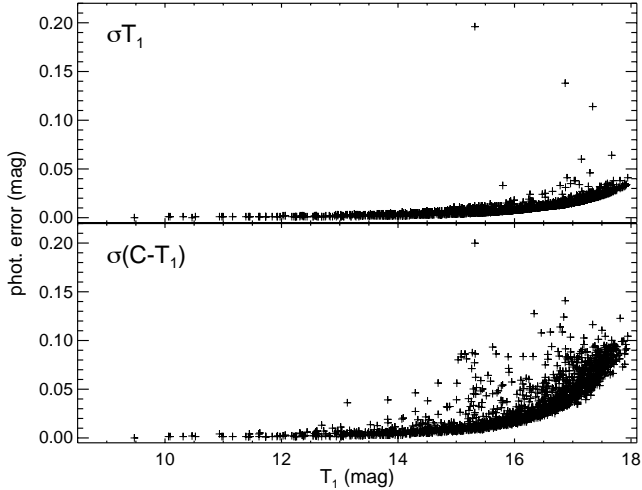}
 \caption{Photometric errors as a function of the $T_1$ mag for stars in the field of NGC\,5381, which
are typical in our photometric catalogues.}
   \label{photerrors_C_T1_NGC5381}
\end{center}
\end{figure}

Finally, the above equations were inverted in order to convert instrumental magnitudes to the standard system and obtain photometric uncertainties properly propagated into the final magnitudes, according to the STARFINDER algorithm. This step was implemented via the {\fontfamily{ptm}\selectfont IRAF INVERTFIT} task. For each OC, the photometric catalogue consists of an identifier for each star (ID), equatorial coordinates ($\rmn{RA}$ and $\rmn{DEC}$), magnitudes in filters $C$ and $T_1$ with their respective photometric uncertainties. An excerpt of the final table for the OC NGC\,5381 is presented in Table \ref{excerpt_NGC5381}. Our typical uncertainties are illustrated in Figure \ref{photerrors_C_T1_NGC5381}. We also derived the completeness level of our photometry at different magnitudes. To accomplish this, we performed artificial star tests on our images. The detailed procedure and results are described in section 2 of \cite{Angelo:2018}.

\section{Method}
\label{method}

\subsection{Structural parameters determination}
\label{center_RDPs_struct_params}

The first step in our analysis was to determine the OCs' central coordinates, simultaneously with
their core ($r_c$) and tidal ($r_t$) radii. In each case, we used 
a uniformly spaced square grid (steps of 0.25\,arcmin), centred on the literature coordinates and with full
extension equal to $\sim$2$-$4 times the informed limiting radius. These grids typically contained $\sim$200$-$400 square cells. We used each of these cells as a putative centre and built radial density profiles (RDPs) by performing stellar counts in concentric rings with widths varying from 0.50 to 1.50\,arcmin, in steps of 0.25\,arcmin. The background levels were estimated by taking into account
the average of the stellar densities corresponding to the external rings, where the stellar densities fluctuate around a nearly constant value. 

The background subtracted RDPs were then fitted using a \cite{King:1962} model:

\begin{equation}
  \sigma(r) \propto \left( \frac{1}{\sqrt{1+(r/r_c)^2}} - \frac{1}{\sqrt{1+(r_t/r_c)^2}} \right)^2
,\end{equation}

\begin{figure*}
\begin{center}

\parbox[c]{1.0\textwidth}
  {
   
    \includegraphics[width=0.333\textwidth]{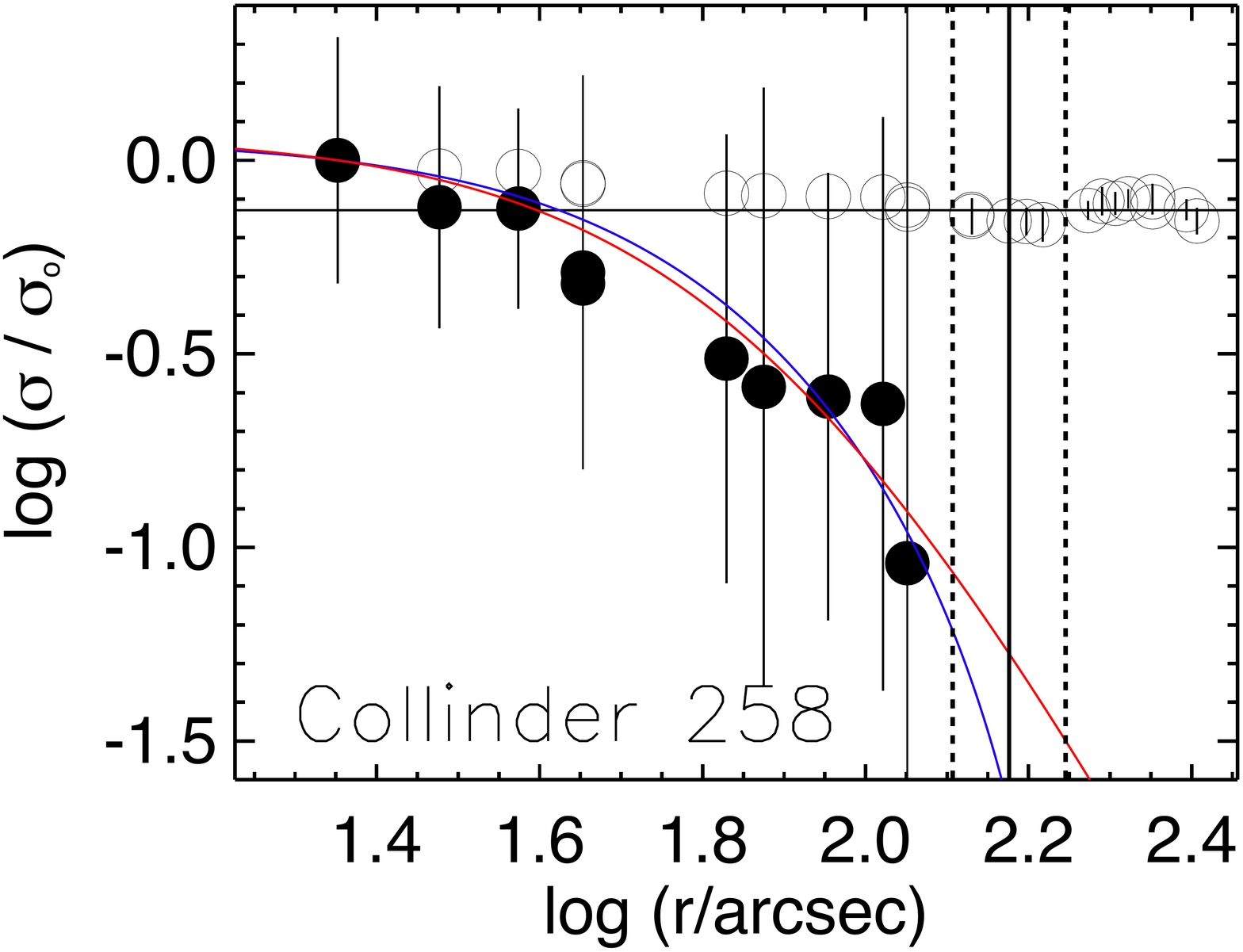}    
    \includegraphics[width=0.333\textwidth]{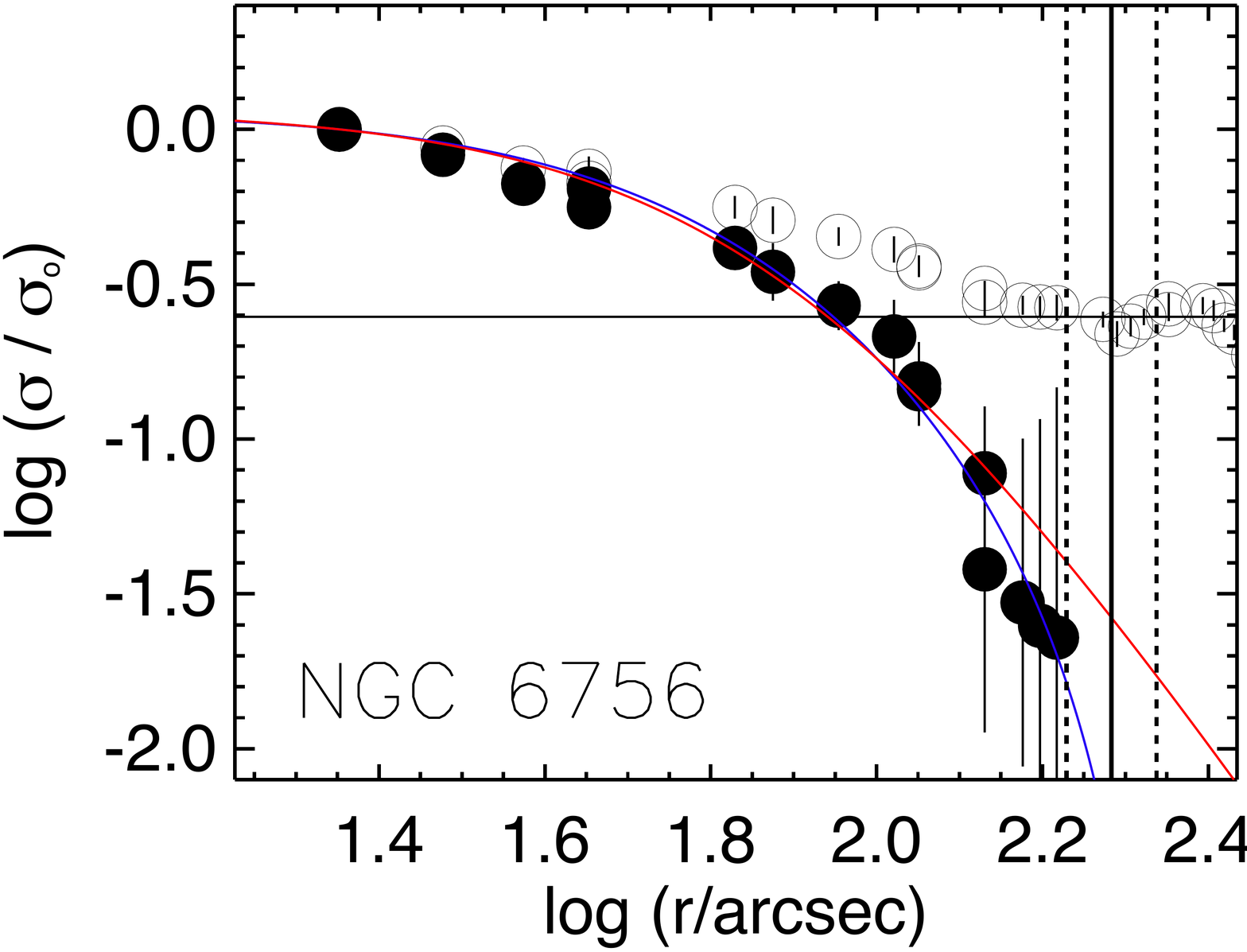}   
    \includegraphics[width=0.333\textwidth]{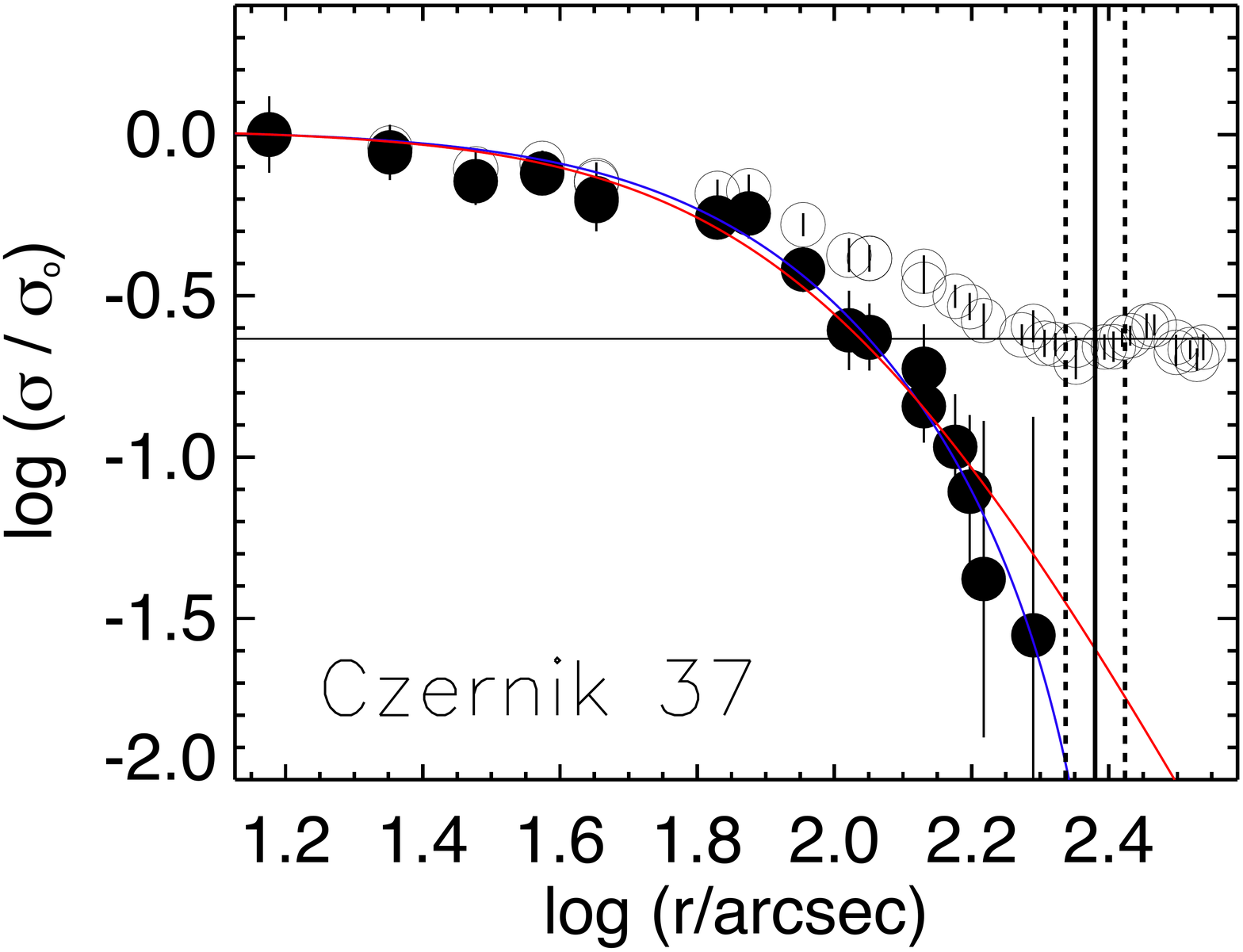}  
    \includegraphics[width=0.333\textwidth]{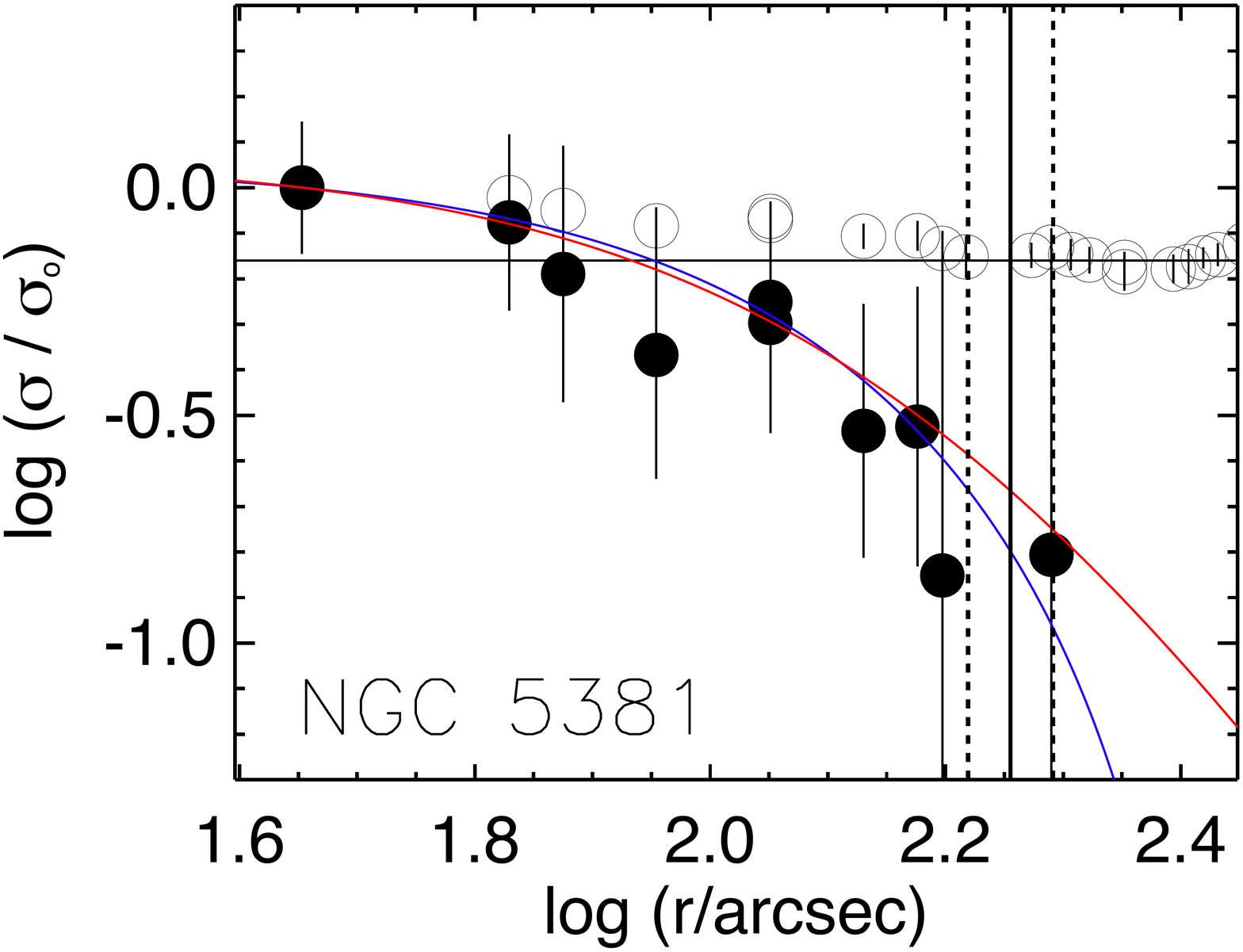}    
    \includegraphics[width=0.333\textwidth]{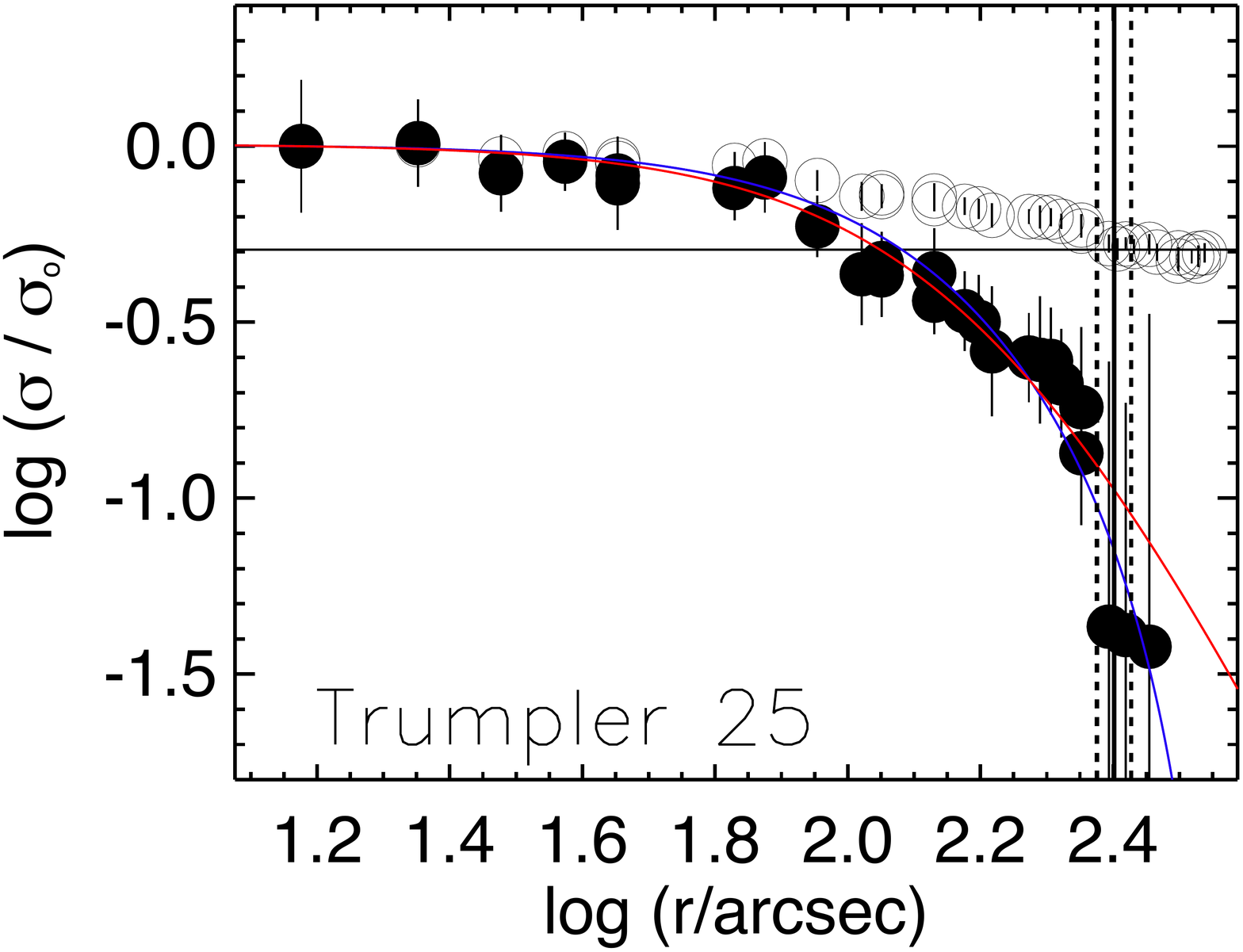}   
    \includegraphics[width=0.333\textwidth]{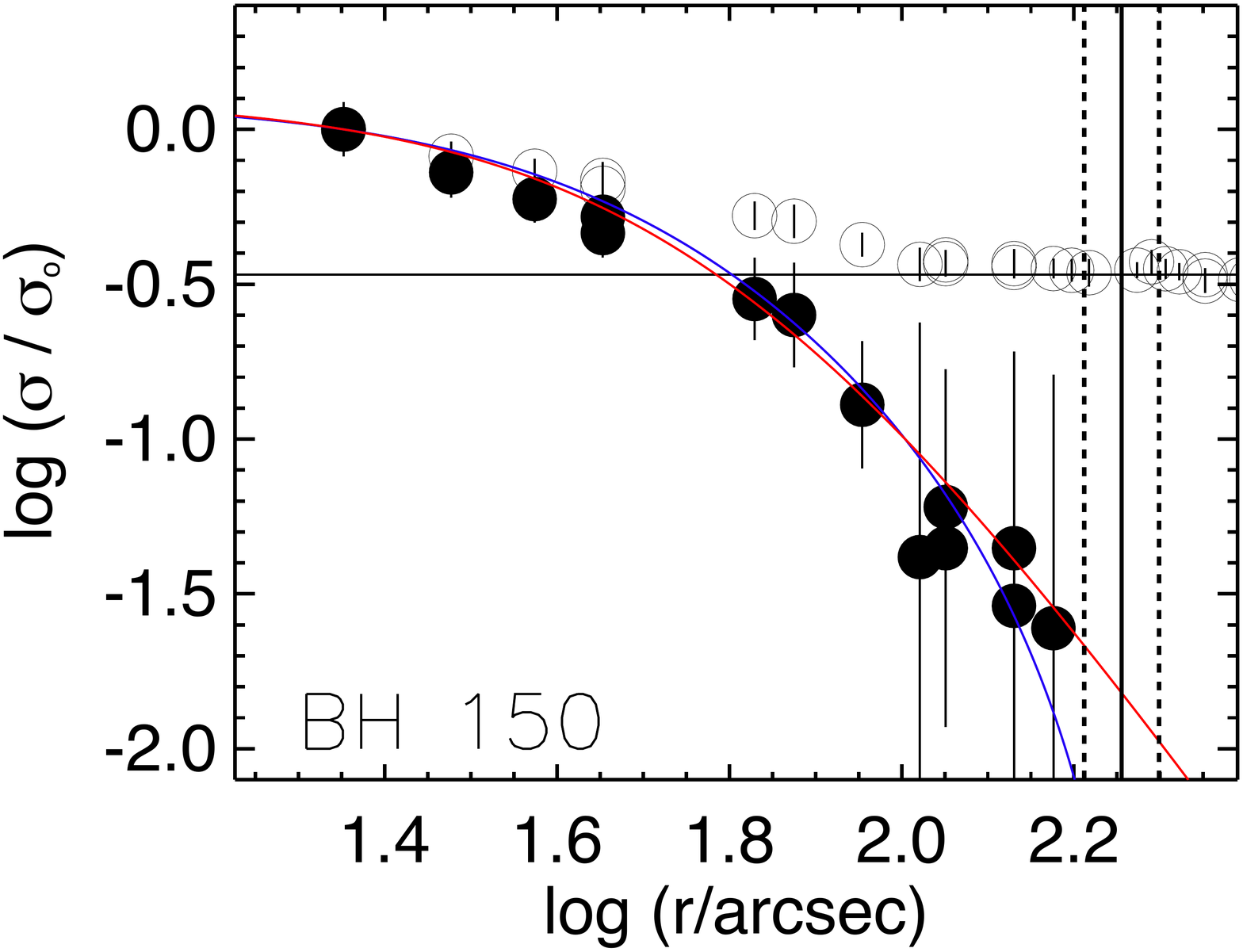}  
    \includegraphics[width=0.333\textwidth]{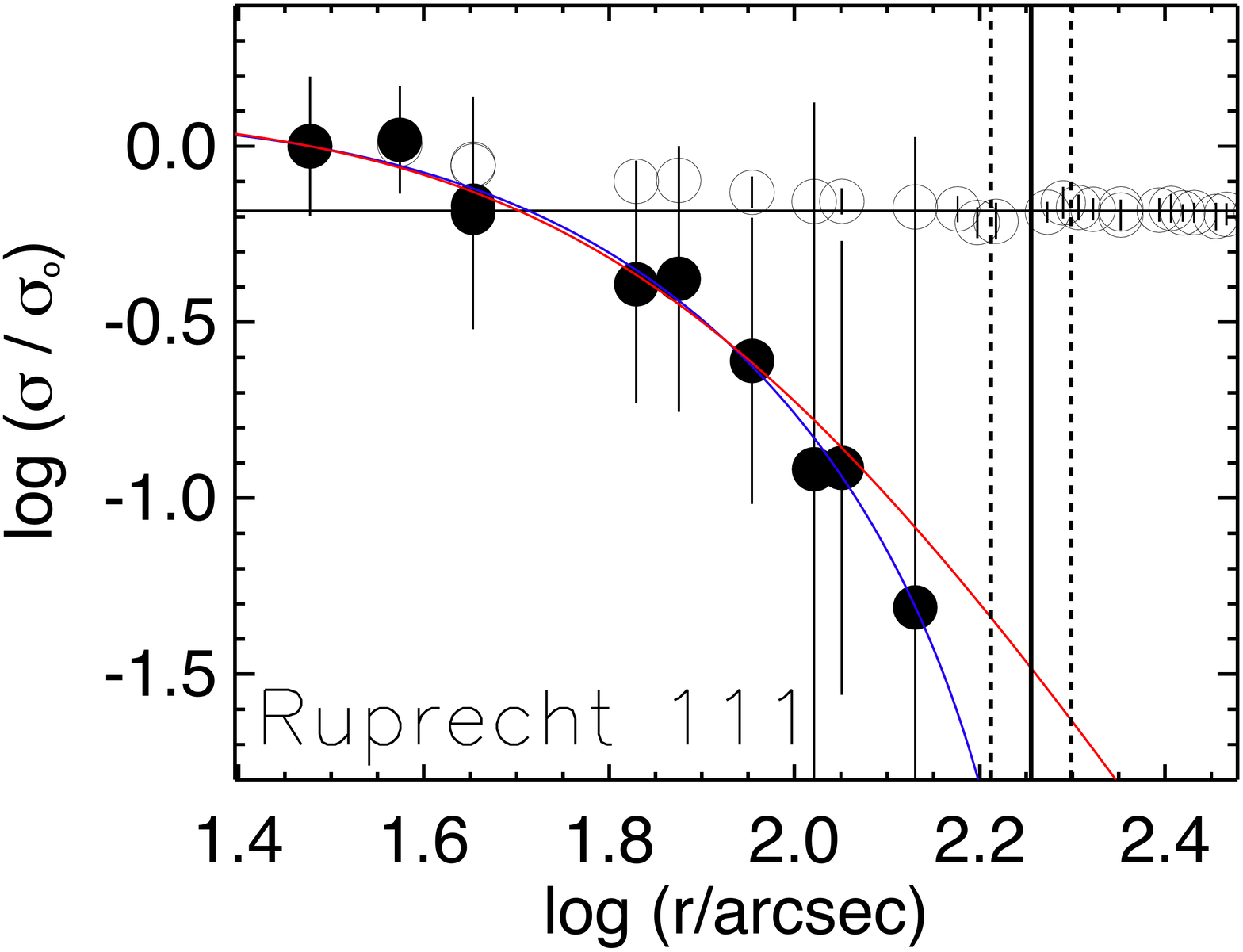}
    \includegraphics[width=0.333\textwidth]{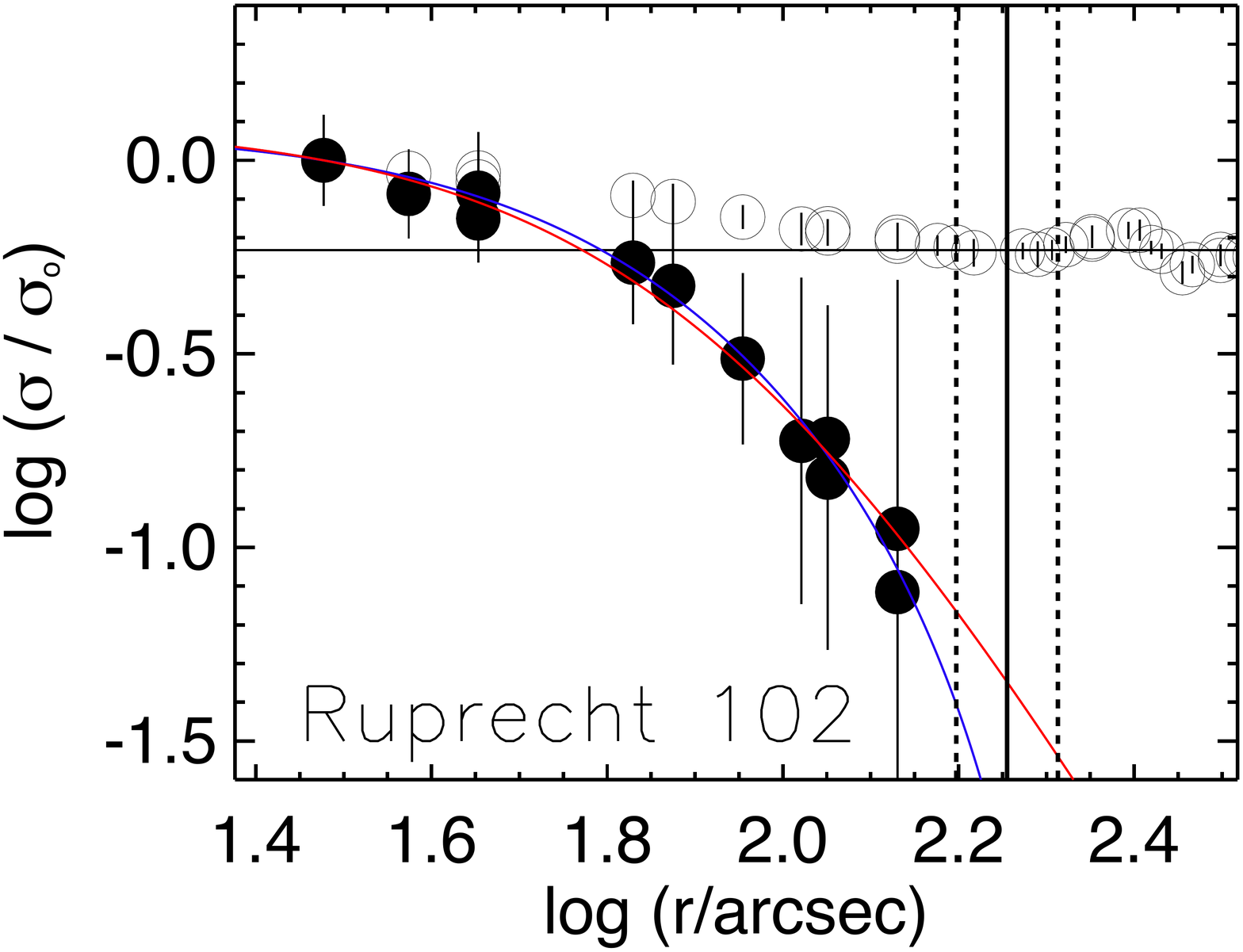}
    \includegraphics[width=0.333\textwidth]{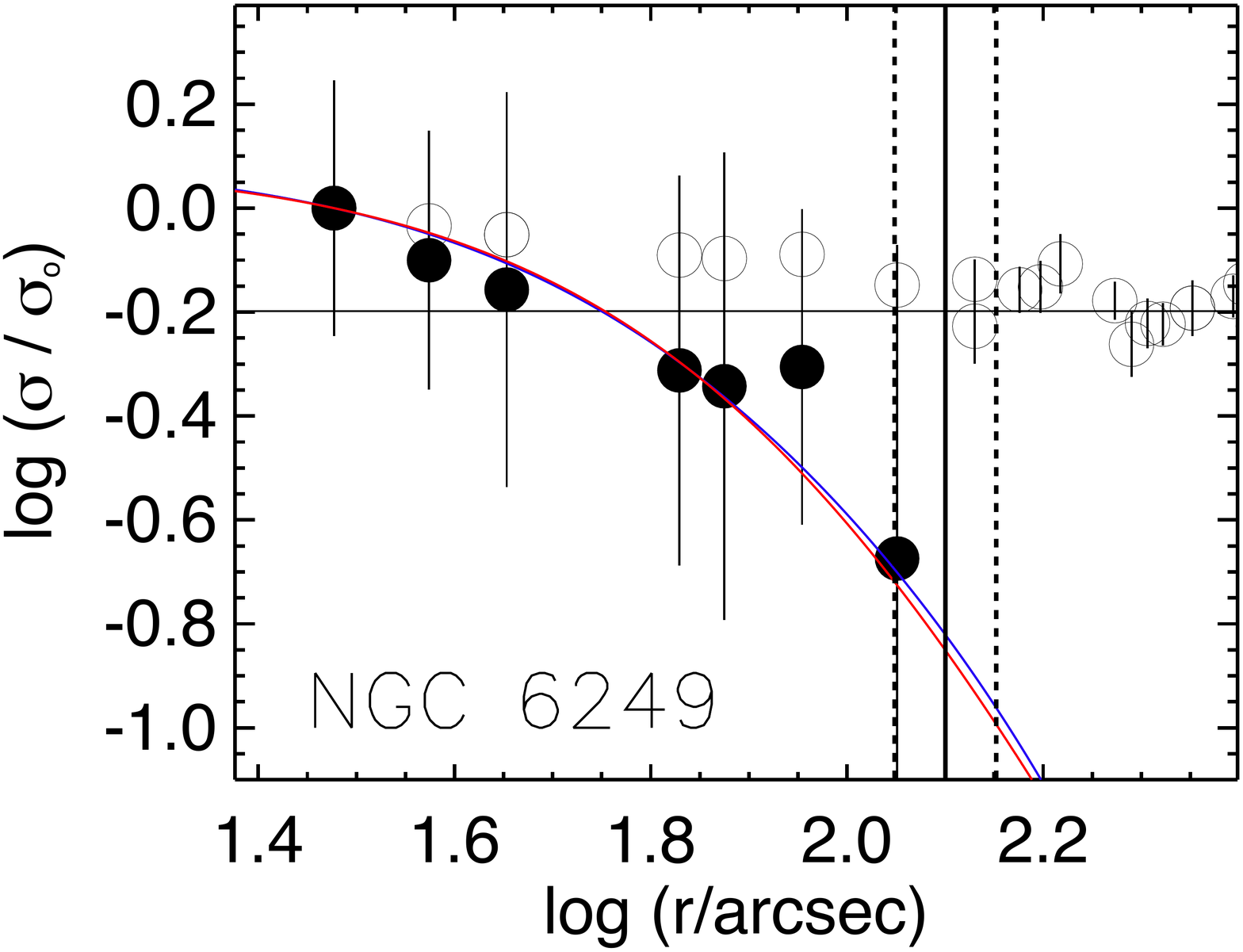}
    \includegraphics[width=0.333\textwidth]{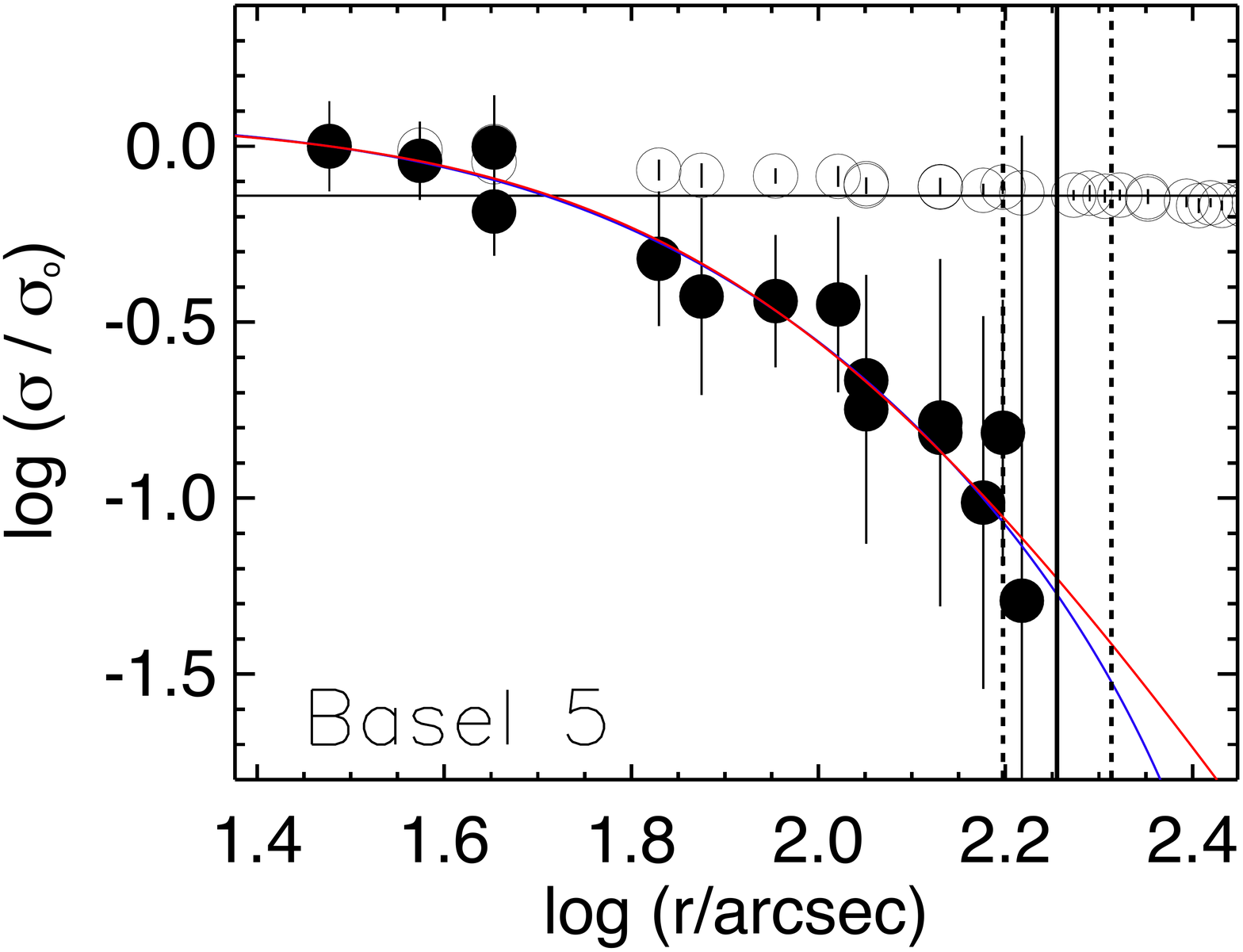}
    \includegraphics[width=0.333\textwidth]{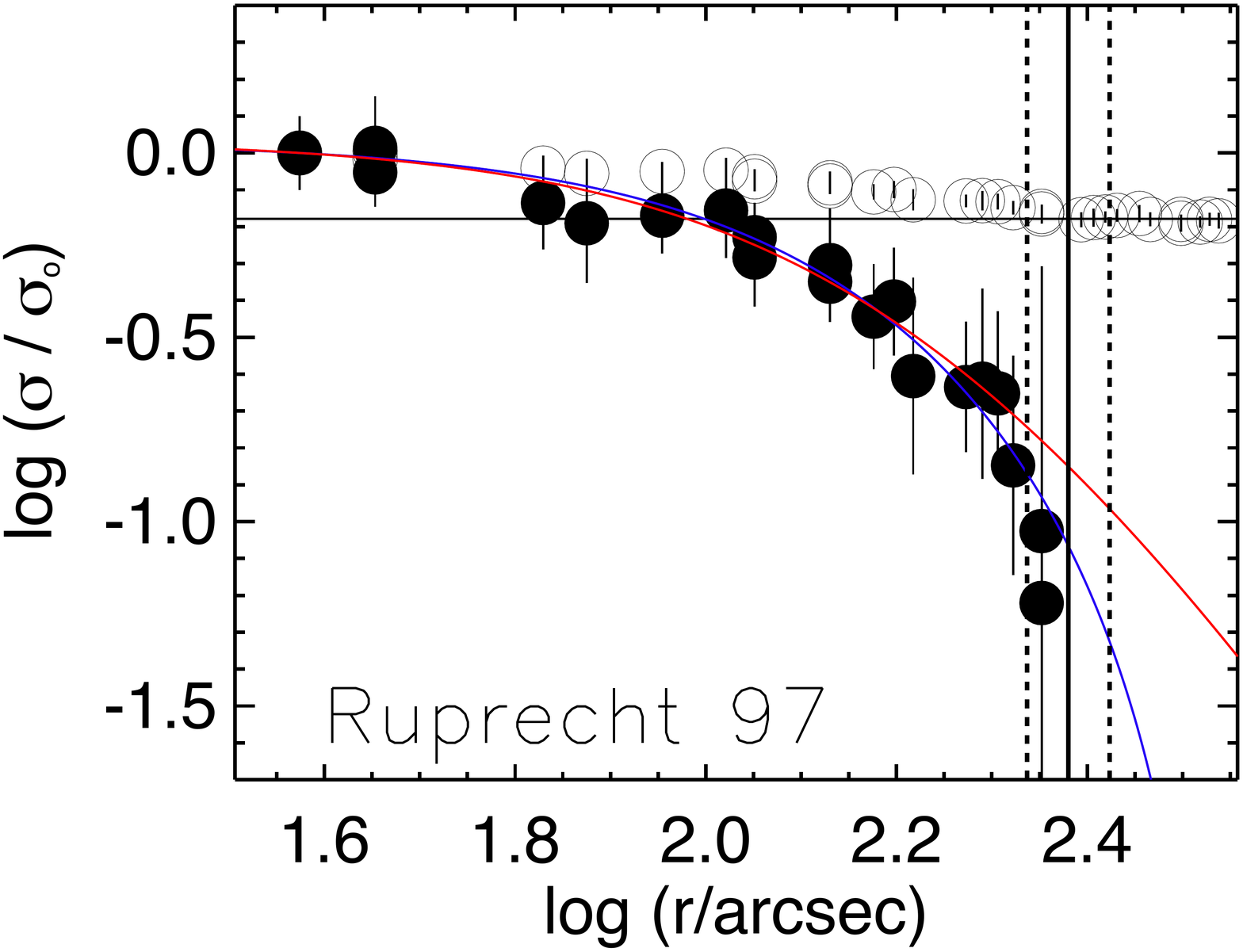}
    \includegraphics[width=0.333\textwidth]{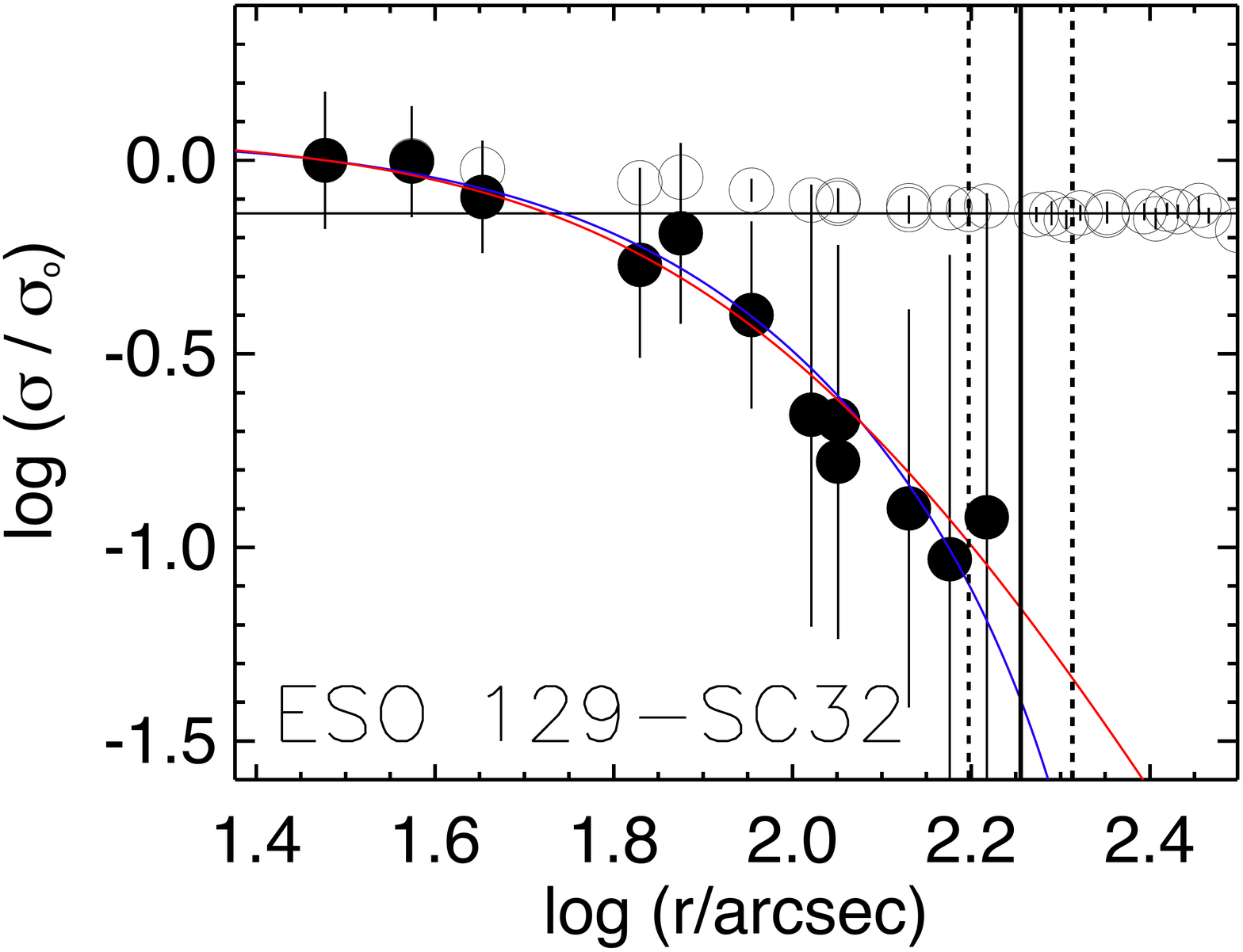}

  }
\caption{ Normalized RDPs before and after background subtraction drawn with open and filled symbols, respectively.
Poisson error bars are showed. The vertical continous and dotted lines represent the OC limiting radius and its uncertainty, respectively. The horizontal continuous line represents the mean backround density. The blue and red curves represent the fitted King\,(1962) and Plummer\,(1911) profiles, respectively.}

\label{RPDs_parte1}
\end{center}
\end{figure*}

%\noindent where $r_t$ and $r_c$ are the core and tidal radii, respectively. 
\noindent
A grid of $r_t$ and $r_c$ values spanning the whole range of radii \citep{Piskunov:2007} was employed and we searched for the values that minimized $\chi^2$. The final centres were assumed as the coordinates that produce the smoothest stellar RDPs and, at the same time, the highest density in the innermost region. This procedure is analogous to that employed by \cite{Bica:2011} in their study of very field-contaminated OCs. The best 
King's model fits are plotted in Fig.~\ref{RPDs_parte1} with blue lines, while the corresponding $r_t$ and $r_c$ values (converted to pc according to the distance moduli; see Section \ref{members_selection}) are 
showed in Table  \ref{struct_params}. The determined central coordinates are also showed in the same table. Additionally, we fitted a \cite{Plummer:1911} profile to each RDP, for comparison purposes:

\begin{equation}
  \sigma(r) \propto \frac{1}{\left[1+(r/a)^2\right]^2}
.\end{equation}    

\noindent
These fits were represented in Fig.~\ref{RPDs_parte1} with red lines. As can be seen, that both profiles are nearly indistinguishable in the inner OCs' region ($r\lesssim r_c$). The $a$ parameter is the Plummer radius which is related to the half-light radius $r_{\textrm{h}}$ by the relation $r_{\textrm{h}}\sim1.3a$.

\begin{table*}
% \scriptsize
% \hskip-2.0cm
 %\begin{minipage}{85mm}
  \caption{  Determined central coordinates, Galactocentric distances and structural parameters of the studied OCs.  }
  \label{struct_params}
 \begin{tabular}{lccccccc}
 
\hline

 Cluster   &$\rmn{RA}$  &$\rmn{DEC}$  & R$_{\textrm{GC}}^{*}$    & $r_c$     & $r_{h}^{\dag}$     & $r_t$     &  R$_J^{\dag\dag}$     \\
               &($\rmn{h}$:$\rmn{m}$:$\rmn{s}$) & ($\degr$:$\arcmin$:$\arcsec$)     &  (kpc)       &    (pc)     &     (pc)      &  (pc)      &  (pc)                            \\                                                

\hline

Collinder\,258      & 12:27:17   & -60:46:45     &7.4\,$\pm$\,0.5    &0.50\,$\pm$\,0.19    &0.65\,$\pm$\,0.12    &1.23\,$\pm$\,0.46     & 4.77\,$\pm$\,0.44  \\ 
NGC\,6756           & 19:08:44   &  04:42:53     &6.6\,$\pm$\,0.5    &0.68\,$\pm$\,0.14    &0.99\,$\pm$\,0.15    &2.10\,$\pm$\,0.40     & 7.73\,$\pm$\,0.65  \\
Czernik\,37         & 17:53:15   & -27:22:53     &6.5\,$\pm$\,0.6    &0.72\,$\pm$\,0.16    &1.01\,$\pm$\,0.15    &2.03\,$\pm$\,0.41     & 7.12\,$\pm$\,0.66  \\
NGC\,5381           & 14:00:45   & -59:35:12     &6.8\,$\pm$\,0.5    &1.88\,$\pm$\,0.49    &2.05\,$\pm$\,0.40    &2.86\,$\pm$\,0.61     & 7.33\,$\pm$\,0.59  \\
Trumpler\,25        & 17:24:29   & -39:00:48     &6.3\,$\pm$\,0.6    &1.57\,$\pm$\,0.30    &1.91\,$\pm$\,0.20    &3.13\,$\pm$\,0.51     &10.04\,$\pm$\,0.89  \\
BH\,150             & 13:38:04   & -63:20:27     &6.6\,$\pm$\,0.5    &0.79\,$\pm$\,0.18    &1.20\,$\pm$\,0.17    &2.90\,$\pm$\,0.70     & 7.82\,$\pm$\,0.69  \\
Ruprecht\,111       & 14:36:04   & -59:59:21     &6.8\,$\pm$\,0.5    &0.67\,$\pm$\,0.22    &0.94\,$\pm$\,0.22    &1.83\,$\pm$\,0.44     & 5.40\,$\pm$\,0.47  \\
Ruprecht\,102       & 12:13:37   & -62:42:55     &7.1\,$\pm$\,0.6    &1.29\,$\pm$\,0.37    &1.73\,$\pm$\,0.24    &3.40\,$\pm$\,0.83     & 6.37\,$\pm$\,0.55  \\
NGC\,6249           & 16:57:38   & -44:48:15     &6.9\,$\pm$\,0.5    &0.37\,$\pm$\,0.10    &0.65\,$\pm$\,0.13    &2.00\,$\pm$\,0.67     & 4.94\,$\pm$\,0.44  \\
Basel\,5            & 17:52:27   & -30:05:32     &6.3\,$\pm$\,0.6    &0.61\,$\pm$\,0.15    &1.05\,$\pm$\,0.20    &2.88\,$\pm$\,0.76     & 5.41\,$\pm$\,0.54  \\
Ruprecht\,97        & 11:57:35   & -62:43:20     &7.2\,$\pm$\,0.5    &3.01\,$\pm$\,0.85    &3.67\,$\pm$\,0.49    &5.55\,$\pm$\,1.22     & 8.54\,$\pm$\,0.65  \\
ESO\,129-SC32       & 11:44:06   & -61:05:56     &7.3\,$\pm$\,0.5    &1.73\,$\pm$\,0.54    &2.39\,$\pm$\,0.42    &4.65\,$\pm$\,1.08     & 7.78\,$\pm$\,0.60  \\

\hline
\multicolumn{8}{l}{ \textit{Note}: To convert 1 arcmin into pc we used the expression $(\pi\,/10800)\times10^{[(m-M)_{0}+5]/5}$, where $(m-M)_{0}$ }\\
\multicolumn{8}{l}{ is the OC distance modulus (see Table \ref{astroph_params}). } \\
\multicolumn{8}{l}{ $^{*}$ The $R_G$ were obtained from the distances in Table \ref{astroph_params}, assuming that the Sun is located at 8.0\,$\pm$\,0.5\,kpc } \\
\multicolumn{8}{l}{ from the Galactic centre \citep{Reid:1993a}. }\\
\multicolumn{8}{l}{ $^{\dag}$ Half-light radius (Section \ref{mass_functions}). } \\
\multicolumn{8}{l}{ $^{\dag\dag}$ Jacobi radius (Section \ref{mass_functions}). }

\end{tabular}
%\end{minipage}
\end{table*}

\subsection{Membership determination}
\label{memberships}

After the structural analysis, we used the Vizier service\footnote[3]{http://vizier.u-strasbg.fr/viz-bin/VizieR} to extract astrometric data from the GAIA DR2 catalogue \citep{Gaia-Collaboration:2018} for stars in a large circular area of radius 20\,arcmin centred on each target. This region is large enough to encompass completely the field of view corresponding to our Washington images. For each OC we cross-matched our photometric catalogues with GAIA and executed a routine that explores the three-dimensional (3D) parameters space of proper motions and parallaxes ($\mu_{\alpha}$, $\mu_{\delta}$, $\varpi$) corresponding to stars in the OC area ($r\lesssim r_t$) and in a control field (stars in the region $r\gtrsim r_t$). The routine is devised to detect and evaluate statistically the overdensity of OC stars in comparison to the field in each part of the parameters space. This was a critical step in our analysis, since the studied OCs are located at low Galactic latitudes ($\vert b\vert\leq2^{\circ}$) and thus projected against dense stellar fields.

The routine is completely described in \cite{Angelo:2019a}. Briefly, the procedure consists in dividing the astrometric space in cells with widths proportional to the sample mean uncertainties (\,$\langle\Delta\mu_{\alpha}\rangle$, $\langle\Delta\mu_{\delta}\rangle$ and $\langle\Delta\varpi\rangle$\,) in each astrometric parameter. Cell widths are typically 1.0\,mas.yr$^{-1}$ and 0.1\,mas for proper motion components and parallax, respectively. These values correspond to $\sim10\times\langle\Delta\mu_{\alpha}\rangle$, $\sim10\times\langle\Delta\mu_{\delta}\rangle$ and $\sim1\times\langle\Delta\varpi\rangle$. These cell sizes allow to accomodate a significant number of stars inside each cell and they are small enough to properly sample the fluctuations across the 3D space.

Inside each cell, we determined membership likelihoods for stars in the OC sample ($l_{\textrm{star}}$) by employing a multivariate gaussian:

\begin{equation}
\begin{aligned}
        l_{\textrm{star}} = \frac{\textrm{exp}\left[-\frac{1}{2}(\boldsymbol{X}-\boldsymbol{\mu})^{\textrm{T}}\boldsymbol{\sum}^{-1}(\boldsymbol{X}-\boldsymbol{\mu})\right]}{\sqrt{(2\pi)^3\vert\boldsymbol{\sum}\vert}}                                   
\end{aligned}   
\label{likelihood_formula}
,\end{equation}

\noindent
where $\boldsymbol{X}$ is the column vector ($\mu_{\alpha}$,\,$\mu_{\delta}$,\,$\varpi$) for a given star and $\boldsymbol{\mu}$ is the mean vector for the sample of OC stars contained within the cell. $\boldsymbol{\sum}$ is the full covariance matrix, which incorporates the uncertainties and the correlations between the astrometric parameters (see equation 2 of \citeauthor{Angelo:2019a}\,\,\citeyear{Angelo:2019a}). Then the same calculation was performed for stars in the control field. For each sample inside a given cell, the total likelihood was taken multiplicatively: $\mathcal{L}=\prod_{i}^{} l_i$.

In order to compare statistically the dispersion of data for both samples (OC and control field) in a given cell, we employed the objective function:

\begin{equation}
   S = -\textrm{log}\,\mathcal{L}
   \label{func_entropia}
.\end{equation}

\noindent
Then we searched for cells for which $S_{\textrm{clu}}<S_{\textrm{field}}$ and the corresponding OC stars were flagged (``1") as member candidates (stars in cells which do not satisfy this criterion were flagged as ``0"). In these cases, the ``entropy" of the parameter space for stars in the OC sample is locally  smaller than that for field stars.

The final membership likelihoods were assigned to stars inside those cells flagged as ``1" according to the relation:

\begin{equation}
   L_{\textrm{star}} \propto\,\textrm{exp}\left(-\frac{\langle N_{\textrm{clu}}\rangle}{N_{\textrm{clu}}}\right)     
   \label{termo_exponencial} 
,\end{equation}

\noindent
where $\langle N_{\textrm{clu}}\rangle$ is the average number of stars inside cells for a given grid size. This exponential factor is necessary to make sure that only stars within cells where $N_{\textrm{clu}}$ is considerably greater than $\langle N_{\textrm{clu}}\rangle$ will receive large membership likelihoods.

Finally, the cell sizes were varied by one third of their original sizes in each direction and the procedure stated above was repeated. For each star the algorithm determines 27 different likelihoods and registers the median of this set of values. The maximum likelihood (equation \ref{termo_exponencial}) for the complete sample of stars in the OC region is then normalized to unity.

\section{Results}
\label{results}

\subsection{Members selection}
\label{members_selection}

\begin{figure*}
\begin{center}

\parbox[c]{1.0\textwidth}
  {
   
    \includegraphics[width=0.333\textwidth]{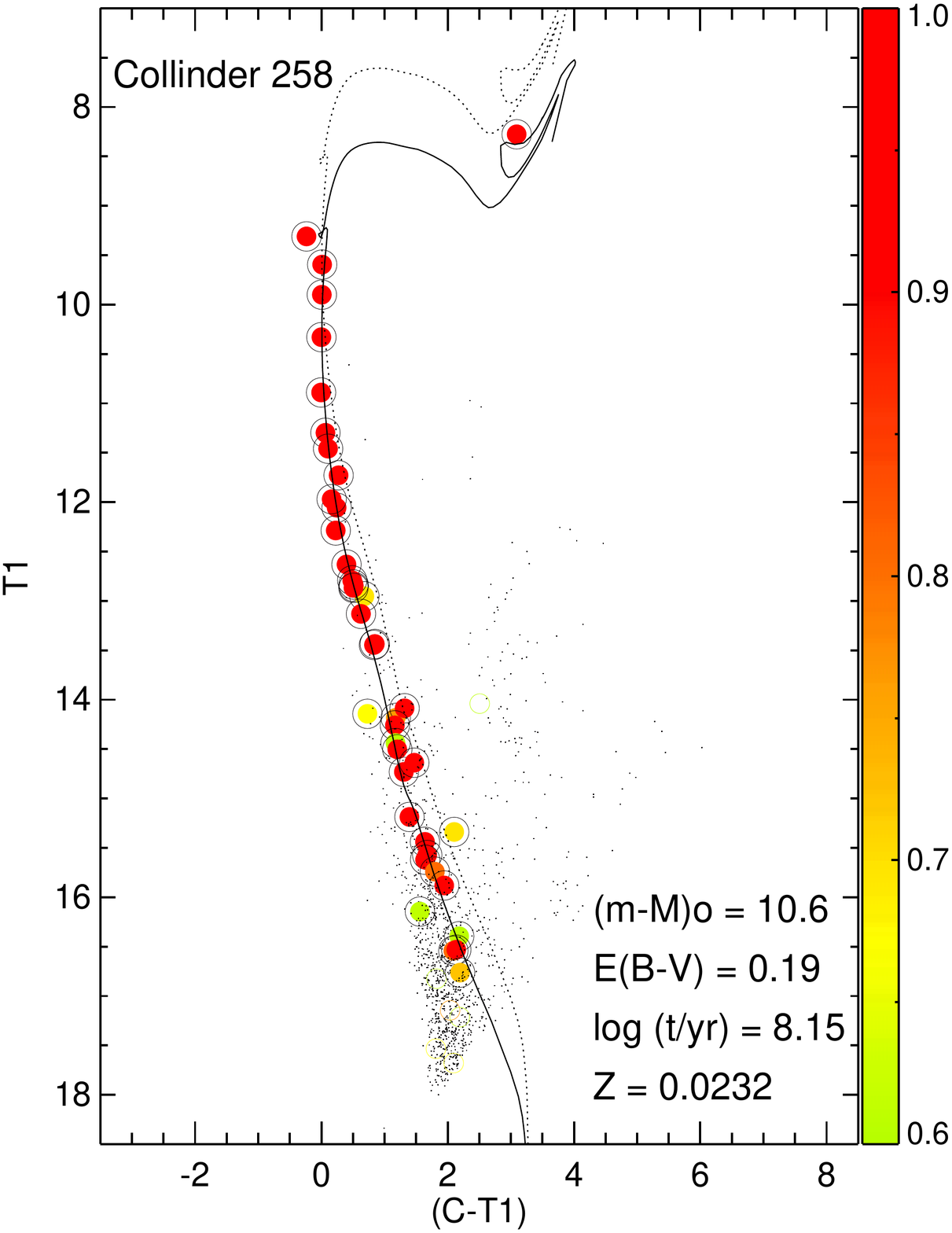}
    \includegraphics[width=0.333\textwidth]{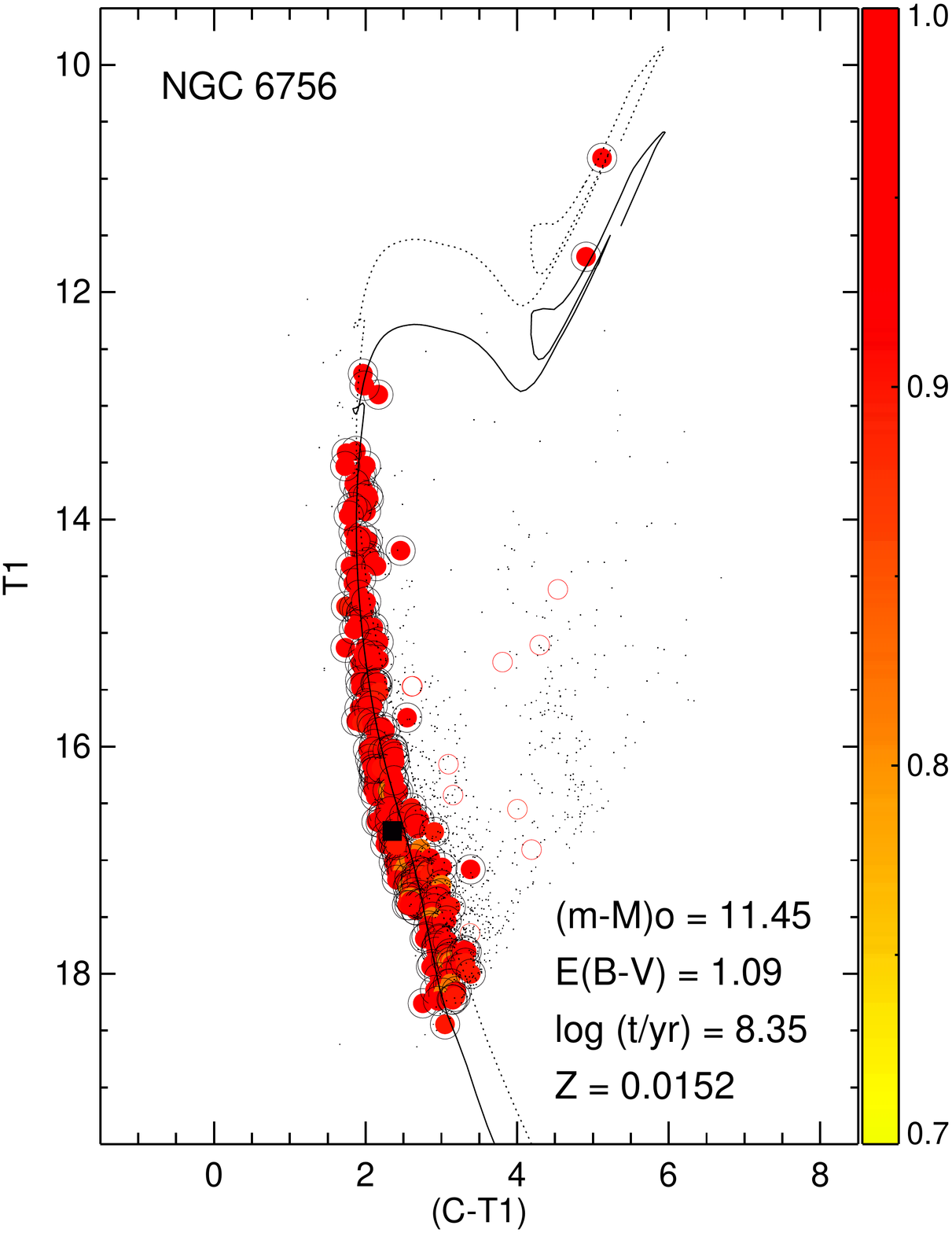}
    \includegraphics[width=0.333\textwidth]{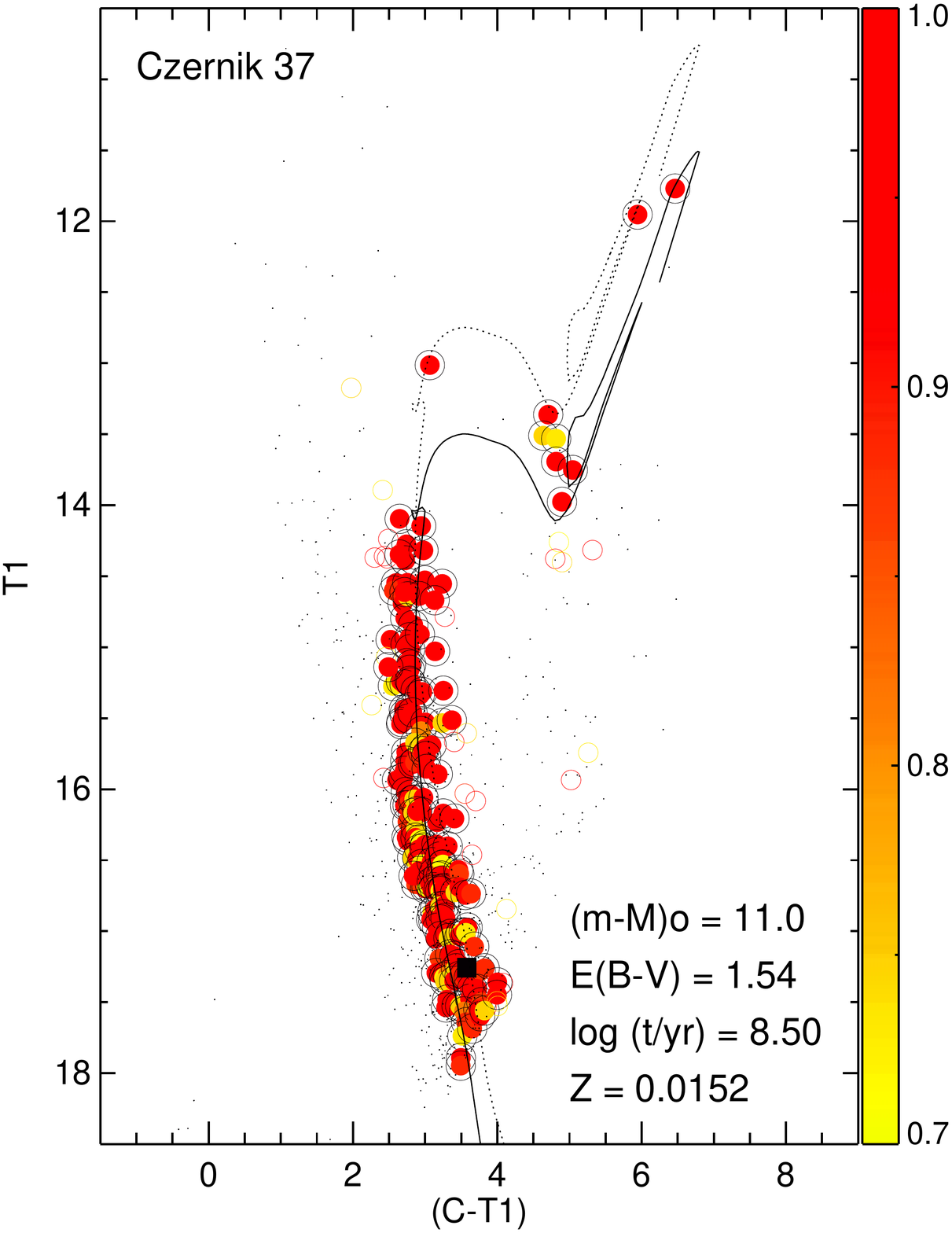}
    \includegraphics[width=0.333\textwidth]{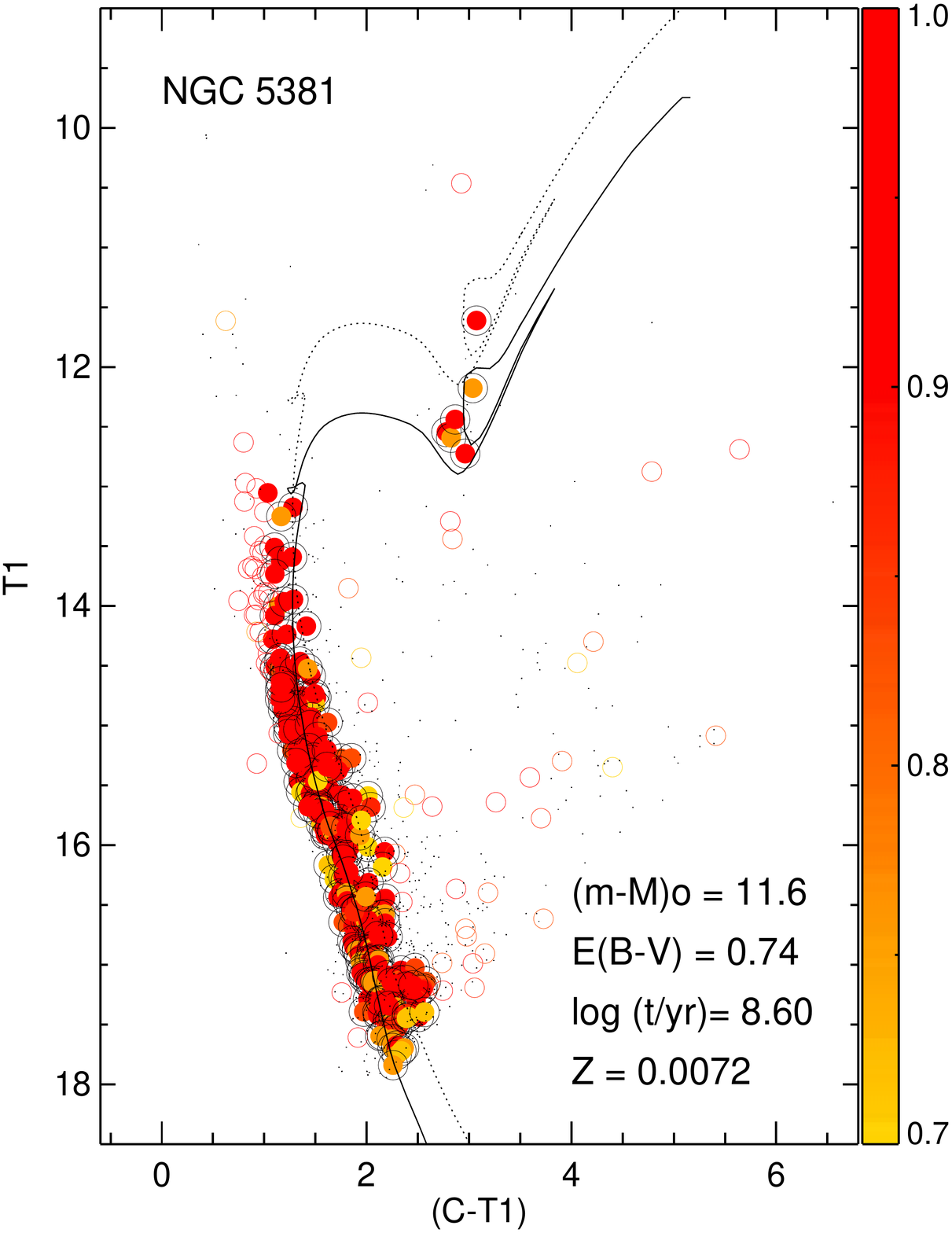}
    \includegraphics[width=0.333\textwidth]{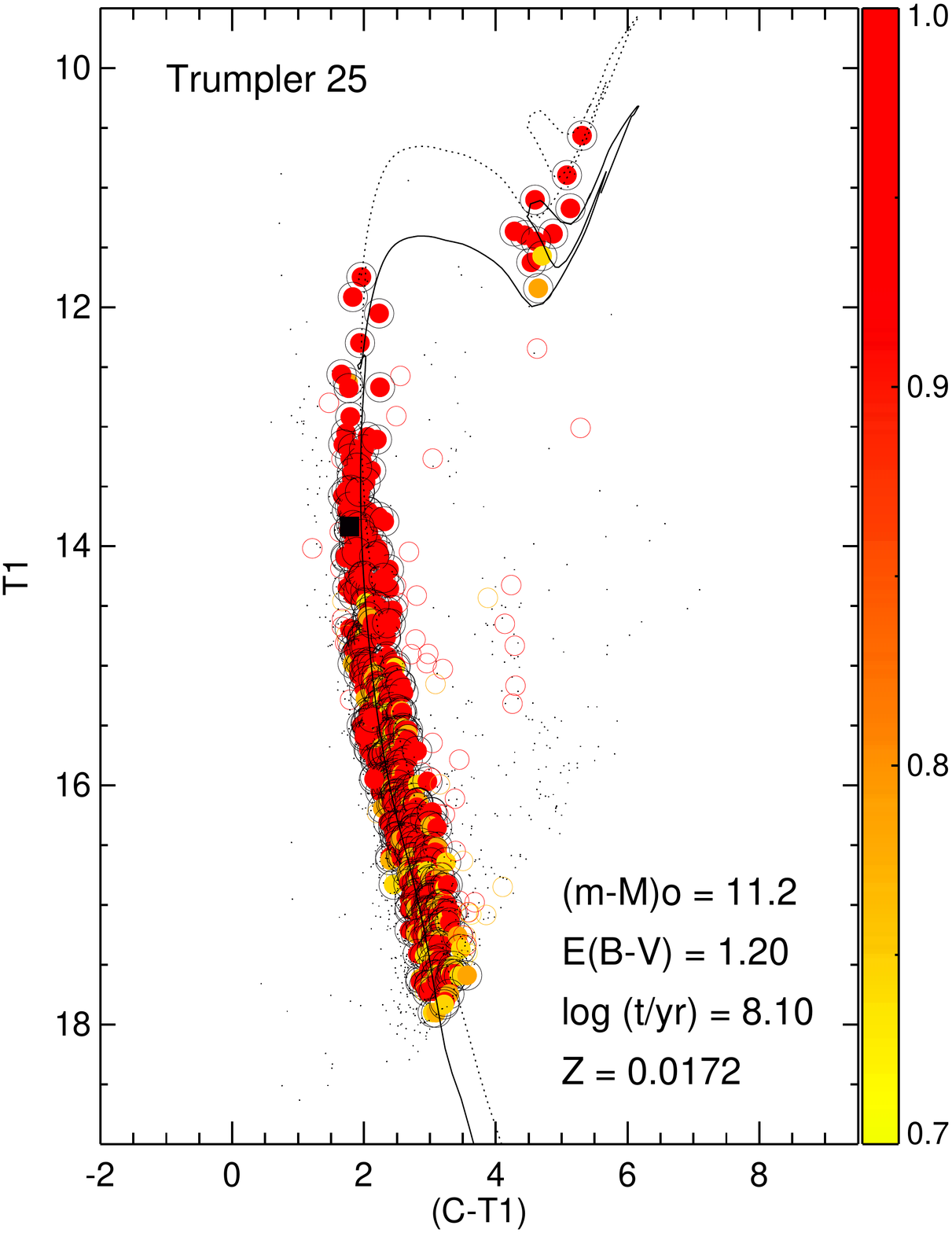}
    \includegraphics[width=0.333\textwidth]{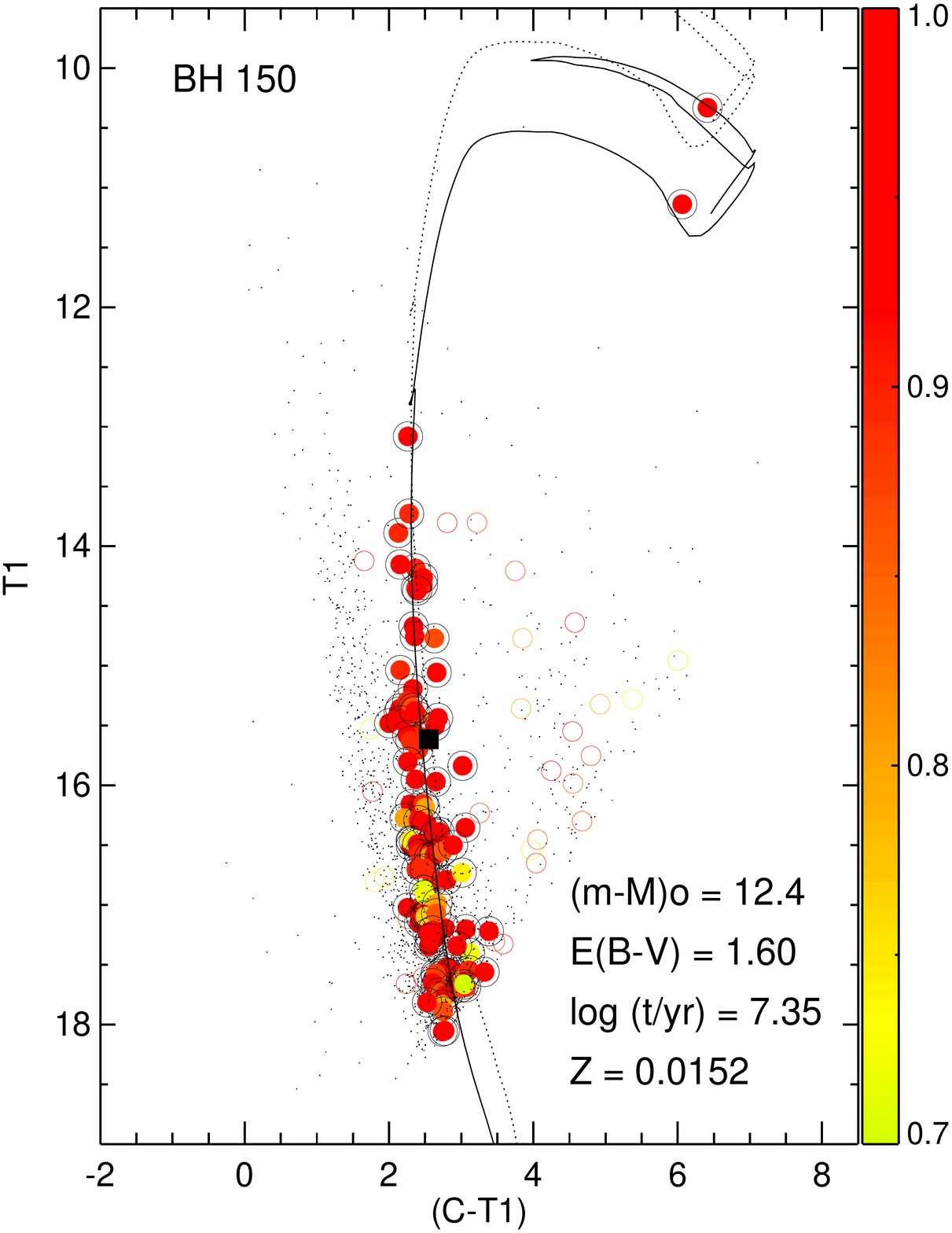}

  }
\caption{ $T_{1}\,\times\,(C-T_1)$ CMDs after applying the decontamination procedure described in Section \ref{memberships}. Filled and open symbols represent member and non-member stars, respectively. Colours were assigned according to the 
astrometric membership scale, as indicated by the colour bars. Small black dots represent stars in a control field. The continuous lines are the best-fitted PARSEC isochrones and the dotted ones are shifted by -0.75 mag in $T_1$ to match to loci of unresolved binaries with equal mass components. The basic astrophysical parameters are indicated. The filled black squares represent stars without GAIA data and with $\mathcal{L_{\textrm{phot}}}\ge0.1$ (see text for details). }

\label{CMDs_parte1}
\end{center}
\end{figure*}

\setcounter{figure}{3}
\begin{figure*}
\begin{center}

\parbox[c]{1.0\textwidth}
  {
   
    \includegraphics[width=0.333\textwidth]{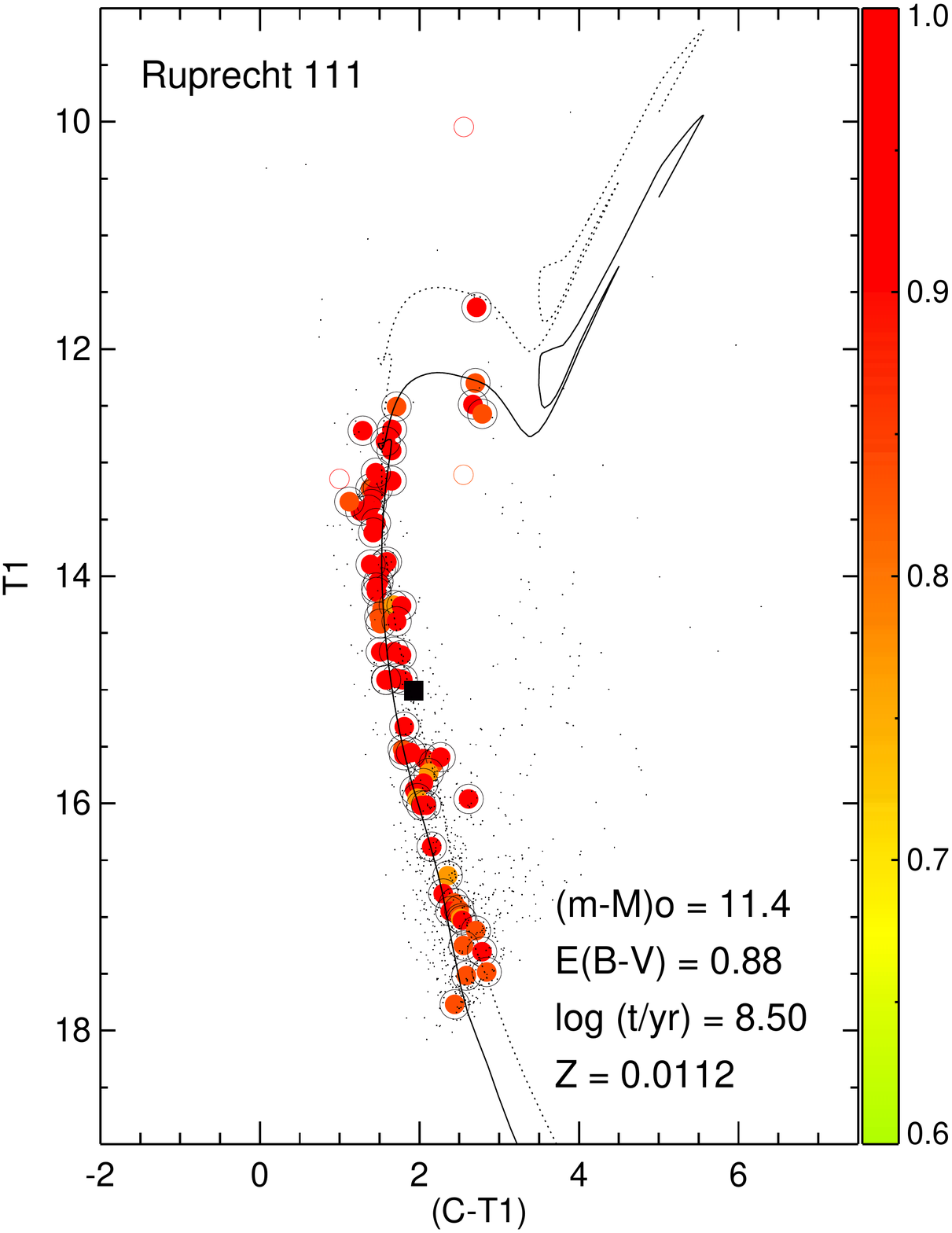}  
    \includegraphics[width=0.333\textwidth]{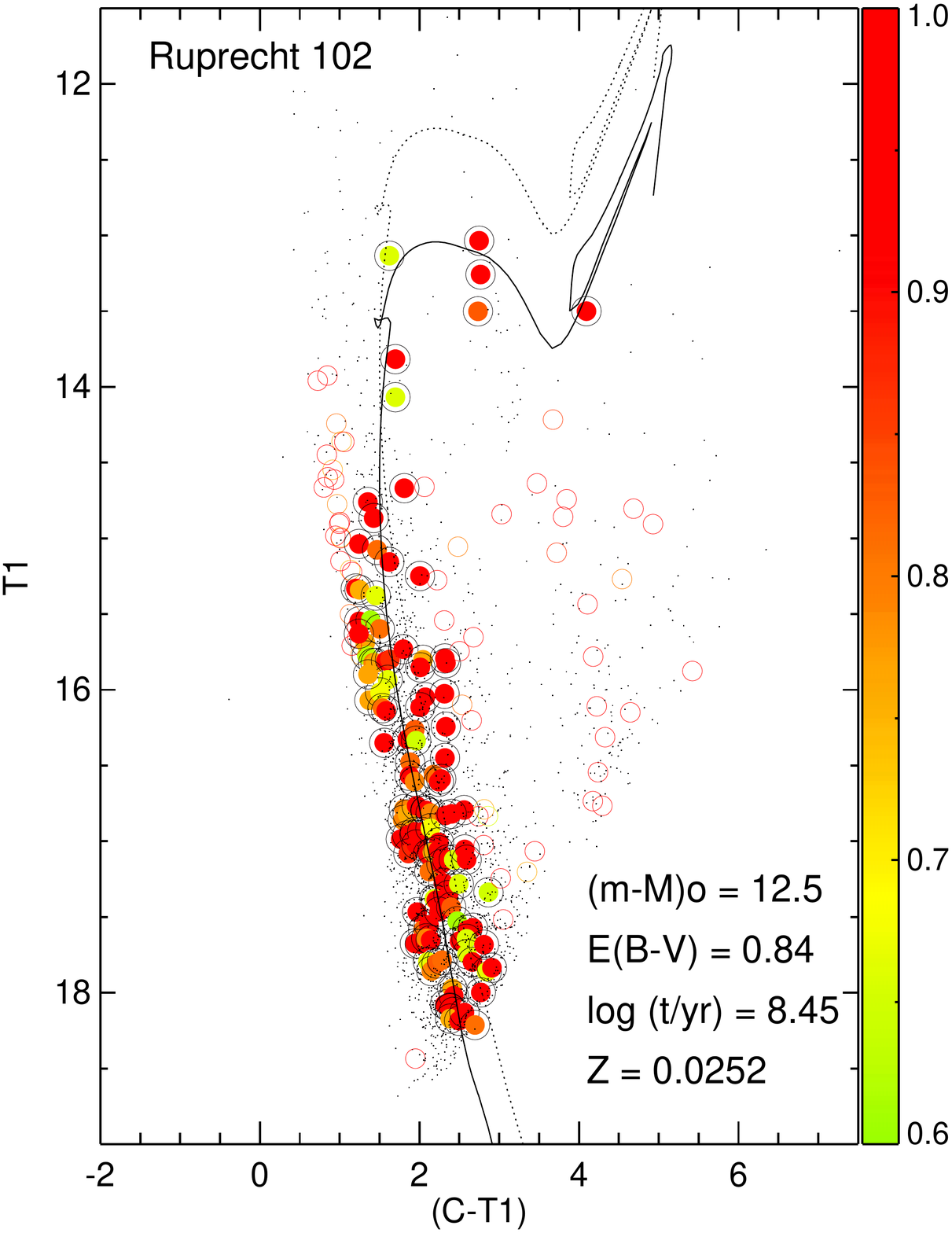} 
    \includegraphics[width=0.333\textwidth]{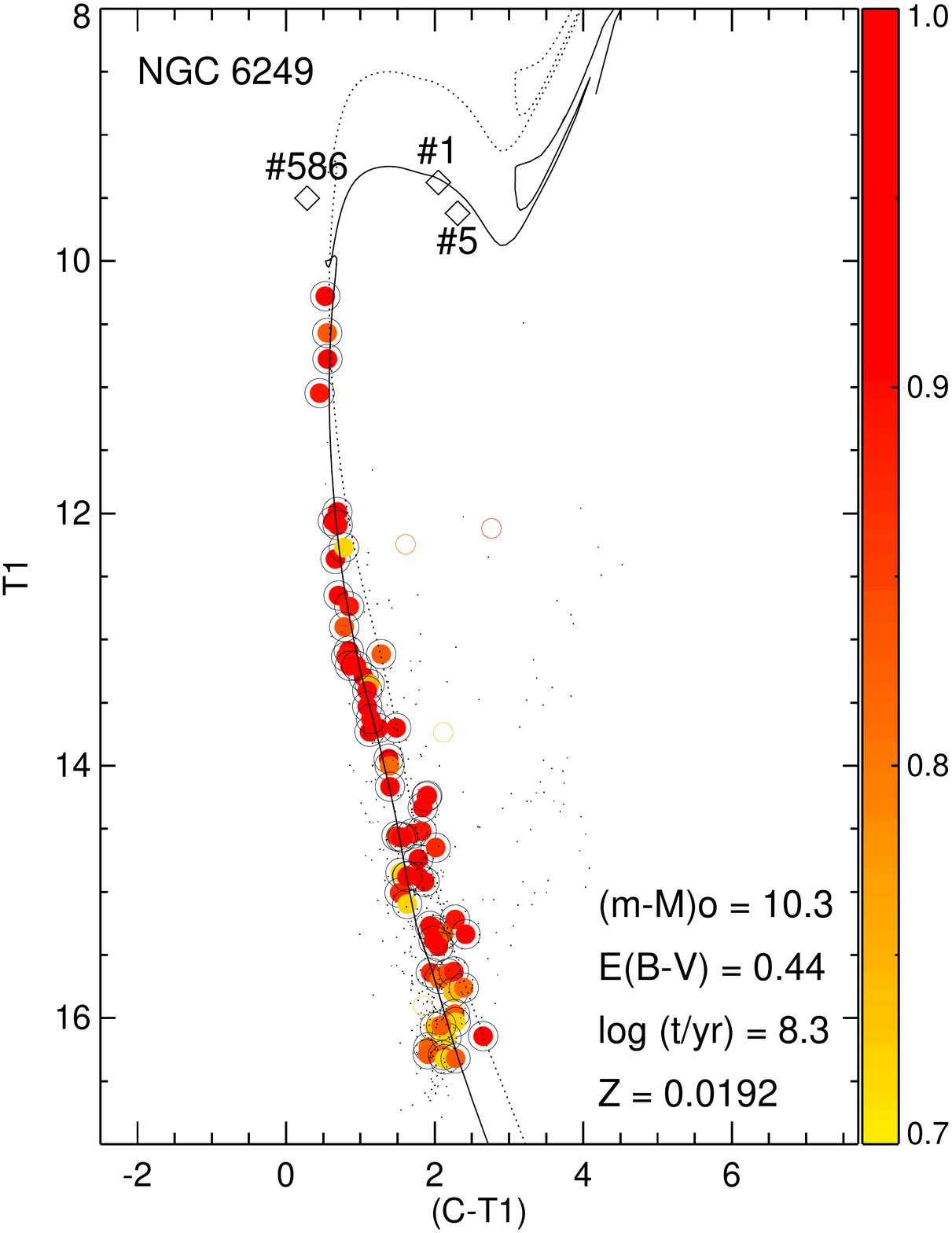}       
    \includegraphics[width=0.333\textwidth]{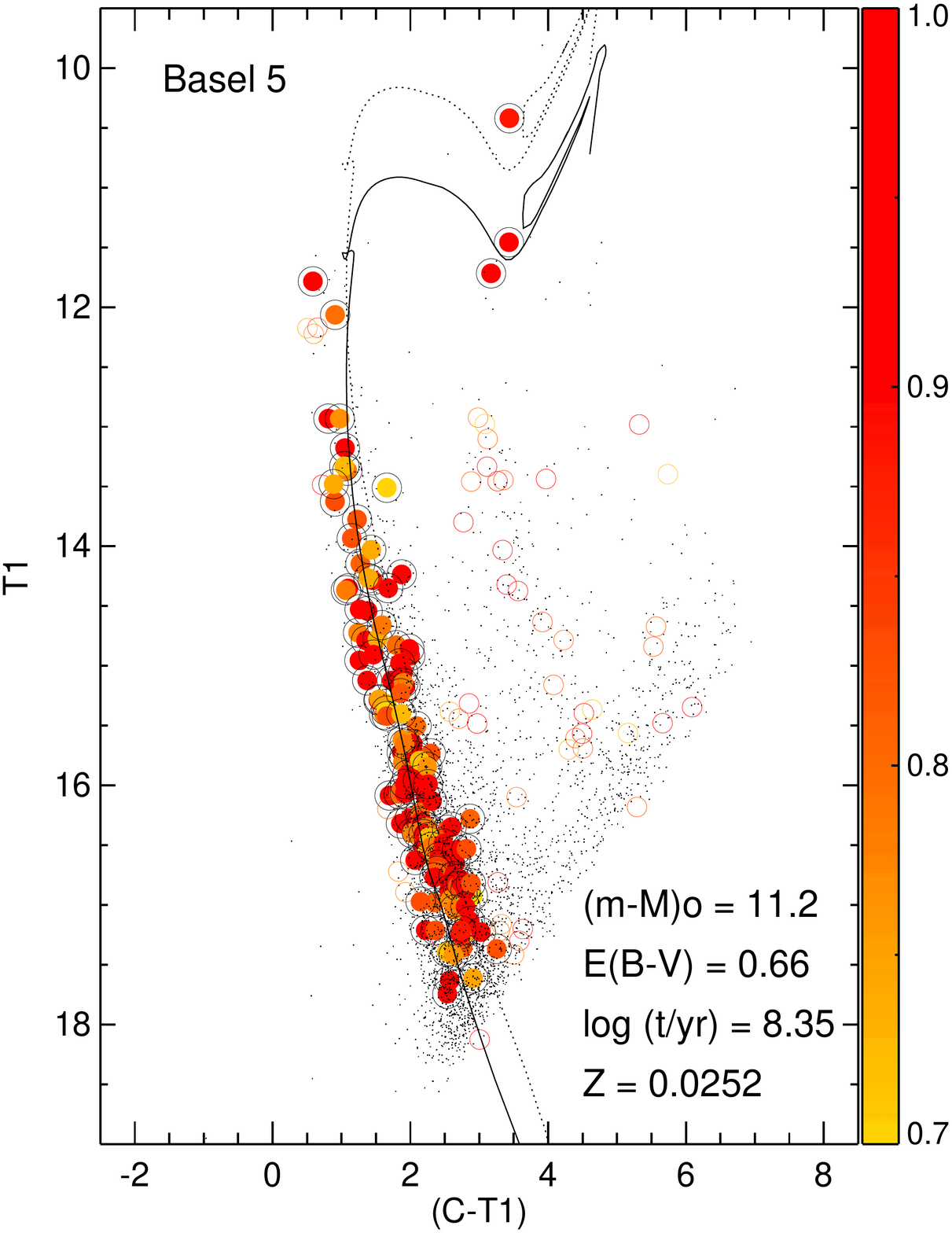}        
    \includegraphics[width=0.333\textwidth]{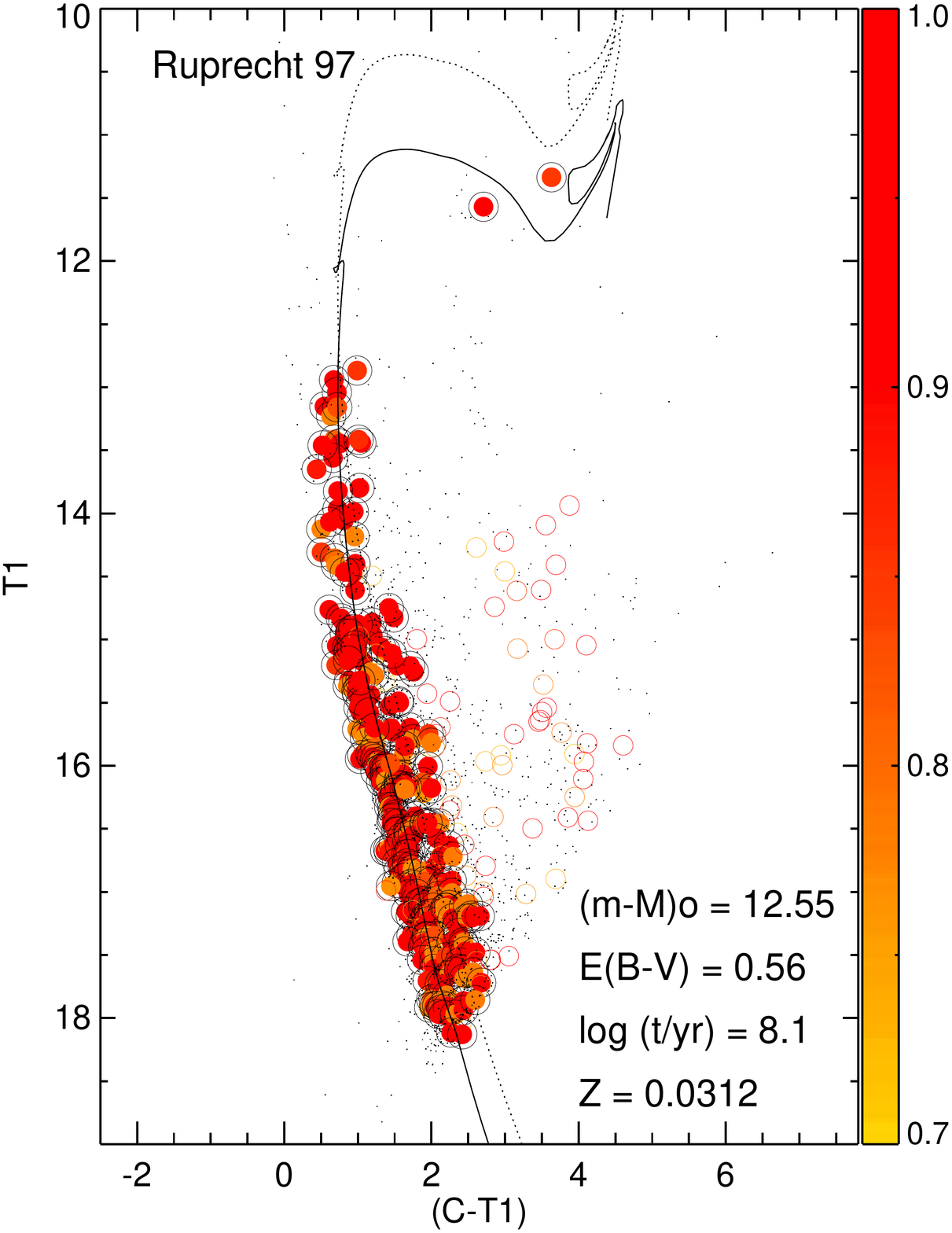}       
    \includegraphics[width=0.333\textwidth]{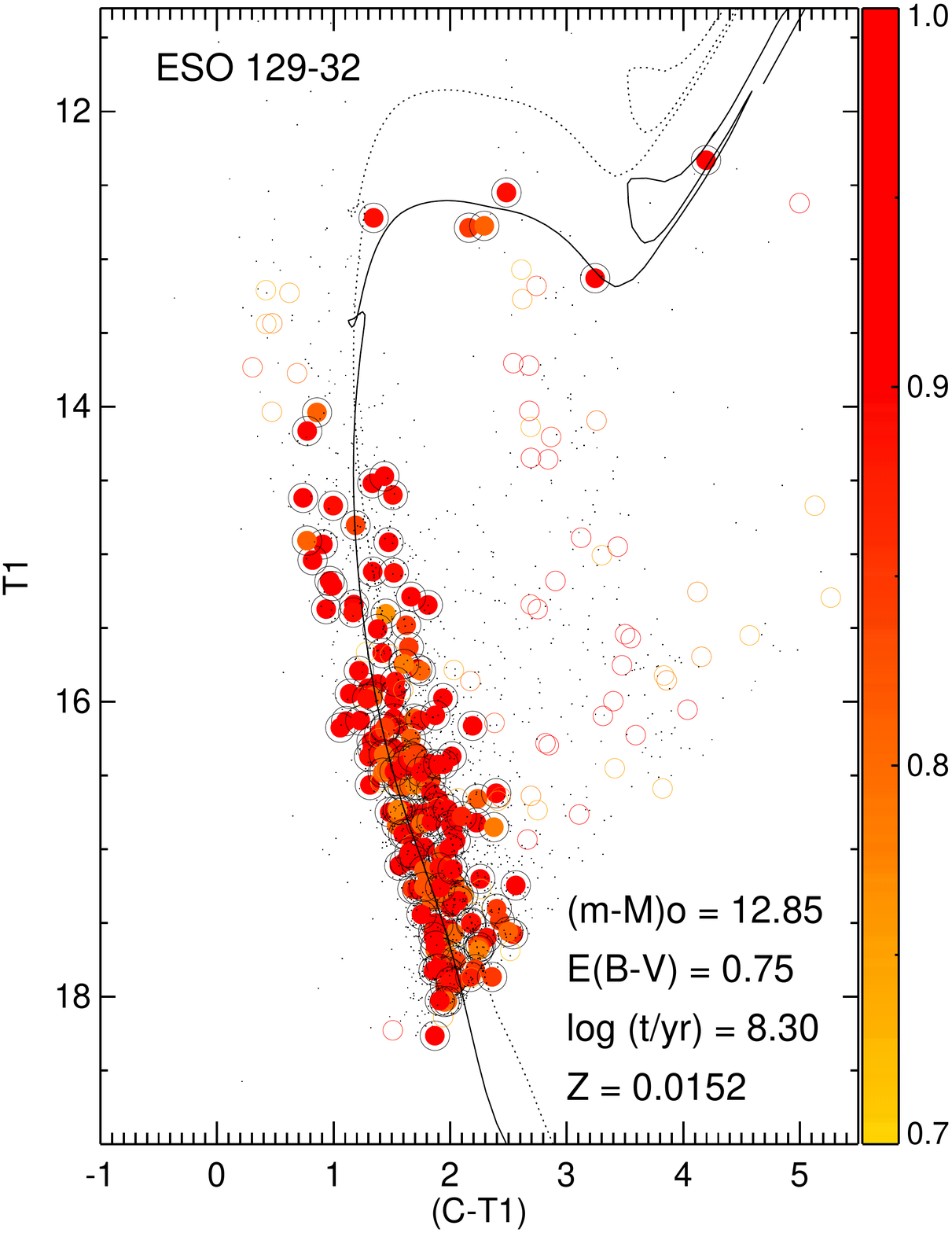}  

  }
\caption{(continued).}

\label{CMDs_parte2}
\end{center}
\end{figure*}

\begin{figure*}
\begin{center}

%\parbox[c]{1.0\textwidth}
%  {   
    \includegraphics[width=0.31\textwidth]{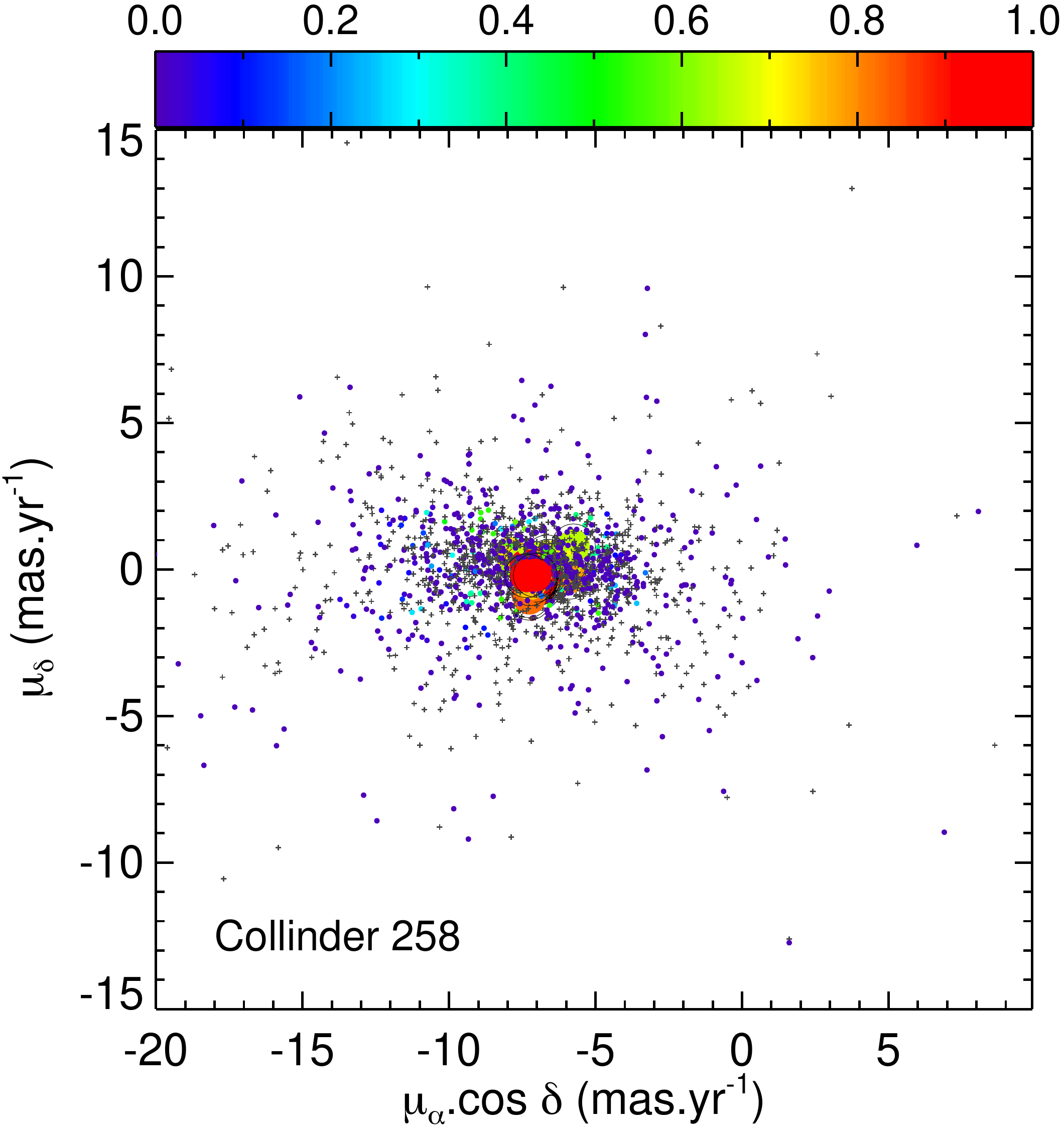}
    \includegraphics[width=0.31\textwidth]{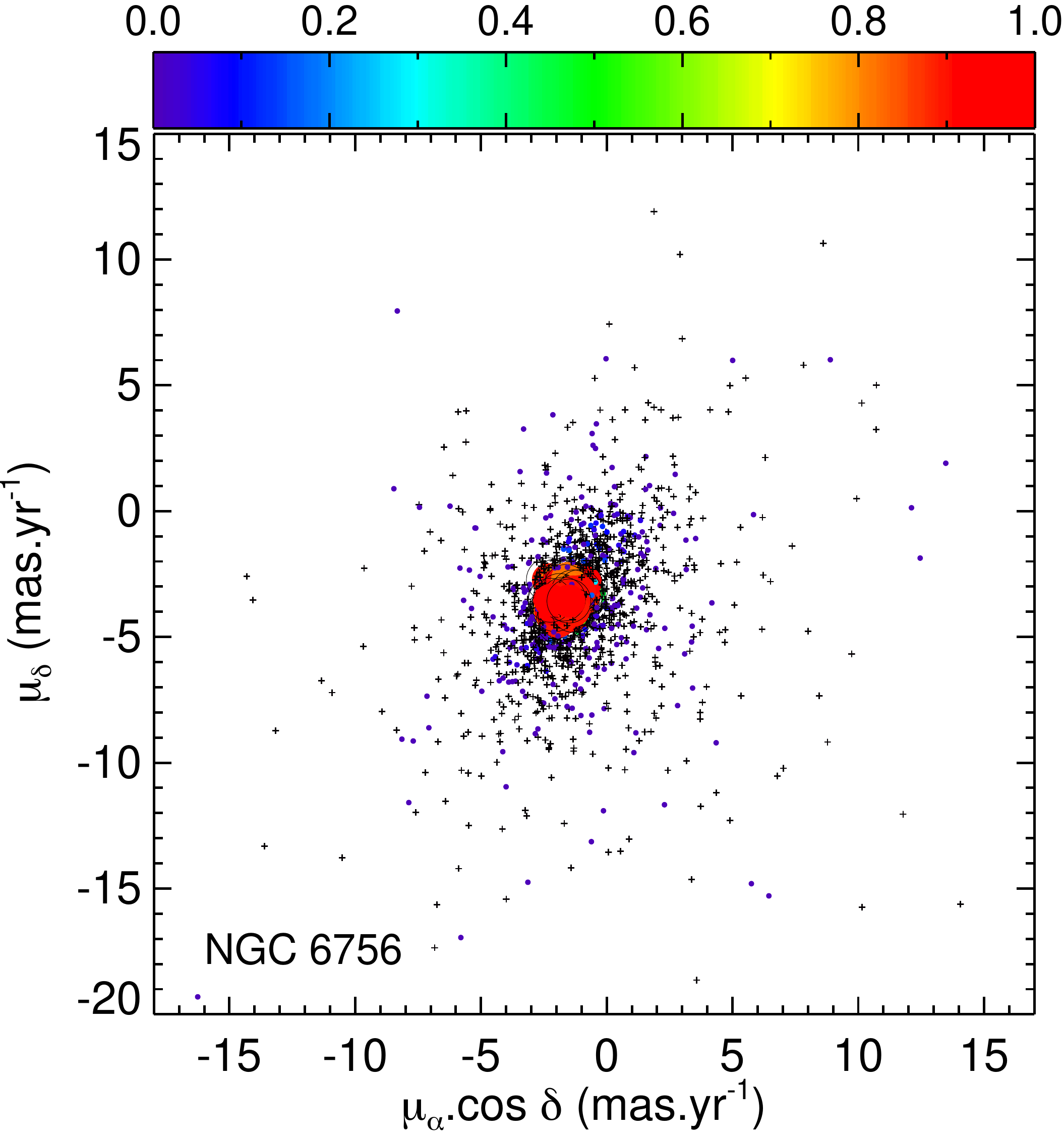}
    \includegraphics[width=0.31\textwidth]{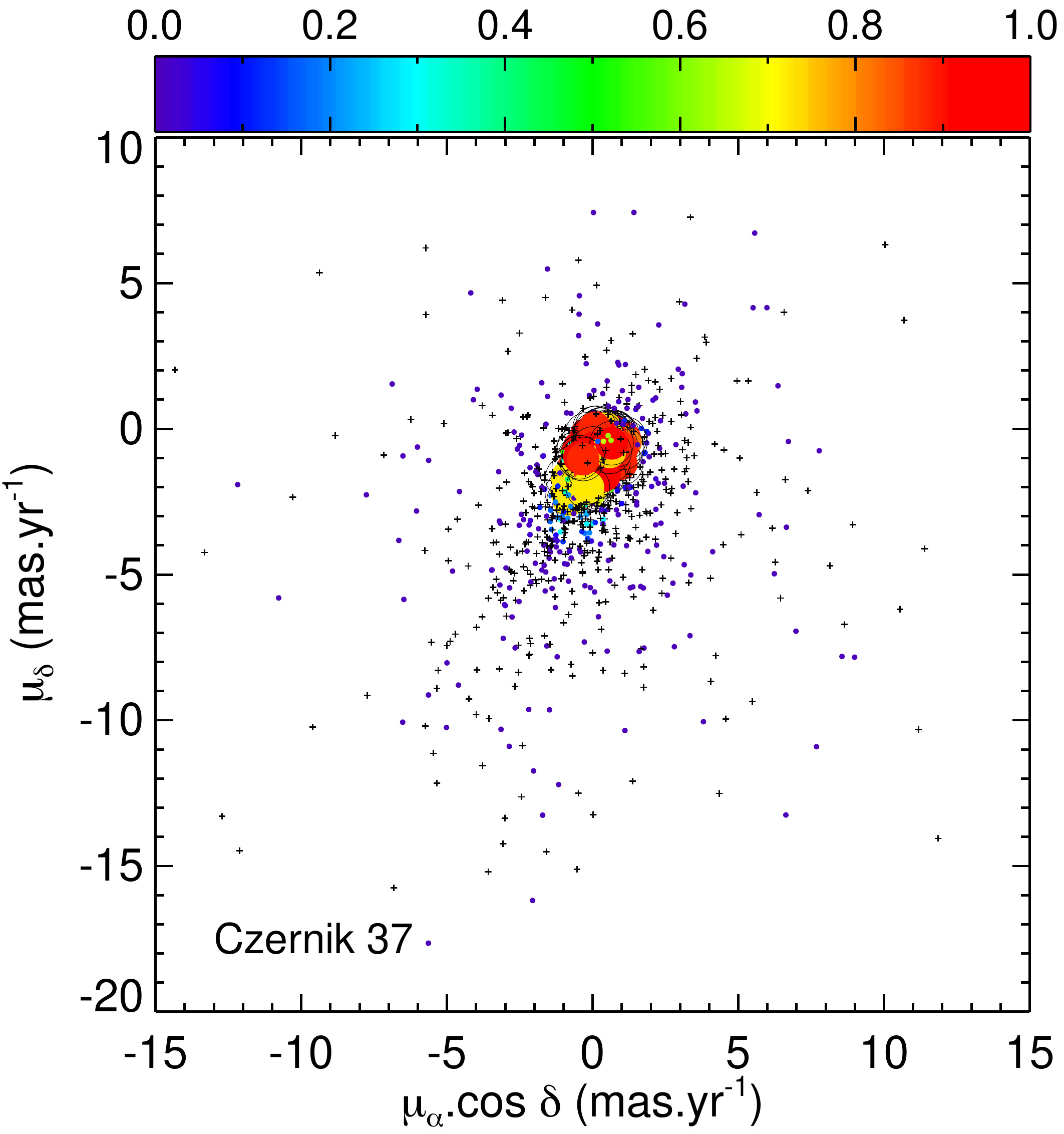}
    \includegraphics[width=0.31\textwidth]{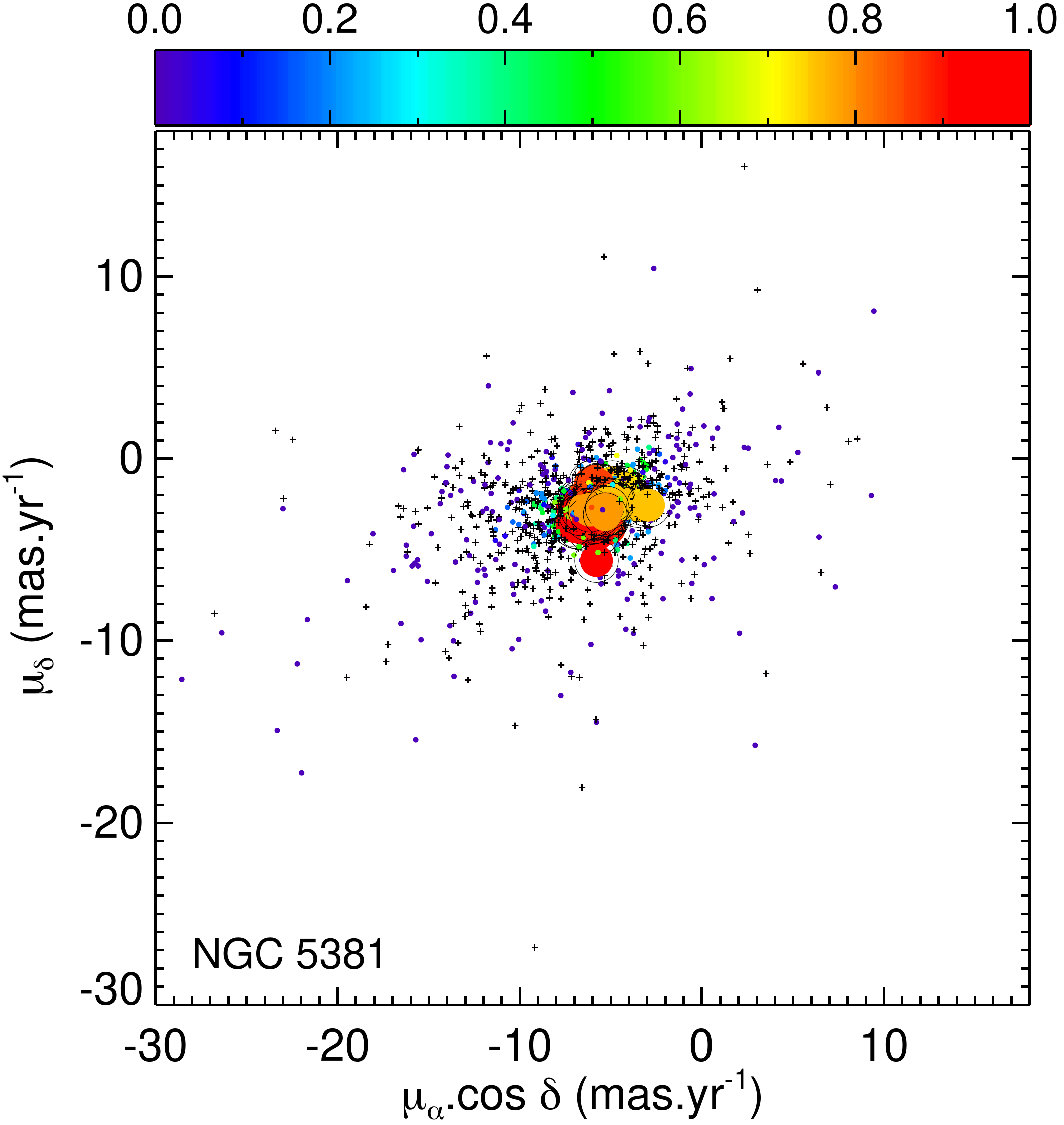}
    \includegraphics[width=0.31\textwidth]{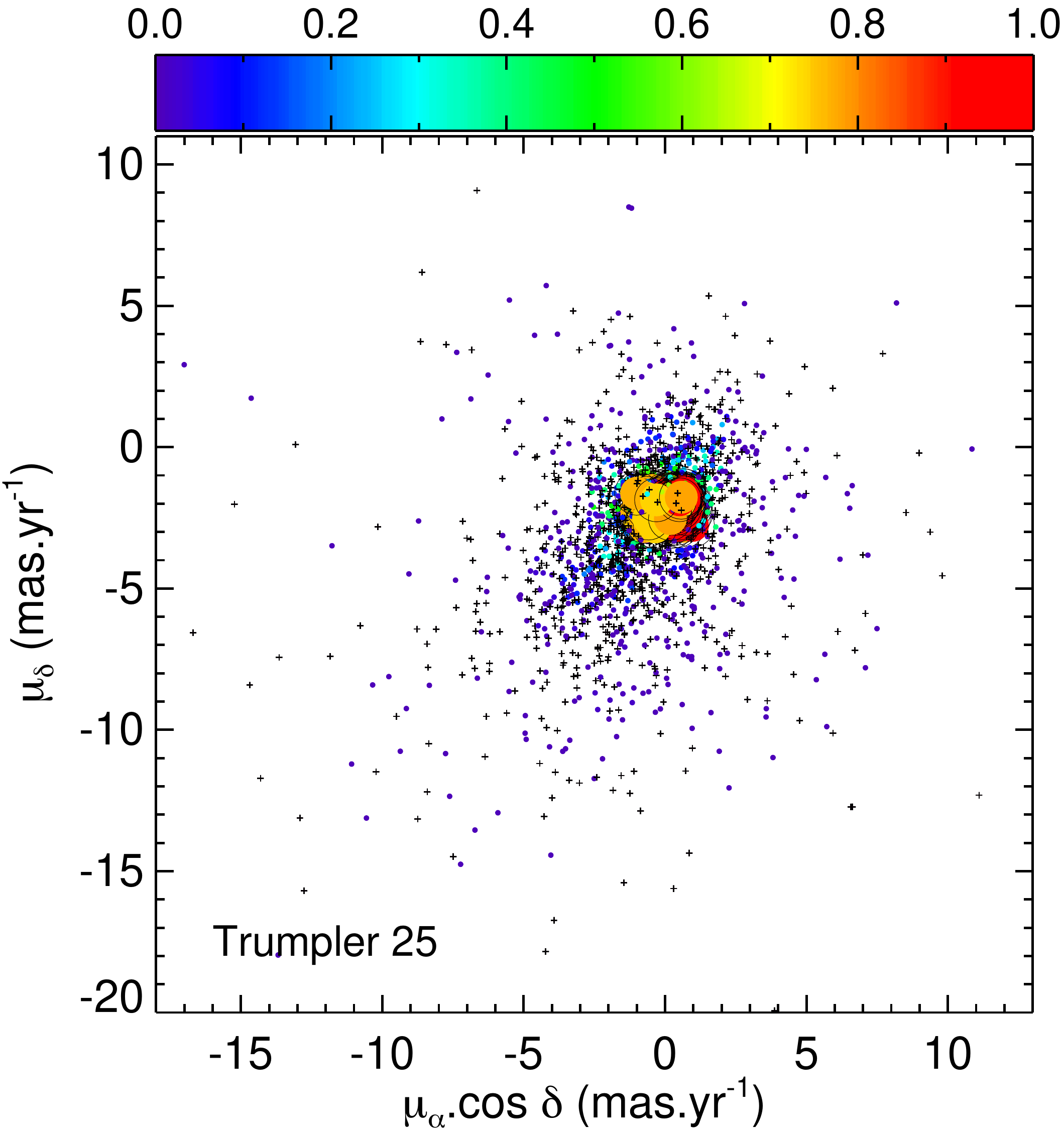}
    \includegraphics[width=0.31\textwidth]{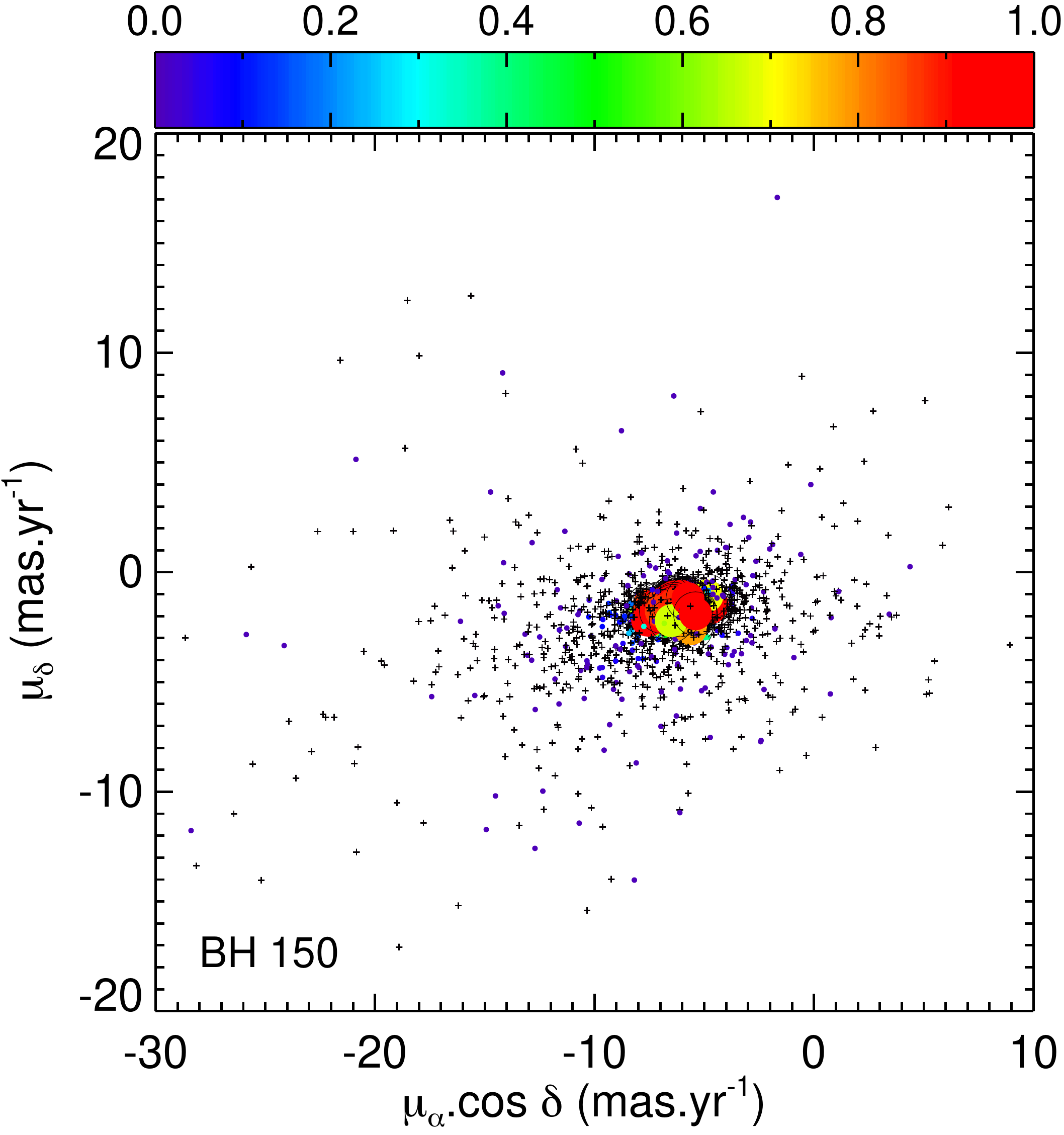}
    \includegraphics[width=0.31\textwidth]{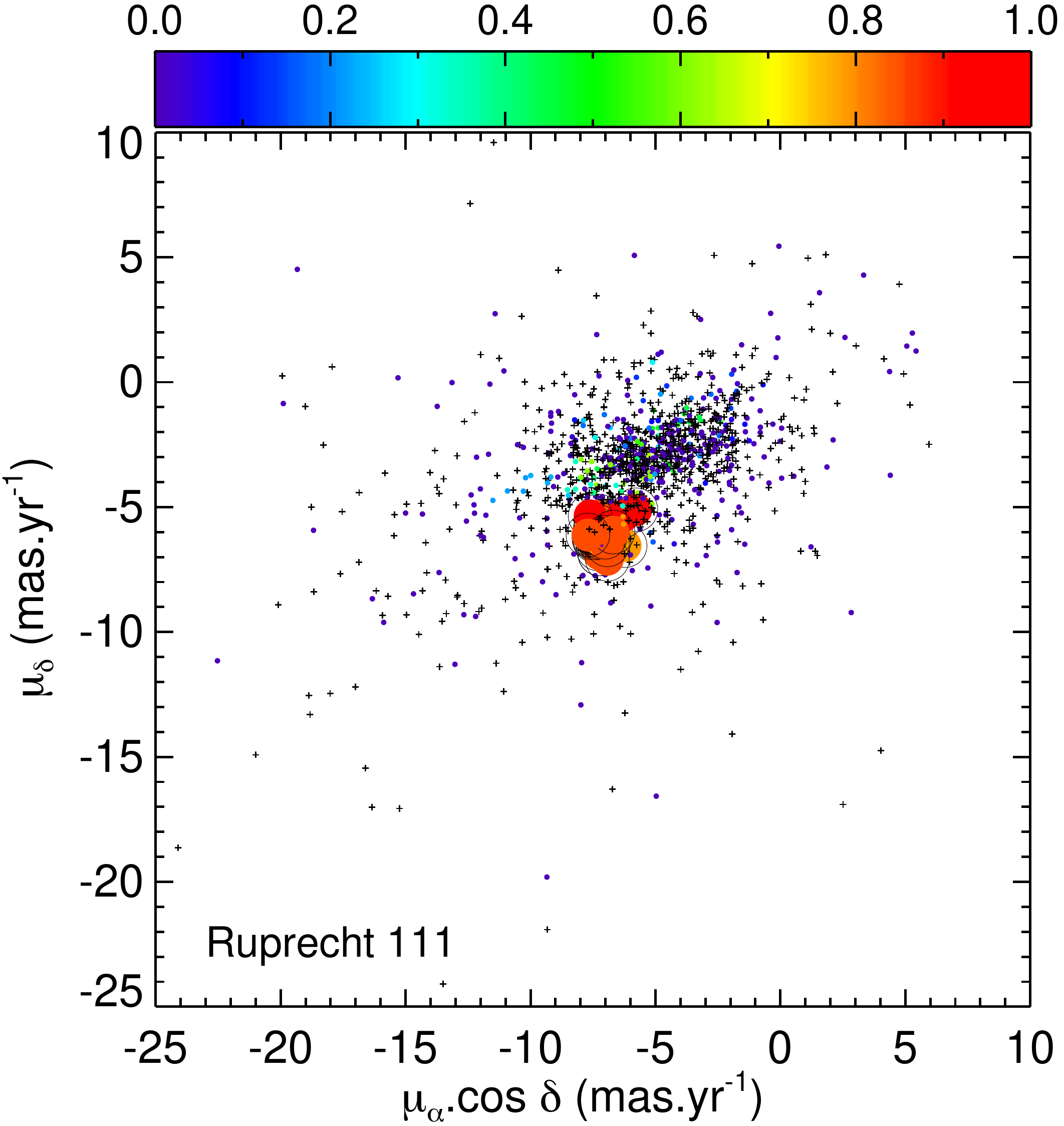}  
    \includegraphics[width=0.31\textwidth]{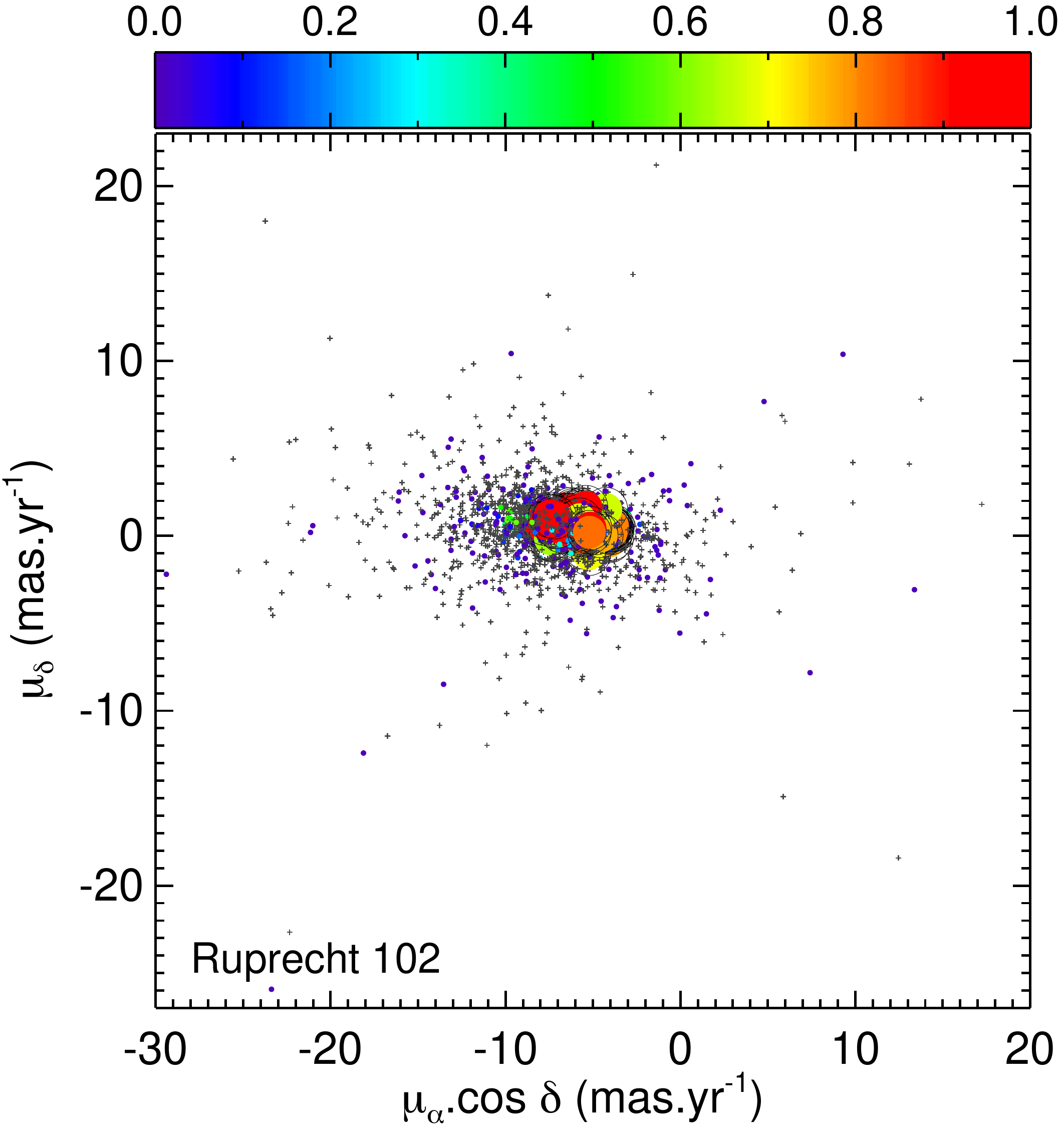}  
    \includegraphics[width=0.31\textwidth]{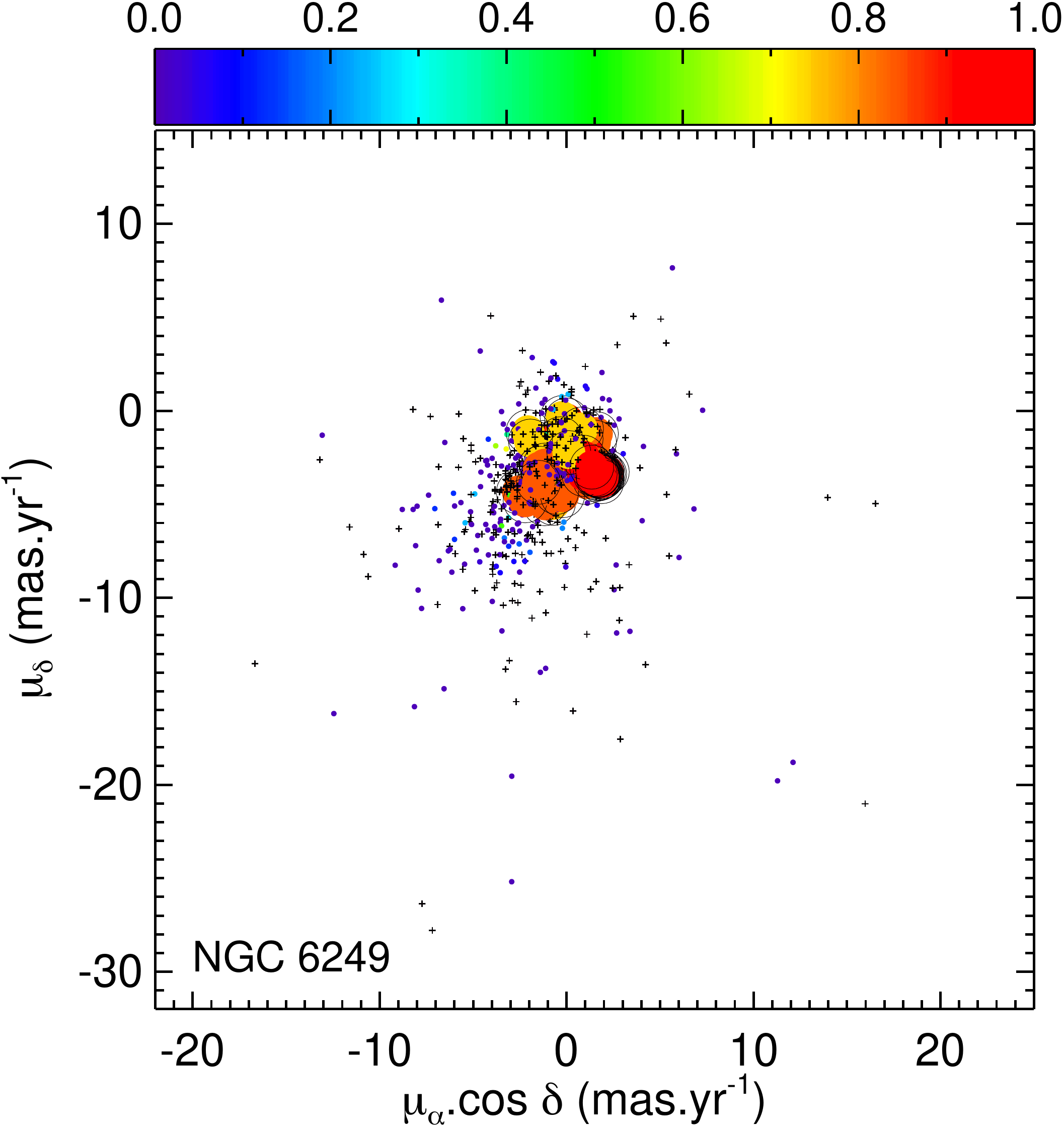}  
    \includegraphics[width=0.31\textwidth]{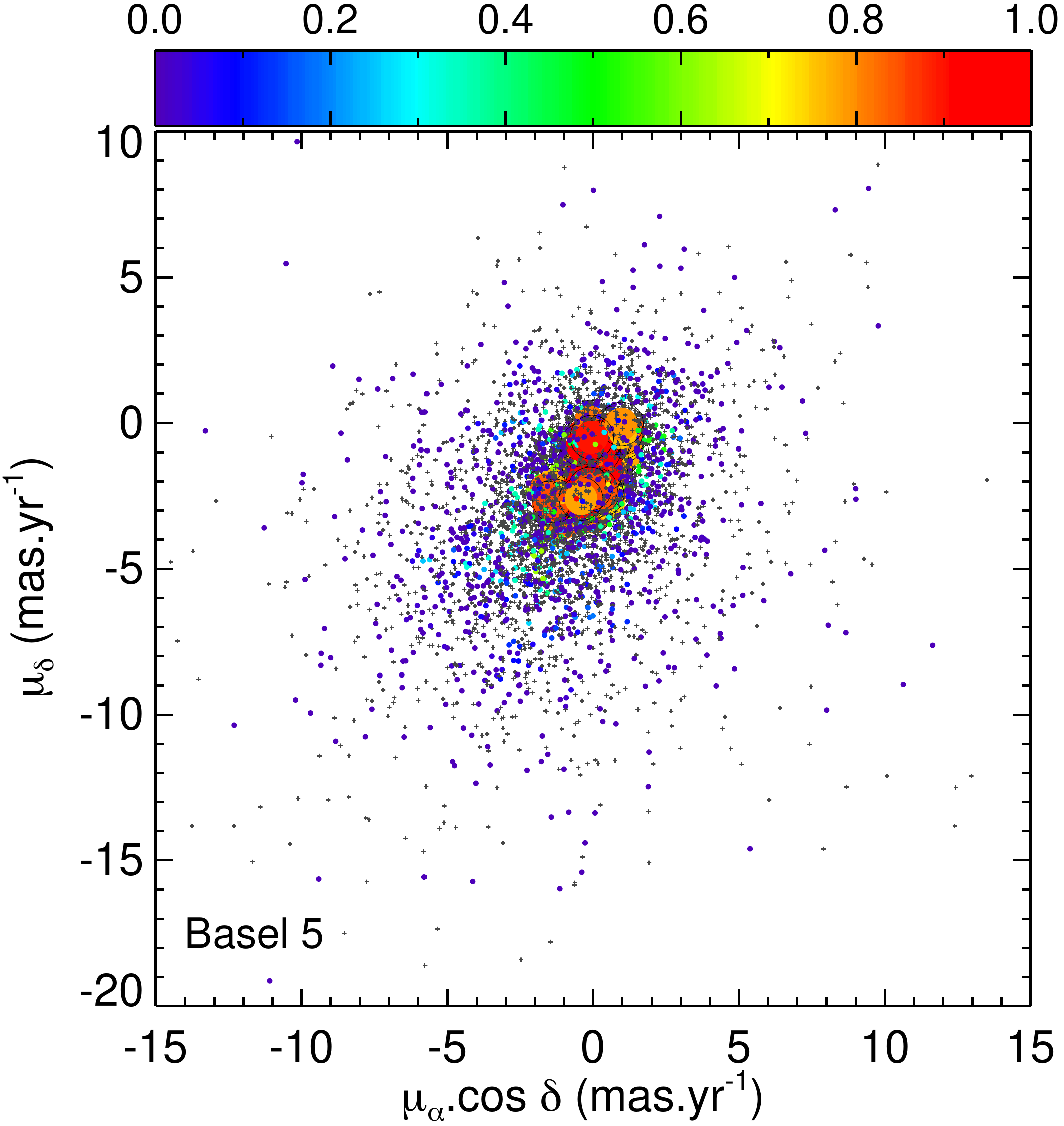}  
    \includegraphics[width=0.31\textwidth]{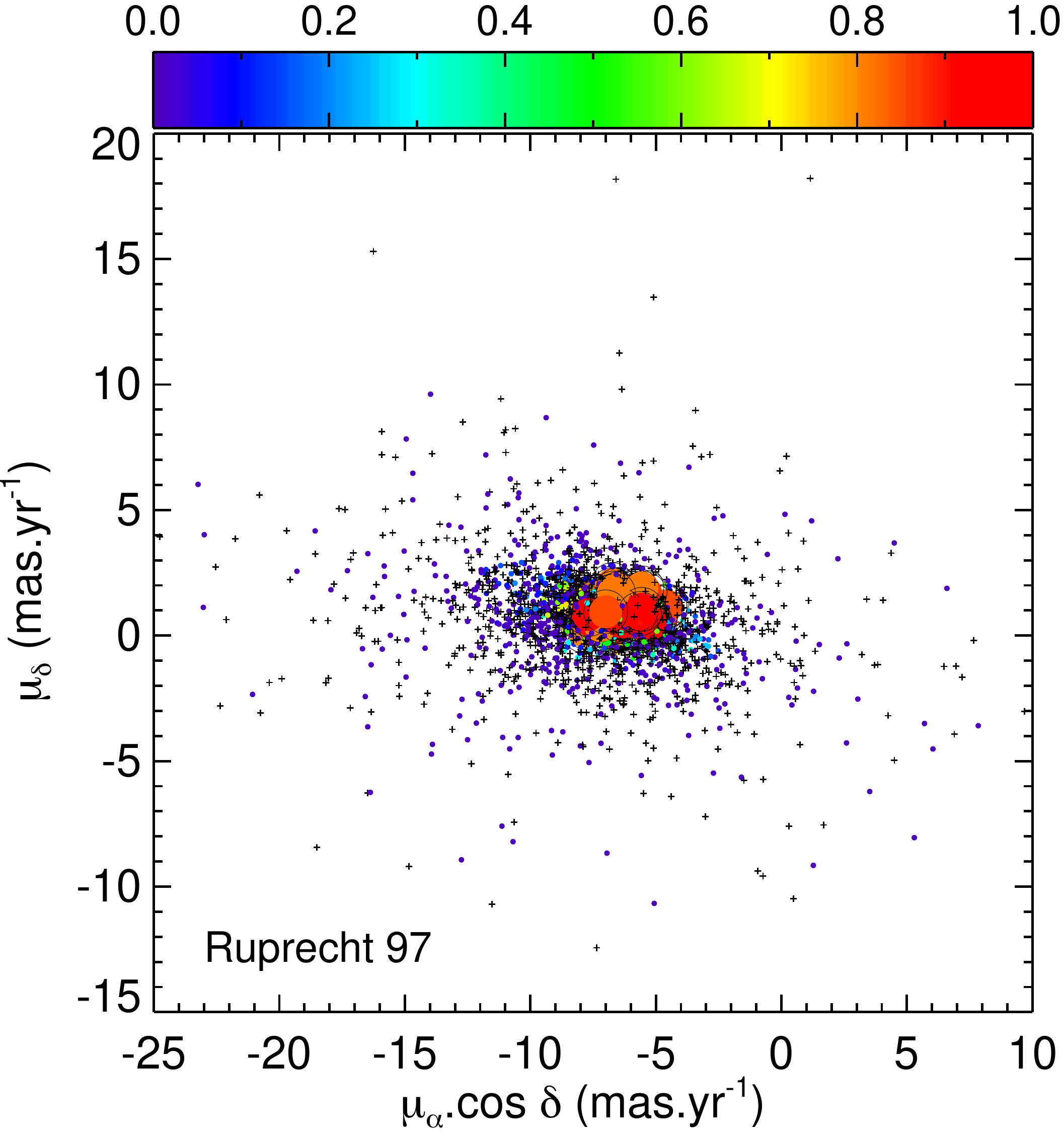}  
    \includegraphics[width=0.31\textwidth]{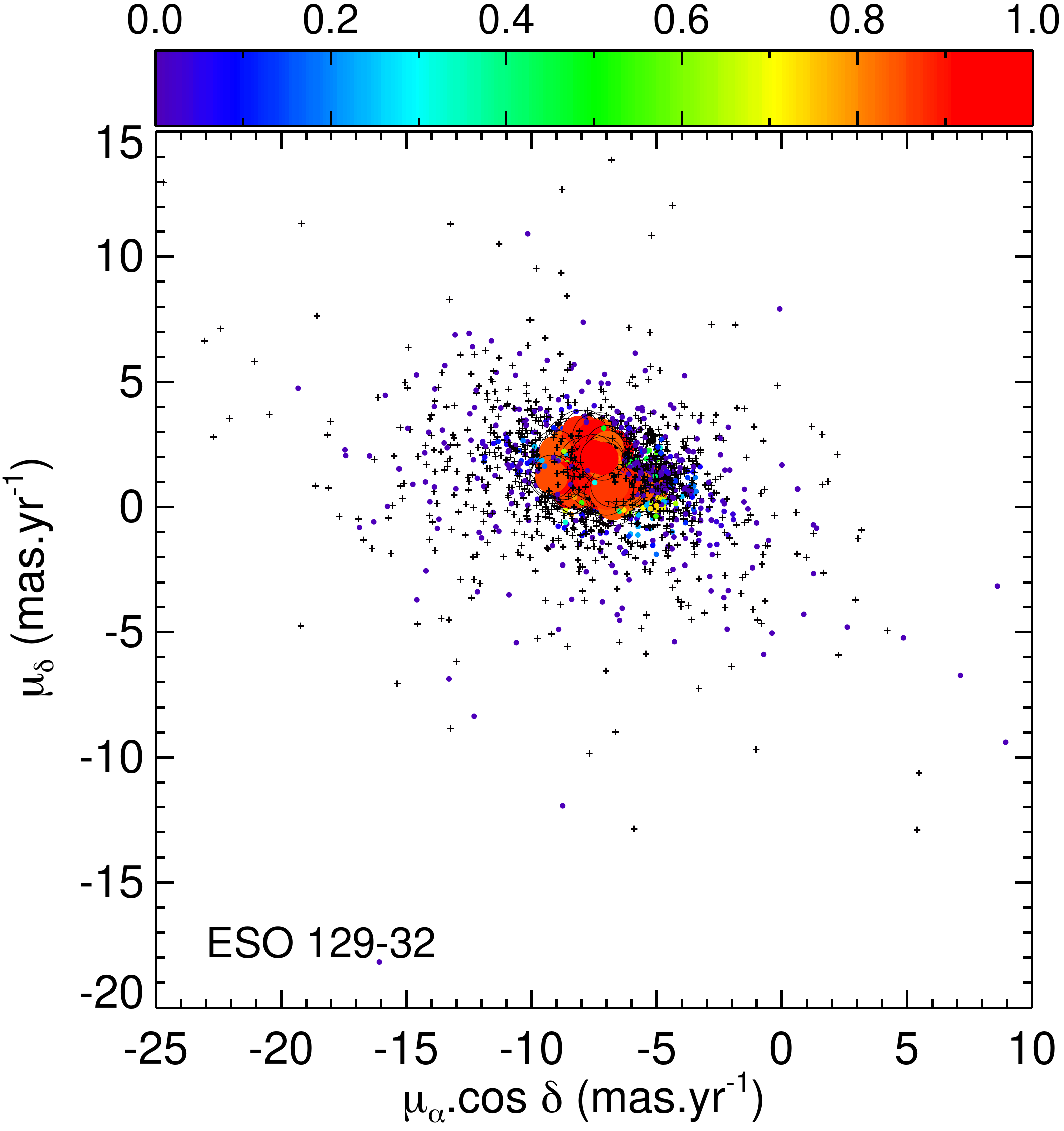}

%  }
\caption{VPDs after applying the decontamination procedure of Section \ref{memberships}. Colours were assigned according to the membership scale and big filled circles represent member stars. }

\label{VPDs_T1plx_parte1}
\end{center}
\end{figure*}

\begin{figure*}
\begin{center}

%\parbox[c]{1.0\textwidth}
%  {
    \includegraphics[width=0.3\textwidth]{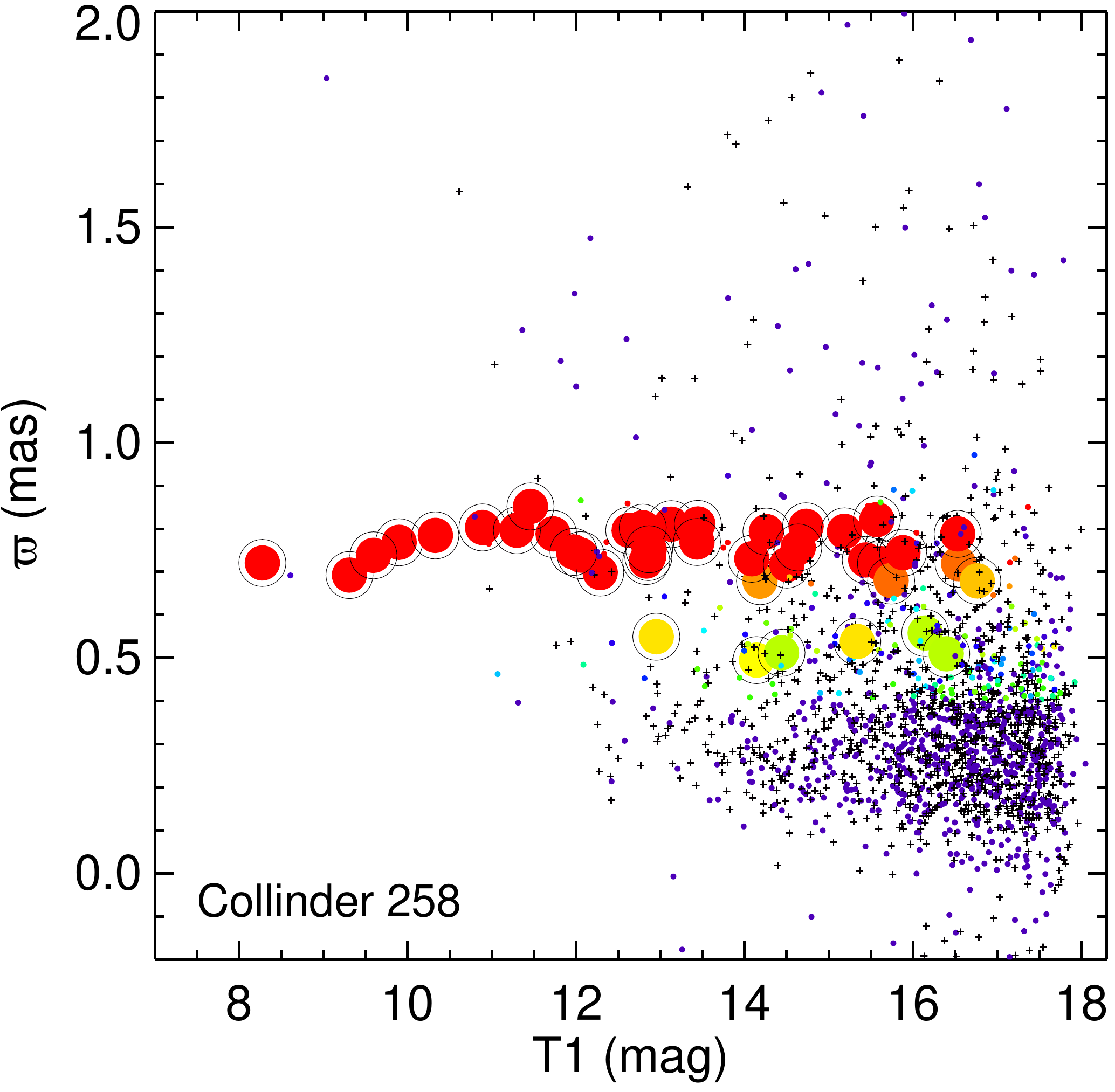}       
    \includegraphics[width=0.3\textwidth]{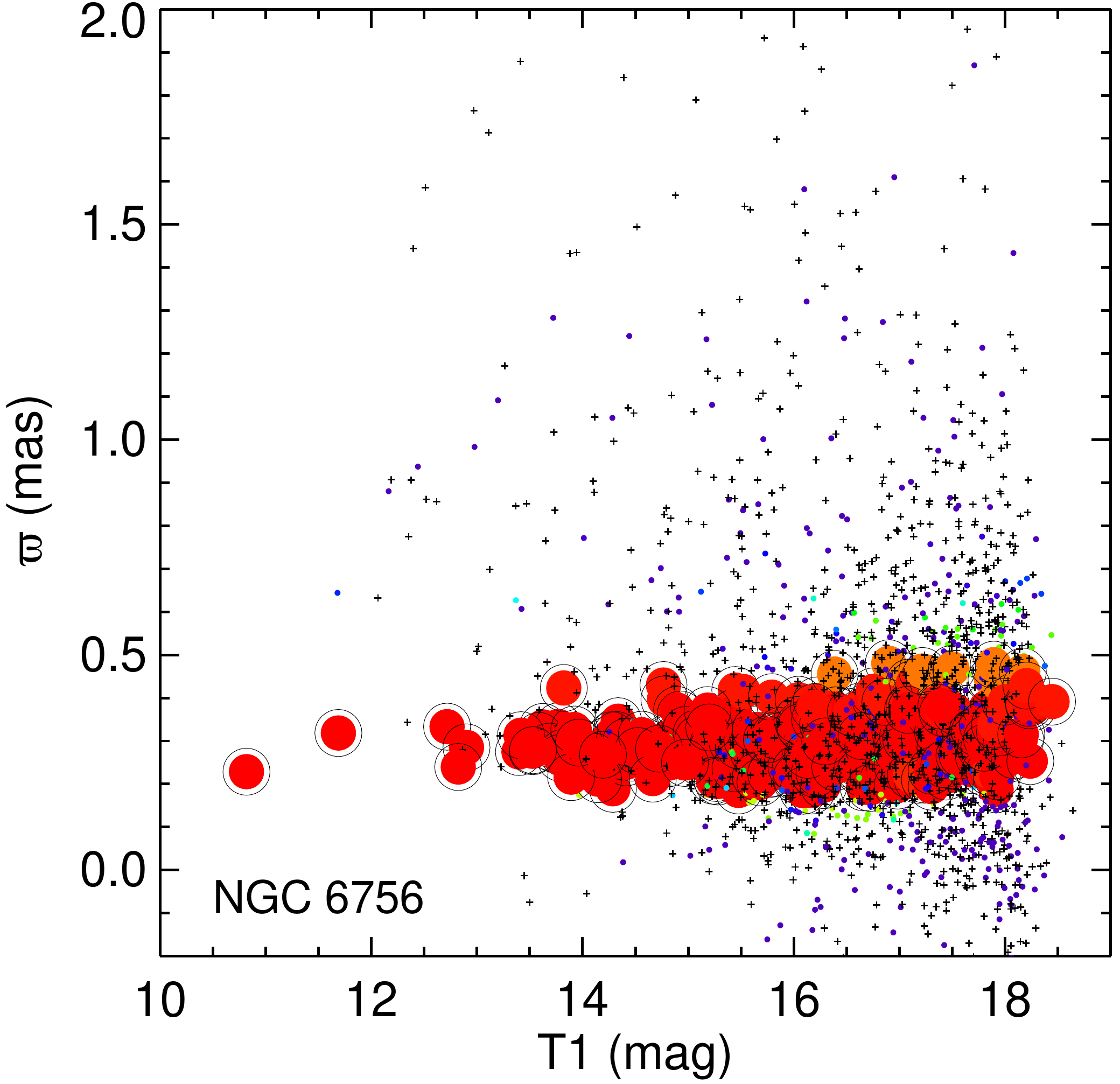}       
    \includegraphics[width=0.3\textwidth]{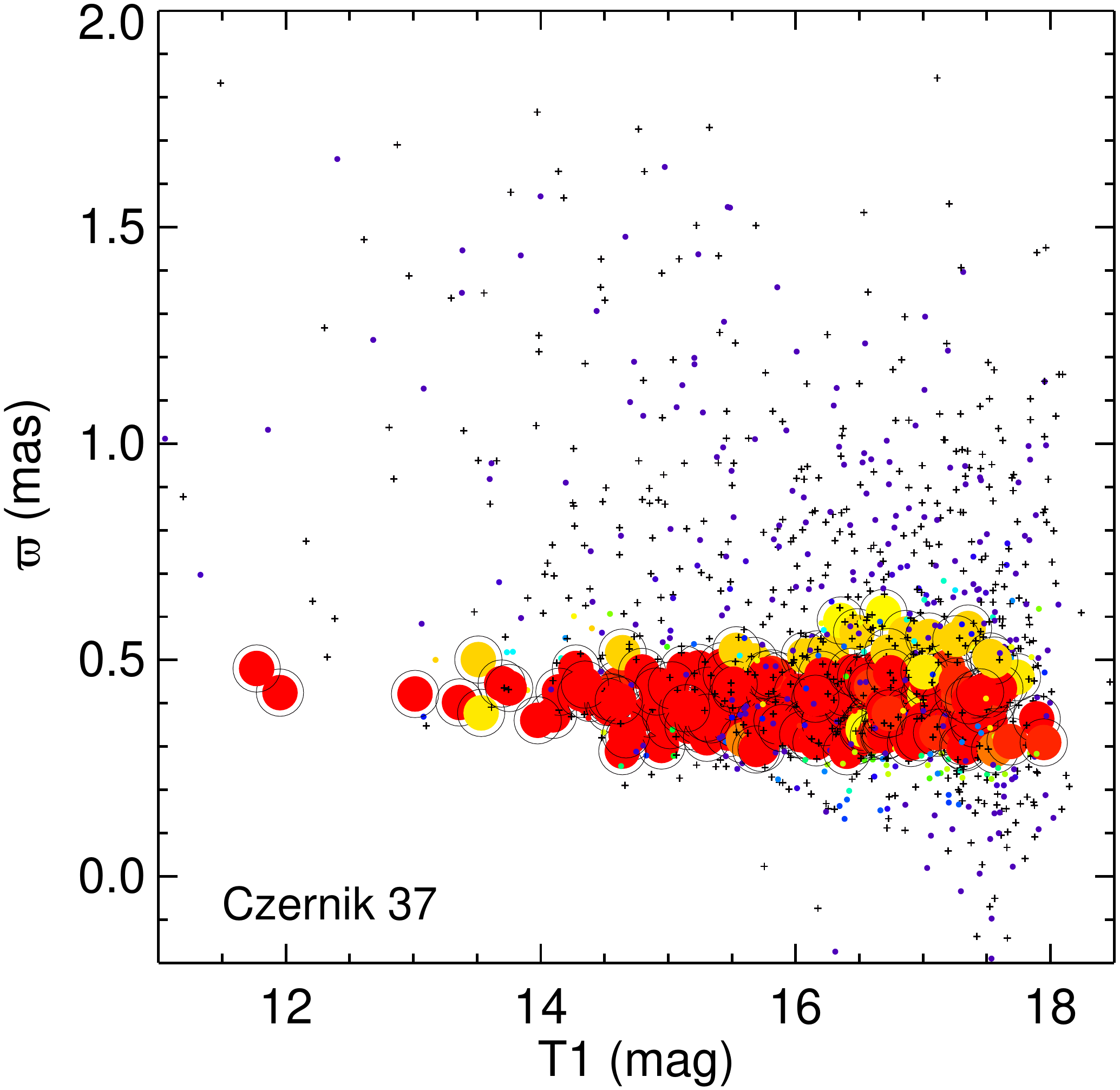}    
    \includegraphics[width=0.3\textwidth]{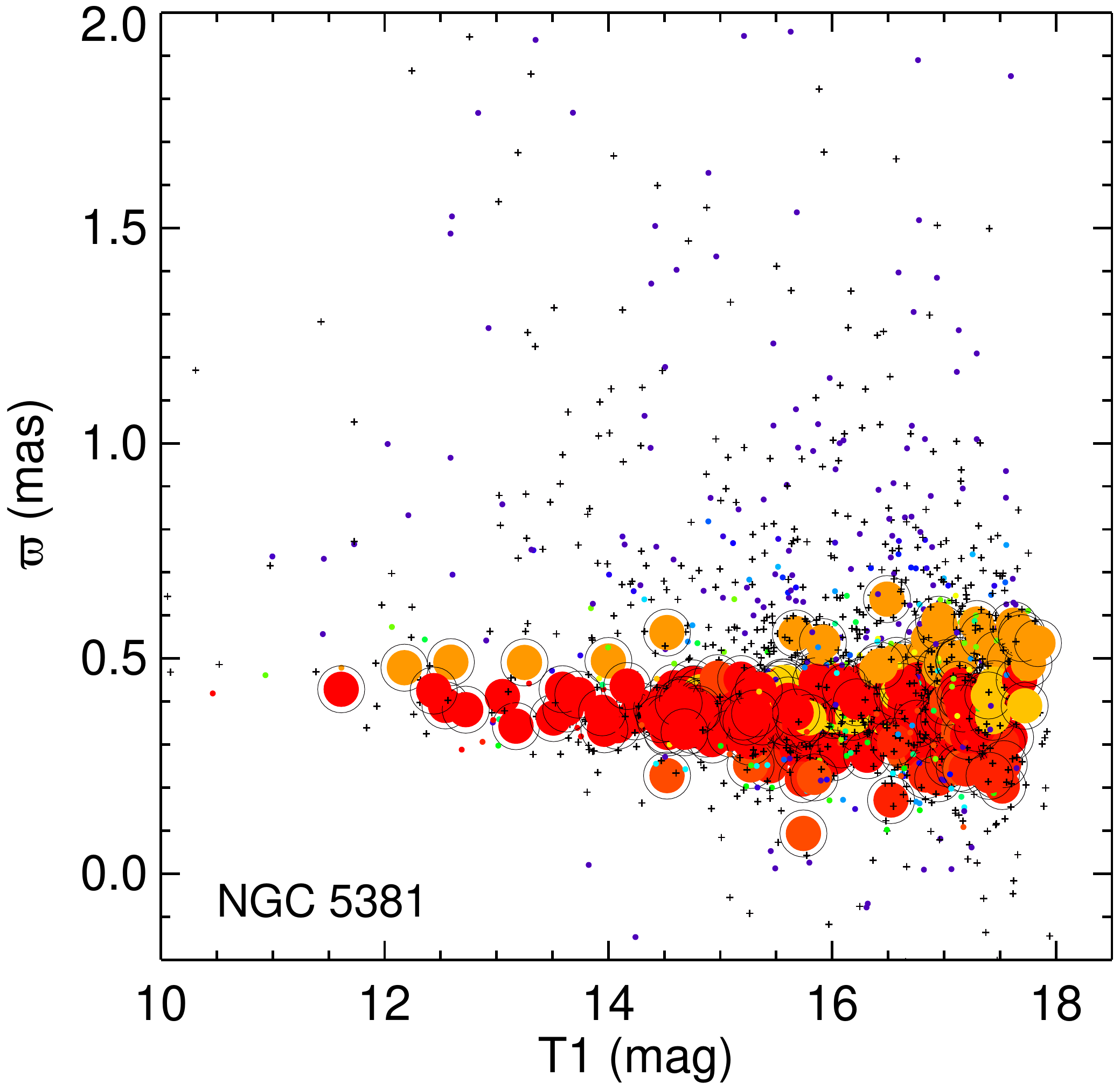}       
    \includegraphics[width=0.3\textwidth]{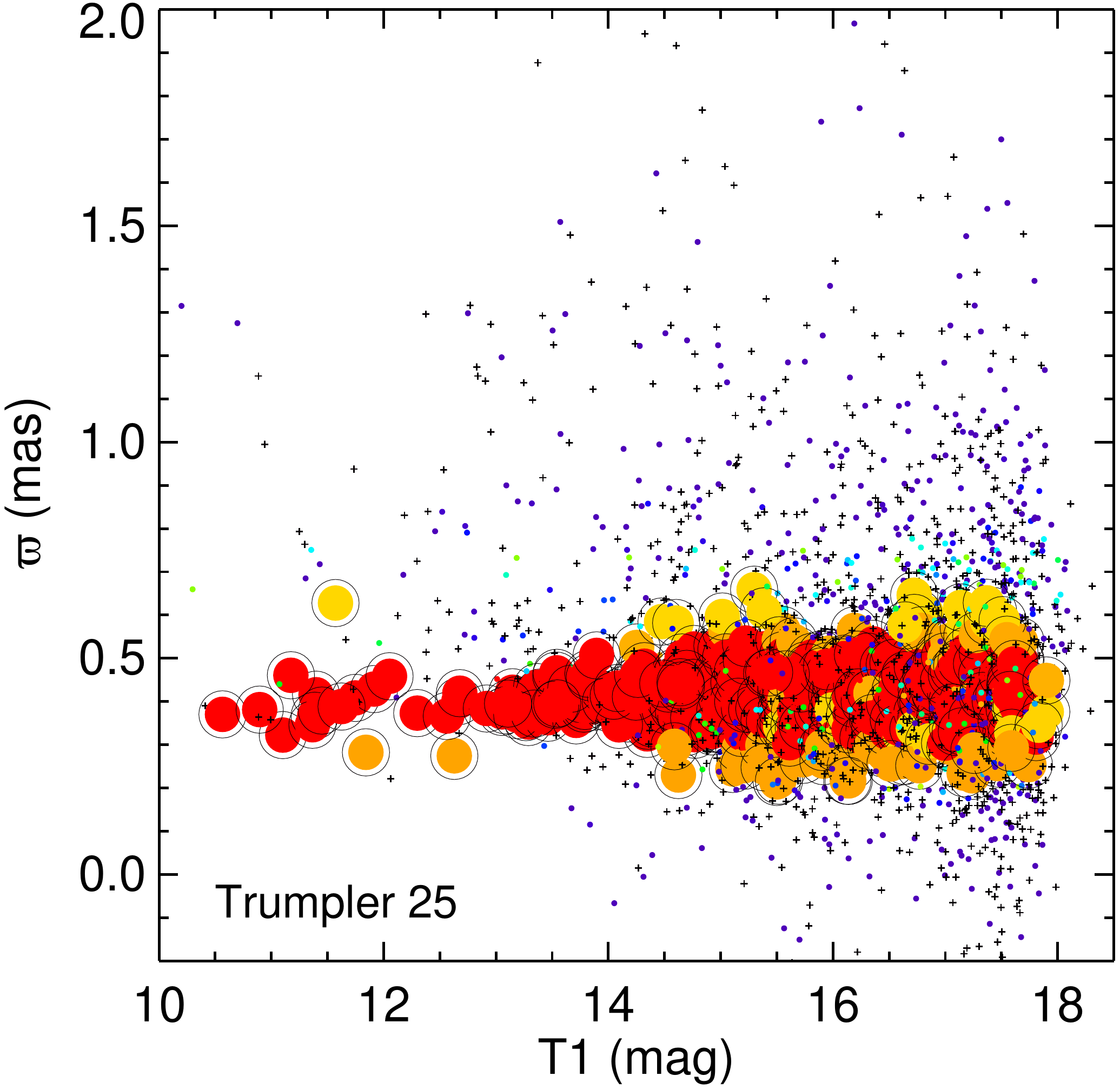}    
    \includegraphics[width=0.3\textwidth]{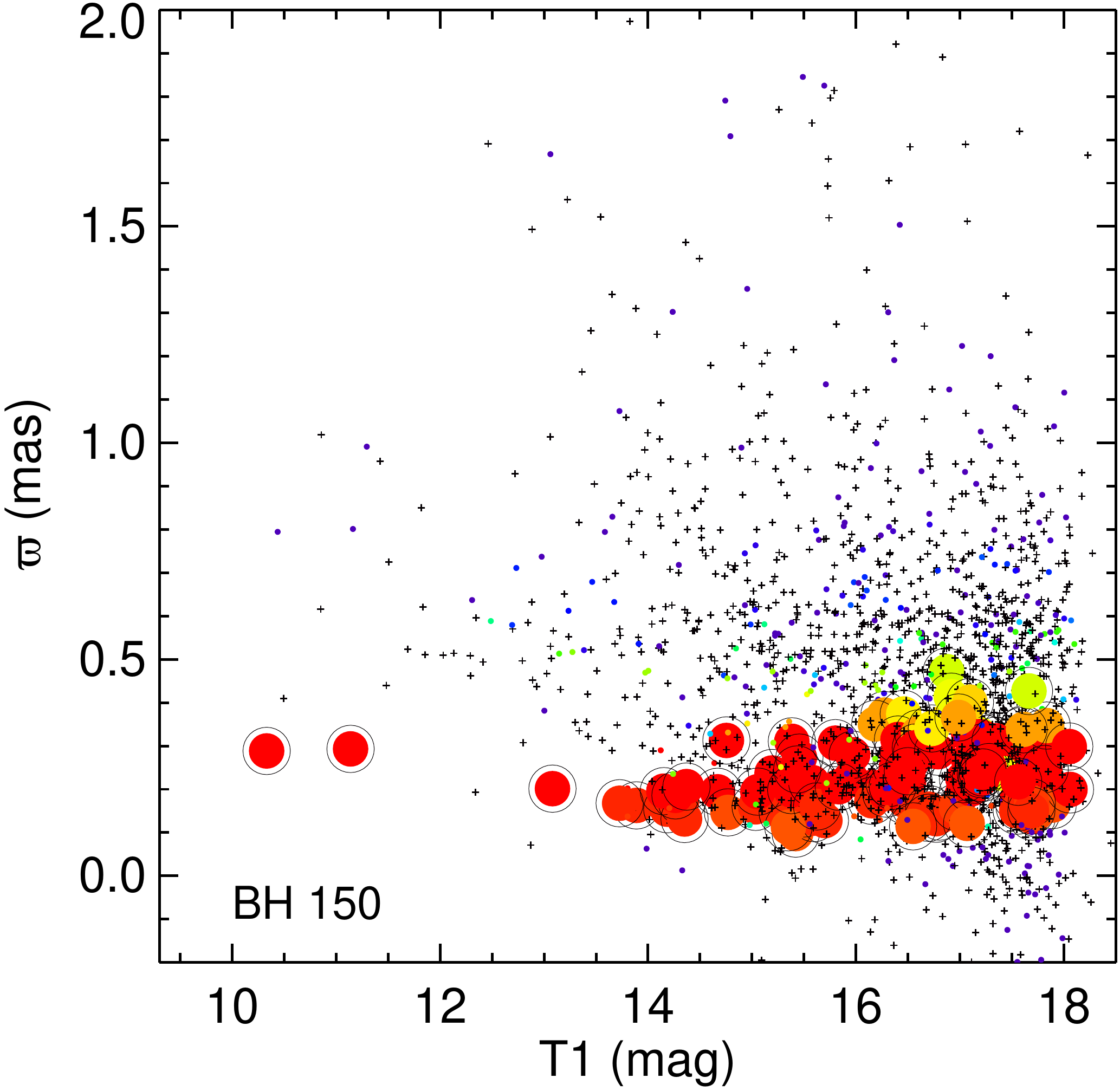}  
    \includegraphics[width=0.3\textwidth]{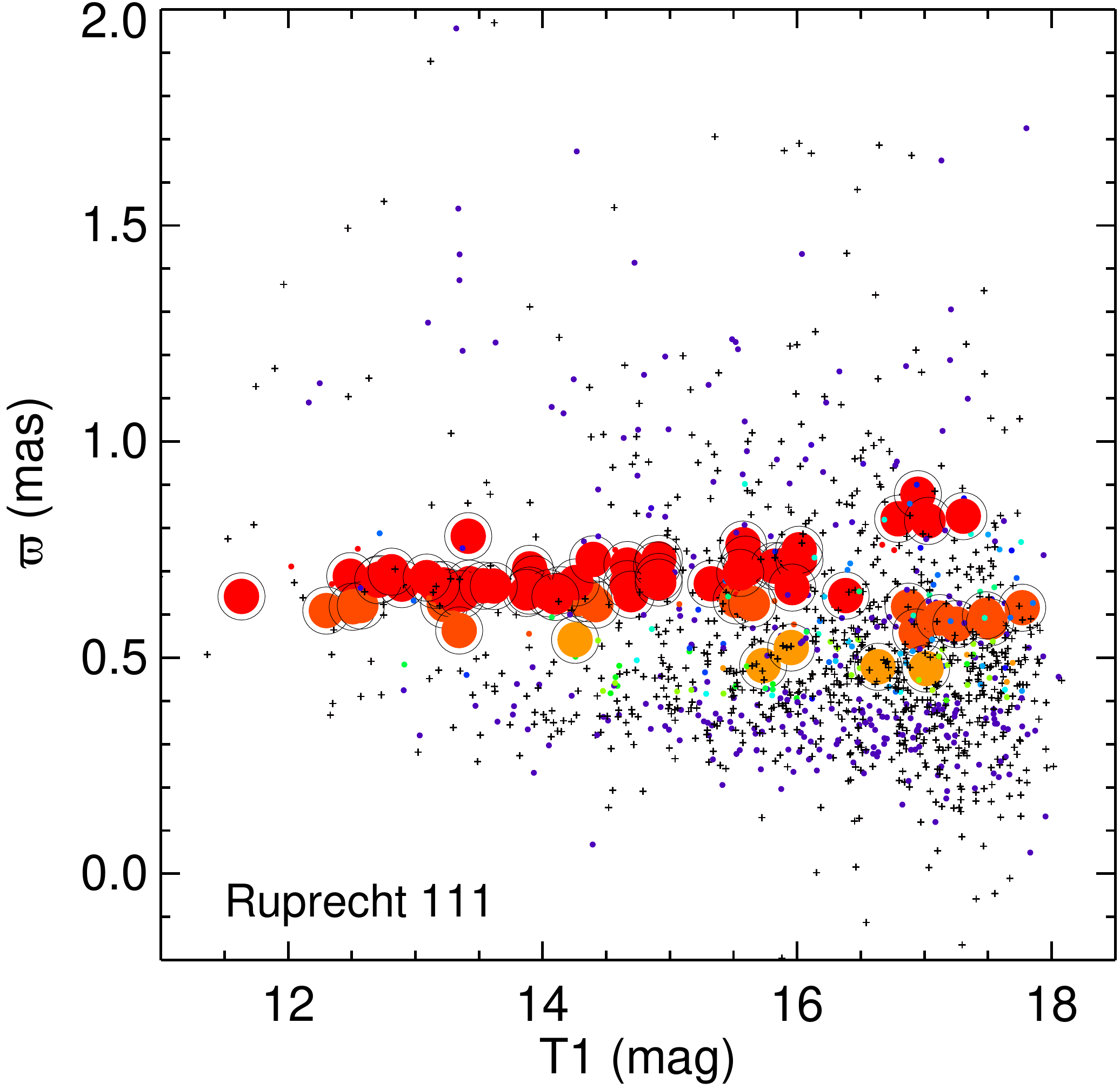}  
    \includegraphics[width=0.3\textwidth]{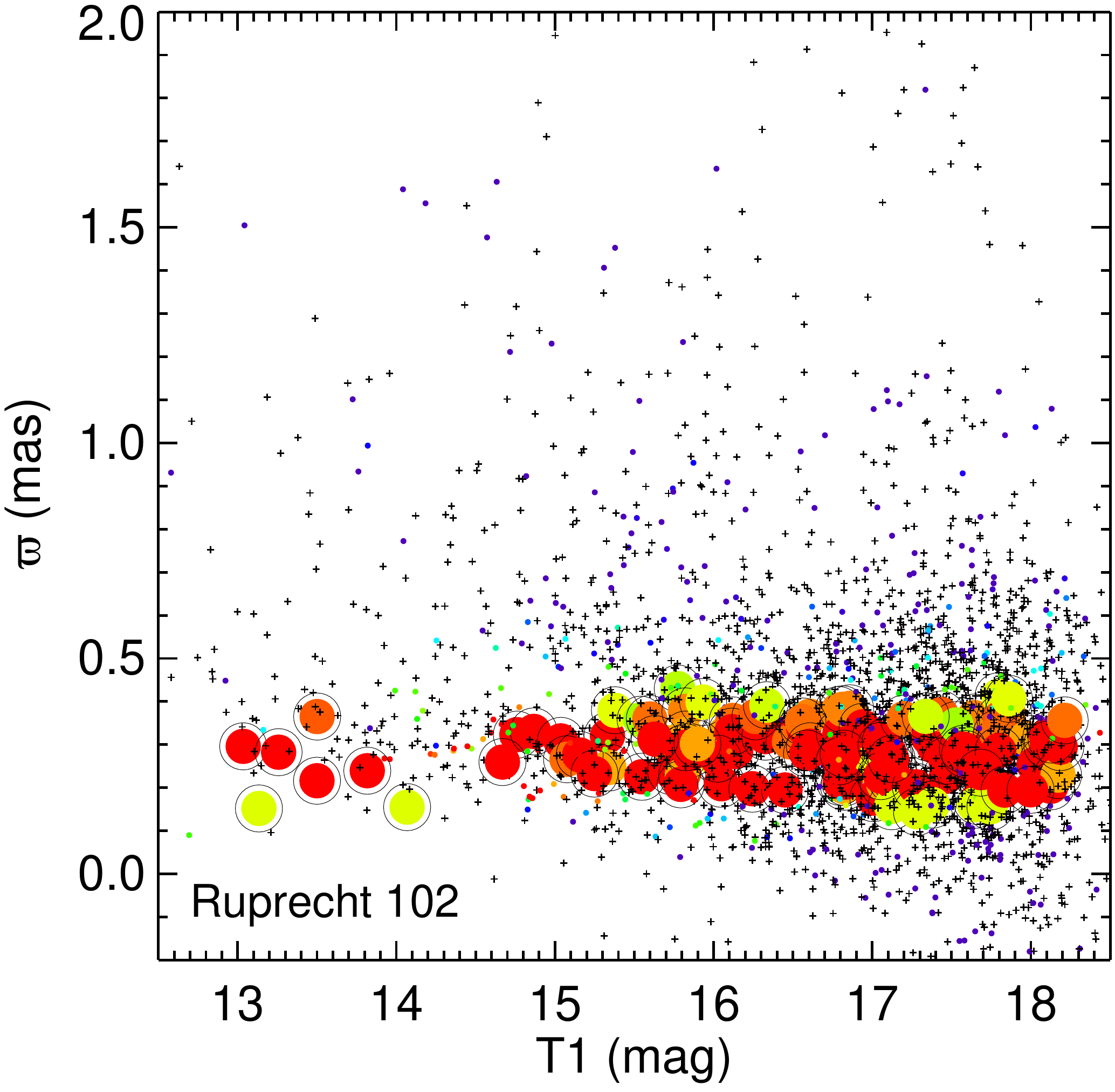}  
    \includegraphics[width=0.3\textwidth]{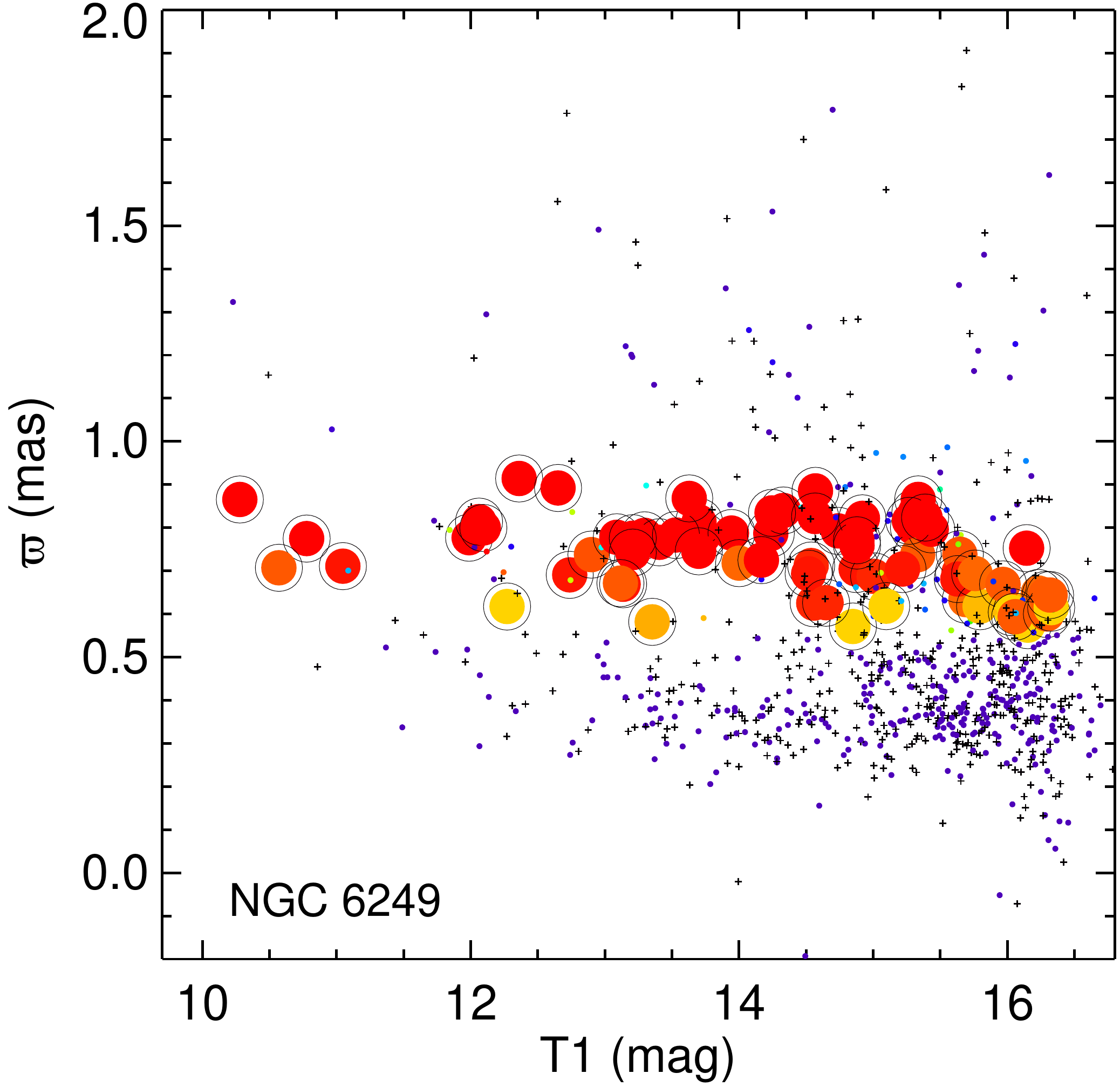}  
    \includegraphics[width=0.3\textwidth]{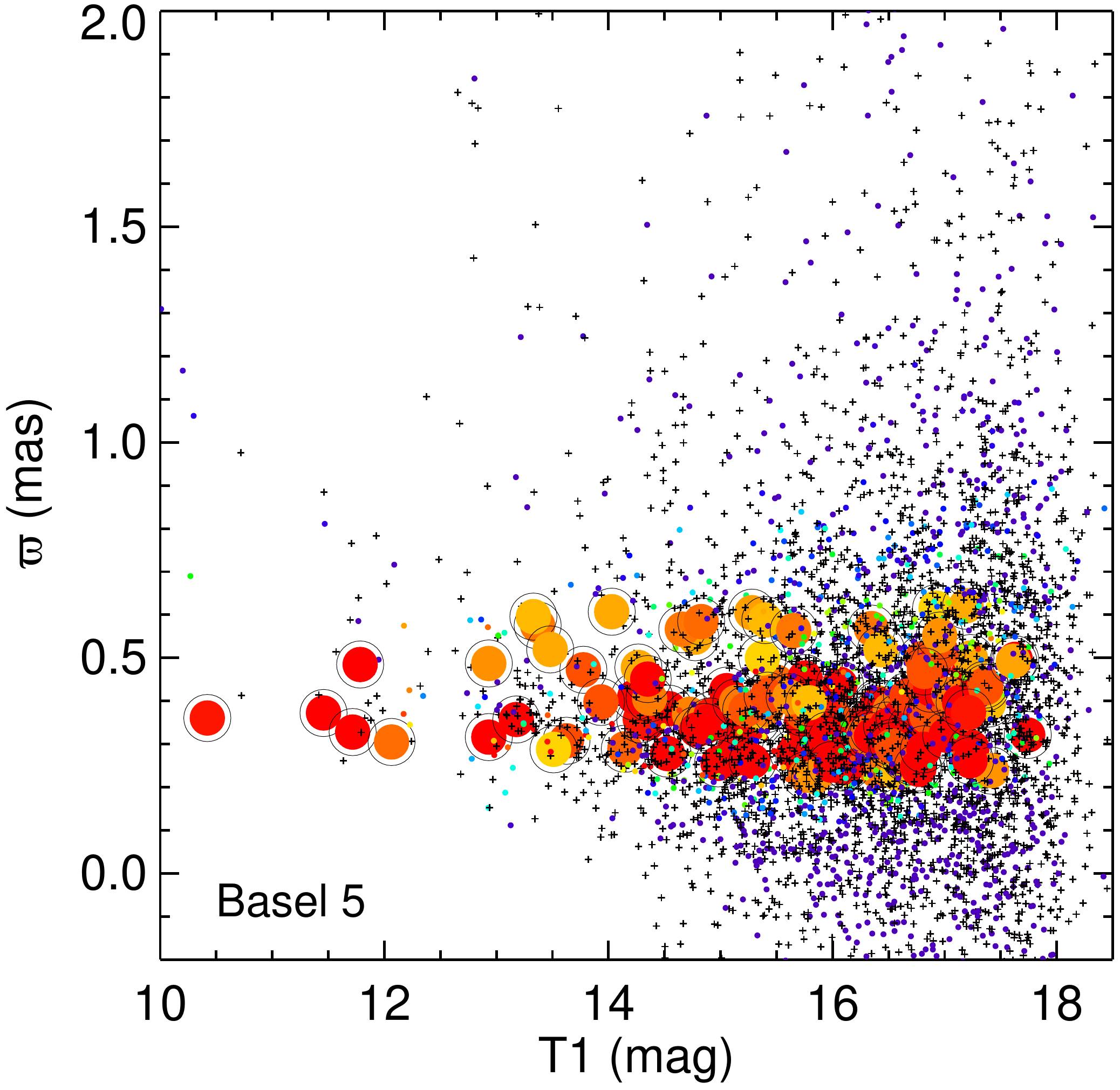}  
    \includegraphics[width=0.3\textwidth]{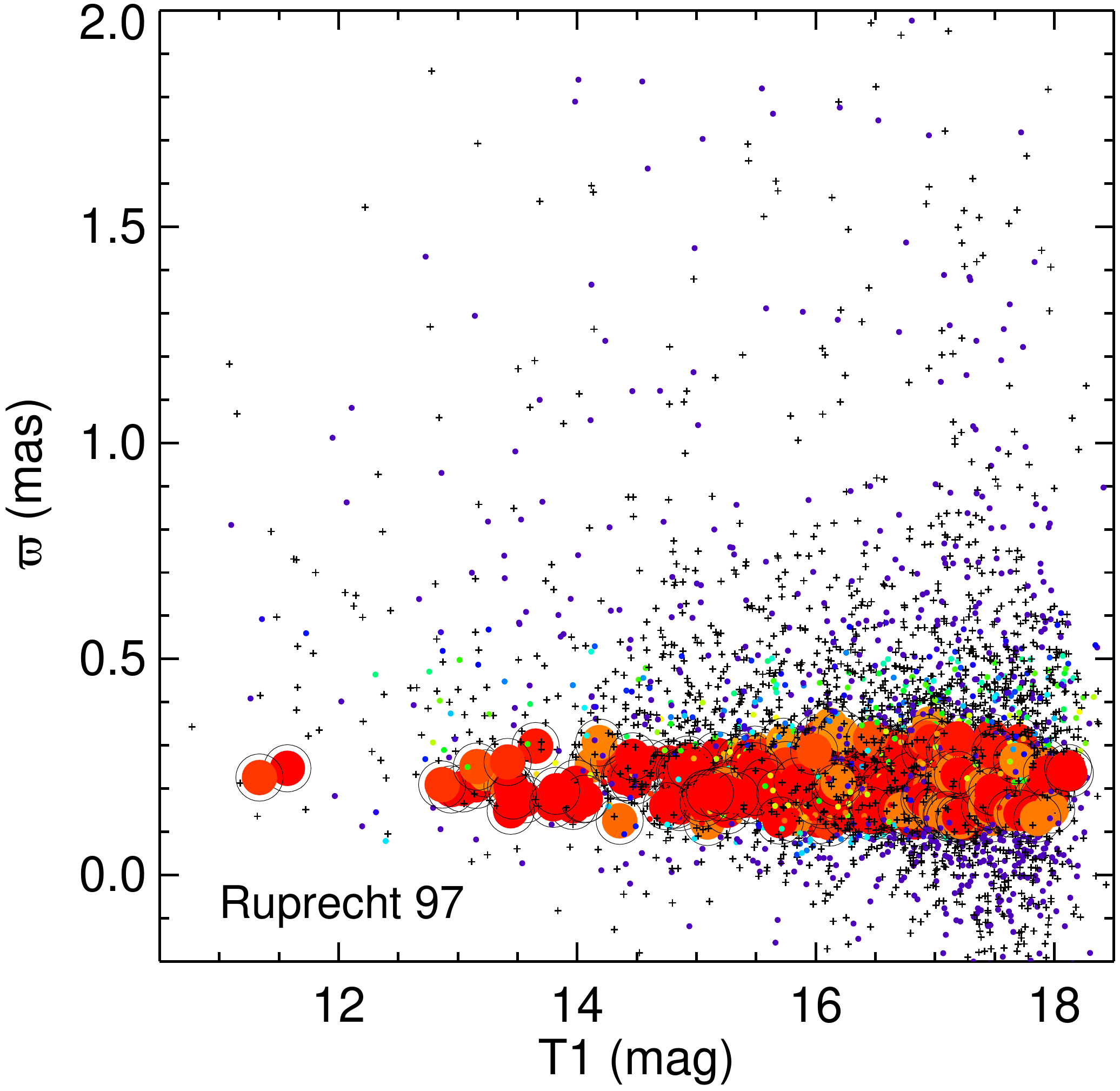}  
    \includegraphics[width=0.3\textwidth]{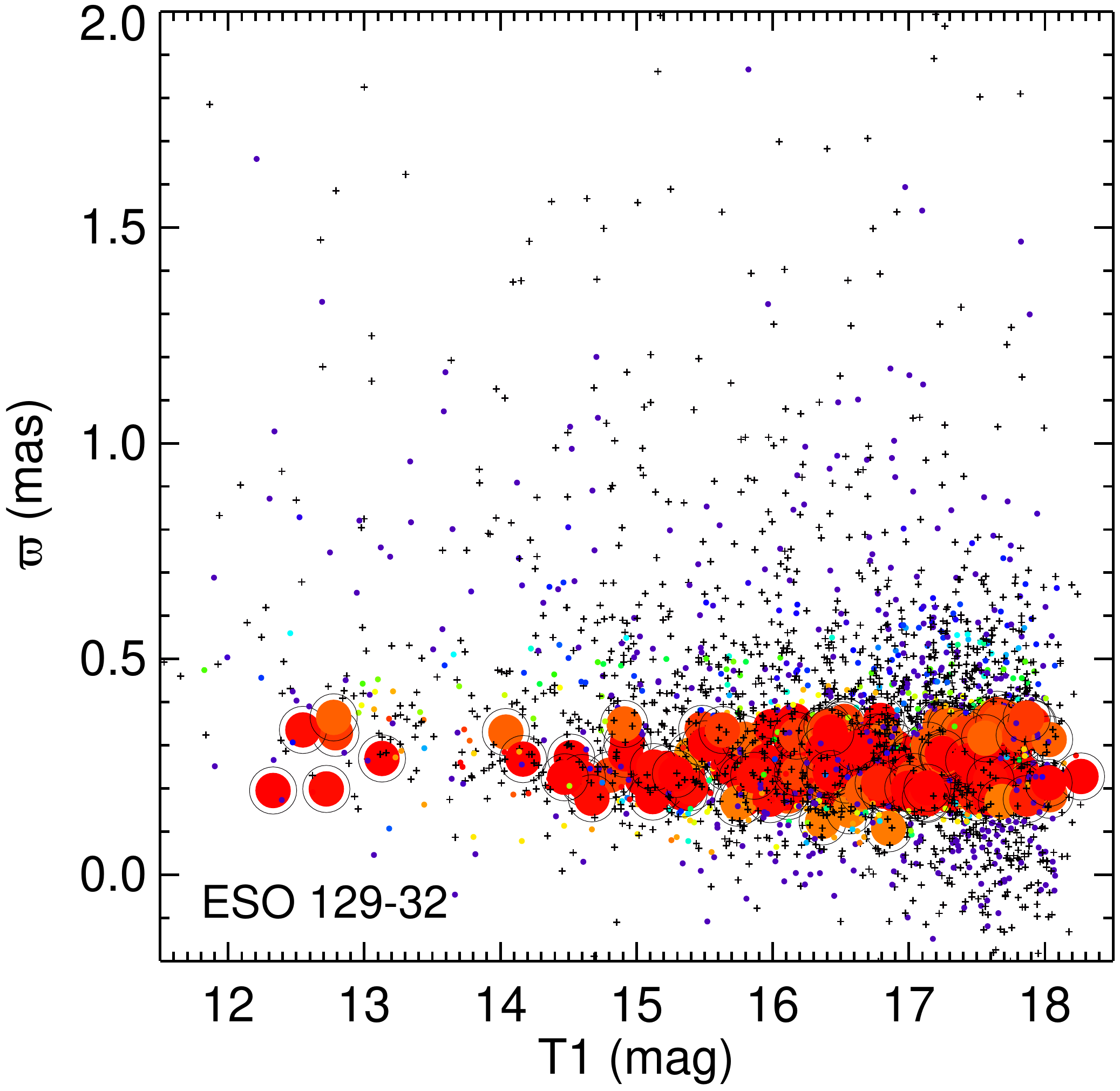}  

%  }
\caption{ $\varpi$\,versus $T_1$ plots for the studied OCs. Symbols and colours are the same of those in Fig.~\ref{VPDs_T1plx_parte1}.  }

\label{T1mag_plx}
\end{center}
\end{figure*}

After applying the procedure described in the previous section, we built decontaminated CMDs  for the whole studied OC sample, as showed in Fig.~\ref{CMDs_parte1}. We also built the vector-point diagrams (VPDs) and the $\varpi\,$versus $T_1\,$ plots for the complete sample (Figs.~\ref{VPDs_T1plx_parte1} and \ref{T1mag_plx}, respectively). We can see that higher membership stars (typically $P\gtrsim0.6-0.7$) define recognizable evolutionary sequences in the CMDs, although some interlopers remain. These stars also define ``clumps" of data in the astrometric plots, since they share common movements and compatible parallaxes.

We then took the photometric data for higher membership stars and, in a first step, we performed preliminary estimates of the basic astrophysical parameters ($(m-M)_0$,\,log($t/$yr) and $E(B-V)$\,) for each OC. To accomplish this first step, we performed visual fits of solar metallicity PARSEC isochrones \citep{Bressan:2012} to each decontaminated CMD. A first guess for the distance modulus was obtained from the relation $(m-M)_{0}=5\times\,$log(100/$\langle\varpi\rangle$), where $\langle\varpi\rangle$ is the mean parallax taken over high membership stars. Then the isochrone was reddened and vertically shifted in order to provide reasonable matches to key evolutionary CMD features, such as the main sequence, the subgiant branch, the red clump (when present) and the red giant branch.

In a second step, we used a set of PARSEC isochrones covering metallicty values $[Fe/H]$ in the range [-1.0\,,\,+0.65]\,dex (the upper limit corresponding to the maximum value covered by the models) and run the ASTECA code \citep{Perren:2015} in order to perform automatic isochrone fits by employing the parameters estimated visually as a first guess and the Washington photometry of high-membership stars ($L\gtrsim0.6-0.7$; equation \ref{termo_exponencial}). The isochrone fitting process in ASTECA is based on the generation of synthetic OCs from theoretical isochrones and selection of the best fit through a genetic algorithm (see section 2.9.1 of \citeauthor{Perren:2015}\,\,\citeyear{Perren:2015} and references therein). We restricted the parameters space covered by the models according to the following intervals: 

\begin{itemize}
     \item ($(m-M)_{0,\textrm{pre}}-0.5)\leq(m-M)_0\leq((m-M)_{0,\textrm{pre}}+0.5$), in steps of 0.1\,mag;
     \item ($E(B-V)_{\textrm{pre}}-0.15)\leq E(B-V)\leq (E(B-V)_{\textrm{pre}}+0.15$), in steps of 0.01\,mag;
     \item (log($t/$yr)$_{\textrm{pre}}$$-$0.2) $\leq$ log($t$/yr) $\leq$ (logt($t$/yr)$_{\textrm{pre}}$$+$0.2);
     \item overall metallicity: 0.007 $\leq$ Z $\leq$ 0.04, in steps of 0.002, 
\end{itemize}

\noindent
where $(m-M)_{0,\textrm{pre}}$, $E(B-V)_{\textrm{pre}}$ and log($t/$yr)$_{\textrm{pre}}$ are our first estimates for distance modulus, interstellar redenning and age, respectively, as described above.

The continuous lines in Fig.~\ref{CMDs_parte1} represent the best-fitted isochrones, while the dotted lines correspond to the same isochrones shifted by -0.75 mag in $T_1$ to match the loci of unresolved binaries with equal mass components.
Filled symbols represent member stars and colours correspond to the membership likelilhoods, as showed in the colour bars. Open circles are non-members and small black dots are stars in a control field.

Additionally, we ran the photometric decontamination procedure described in \cite{Maia:2010} in order to identify possible member candidates without astrometric information in GAIA. For this procedure, we used the photometric data for the same groups of stars (OC and control field) employed in Section \ref{memberships}. 
Stars with photometric membership likelihoods $\mathcal{L}_{\textrm{phot}}\ge0.60$, but without available astrometry, are plotted as filled black squares in the CMDs of  BH\,150, Czernik\,37, NGC\,6756, Ruprecht\,111 and Trumpler\,25. For reference, we circled the member stars (filled symbols) for which $\mathcal{L}_{\textrm{phot}}\ge0.10$. In the case of NGC\,6249, we identified three bright stars, labeled as \#1, \#2 and \#586 in  Fig.~\ref{CMDs_parte2} and located inside the OC's $r_t$, which could have been considered member stars in a purely photometric analysis. The astrometric data $\mu_{\alpha}$(mas\,yr$^{-1}$), $\mu_{\delta}$(mas\,yr$^{-1}$) and $\varpi$(mas) are: (-4.182$\pm$0.088, -7.707$\pm$0.068, 1.7587$\pm$0.0435)$_{\#1}$, (-8.139$\pm$0.099, -15.823$\pm$0.083, 1.4016$\pm$0.0524)$_{\#5}$ and (1.447$\pm$0.103, -3.183$\pm$0.083, 0.9423$\pm$0.0512)$_{\#586}$. These stars received membership likelihoods smaller than 0.3, since their parallaxes are incompatible with the group of members and/or their proper motion components are inconsistent with the bulk motion of the OC. The complete set of astrophysical parameters for the studied sample is showed in Table \ref{astroph_params}. In this table, $M_{\textrm{Kroupa}}$ and $N_{\textrm{Kroupa}}$ are upper limits for OC's mass and number of stars, respectively.

\begin{table*}
 %\begin{minipage}{85mm}
 \begin{center}
  \caption{Fundamental parameters, half-light relaxation times and photometric masses ($M_{\textrm{phot}}$) for the studied OCs. $M_{\textrm{Kroupa}}$ and $N_{\textrm{Kroupa}}$ are upper limits for OC mass and number of stars, respectively.}
  \label{astroph_params}
 \footnotesize
 \begin{tabular}{lccccccccc}

\hline

Cluster          &  (m-M)$_{0}$          & $d$                 &   $E(B-V)$           &   log($t$/yr)   & $[Fe/H]$       &   $t_{rh}$                   & $M_{\textrm{phot}}$             & $M_{\textrm{Kroupa}}$   &  $N_{\textrm{Kroupa}}$    \\

       & (mag)      & (kpc)      & (mag)          &         & (dex)       & (Myr)          & ($M_{\odot}$)      & ($M_{\odot}$)   &     \\                                                                                                                                    
\hline

Collinder\,258  &10.60\,$\pm$\,0.50   &1.32\,$\pm$\,0.30   &0.19\,$\pm$\,0.15   &8.15\,$\pm$\,0.30     & 0.18\,$\pm$\,0.22    & 1.10\,$\pm$\,0.32   &  80\,$\pm$\,15     & 244     & 524  \\
NGC\,6756       &11.45\,$\pm$\,0.30   &1.95\,$\pm$\,0.27   &1.09\,$\pm$\,0.10   &8.35\,$\pm$\,0.20     & 0.00\,$\pm$\,0.17    & 3.13\,$\pm$\,0.70   & 481\,$\pm$\,33     &1596     &3617  \\
Czernik\,37     &10.95\,$\pm$\,0.40   &1.55\,$\pm$\,0.28   &1.54\,$\pm$\,0.10   &8.50\,$\pm$\,0.15     & 0.00\,$\pm$\,0.17    & 3.35\,$\pm$\,0.73   & 403\,$\pm$\,31     &1835     &4202  \\
NGC\,5381       &11.60\,$\pm$\,0.30   &2.09\,$\pm$\,0.29   &0.74\,$\pm$\,0.05   &8.60\,$\pm$\,0.10     &-0.32\,$\pm$\,0.06    &10.72\,$\pm$\,3.10   & 376\,$\pm$\,26     &1016     &2400  \\
Trumpler\,25    &11.20\,$\pm$\,0.30   &1.74\,$\pm$\,0.24   &1.20\,$\pm$\,0.10   &8.10\,$\pm$\,0.10     & 0.05\,$\pm$\,0.15    & 8.80\,$\pm$\,1.37   &1211\,$\pm$\,58     &4111     &8837  \\
BH\,150         &12.40\,$\pm$\,0.30   &3.02\,$\pm$\,0.42   &1.57\,$\pm$\,0.10   &7.35\,$\pm$\,0.10     & 0.00\,$\pm$\,0.23    & 1.97\,$\pm$\,0.42   & 503\,$\pm$\,52     &1962     &3740  \\
Ruprecht\,111   &11.40\,$\pm$\,0.20   &1.91\,$\pm$\,0.17   &0.88\,$\pm$\,0.10   &8.50\,$\pm$\,0.10     &-0.13\,$\pm$\,0.23    & 2.47\,$\pm$\,0.86   & 152\,$\pm$\,19     & 714     &1638  \\
Ruprecht\,102   &12.50\,$\pm$\,0.40   &3.16\,$\pm$\,0.58   &0.84\,$\pm$\,0.15   &8.45\,$\pm$\,0.25     & 0.22\,$\pm$\,0.14    & 6.35\,$\pm$\,1.32   & 221\,$\pm$\,22     & 820     &1880  \\
NGC\,6249       &10.30\,$\pm$\,0.40   &1.15\,$\pm$\,0.21   &0.44\,$\pm$\,0.10   &8.30\,$\pm$\,0.25     & 0.10\,$\pm$\,0.18    & 1.22\,$\pm$\,0.37   & 109\,$\pm$\,15     & 303     & 689  \\
Basel\,5        &11.20\,$\pm$\,0.40   &1.74\,$\pm$\,0.32   &0.66\,$\pm$\,0.10   &8.35\,$\pm$\,0.15     & 0.22\,$\pm$\,0.14    & 2.75\,$\pm$\,0.78   & 193\,$\pm$\,19     & 509     &1136  \\
Ruprecht\,97    &12.55\,$\pm$\,0.30   &3.24\,$\pm$\,0.45   &0.56\,$\pm$\,0.15   &8.10\,$\pm$\,0.15     & 0.31\,$\pm$\,0.11    &18.85\,$\pm$\,3.78   & 512\,$\pm$\,34     &1491     &3234  \\
ESO\,129-SC32   &12.85\,$\pm$\,0.30   &3.71\,$\pm$\,0.51   &0.75\,$\pm$\,0.10   &8.30\,$\pm$\,0.15     & 0.00\,$\pm$\,0.11    &10.49\,$\pm$\,2.78   & 369\,$\pm$\,29     &1141     &2560  \\

\hline

\end{tabular}
%\end{minipage}
\end{center}
\end{table*}

\subsection{Mass functions}
\label{mass_functions}

\begin{figure*}
\begin{center}

\parbox[c]{1.0\textwidth}
  {
   
    \includegraphics[width=0.333\textwidth]{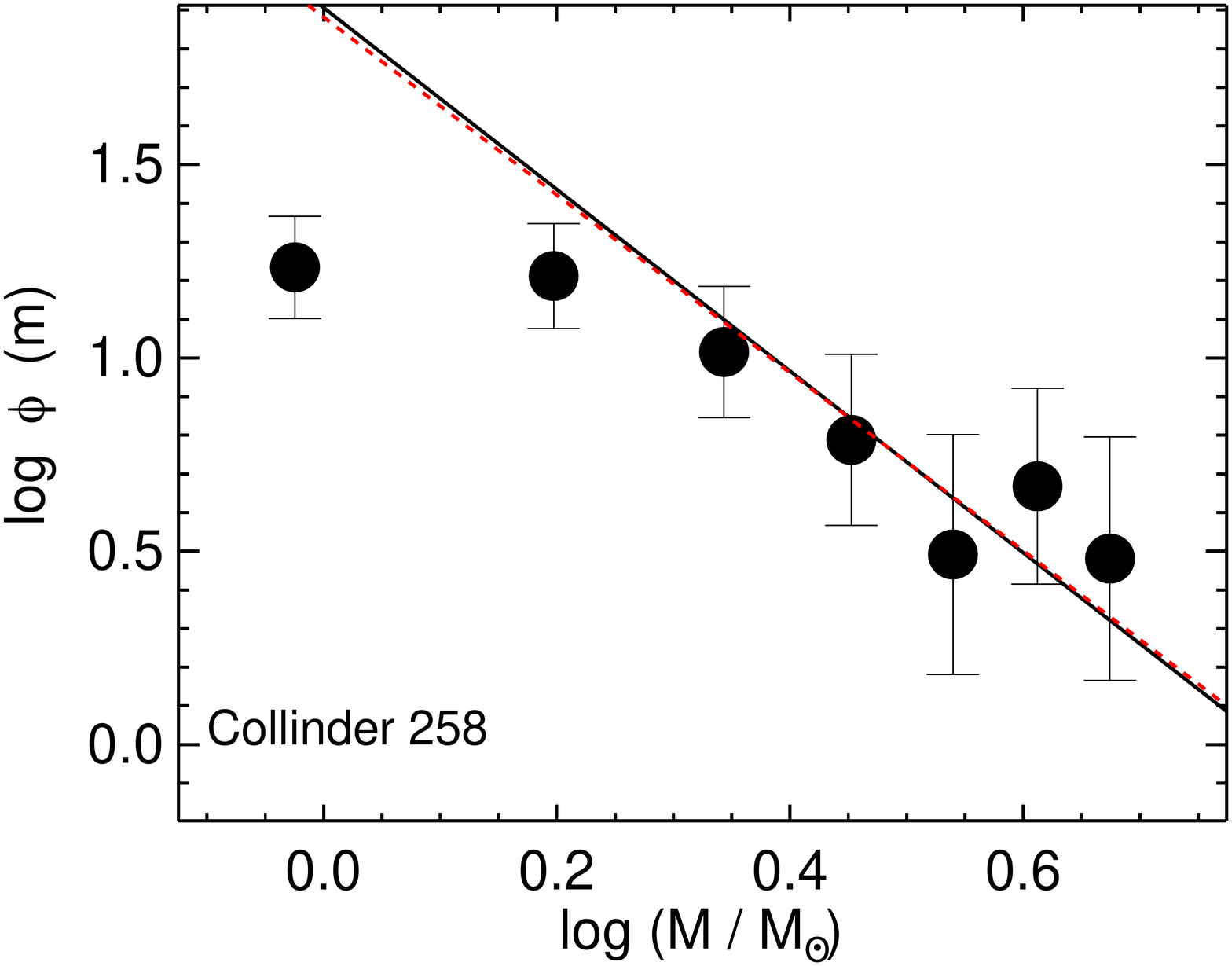}
    \includegraphics[width=0.333\textwidth]{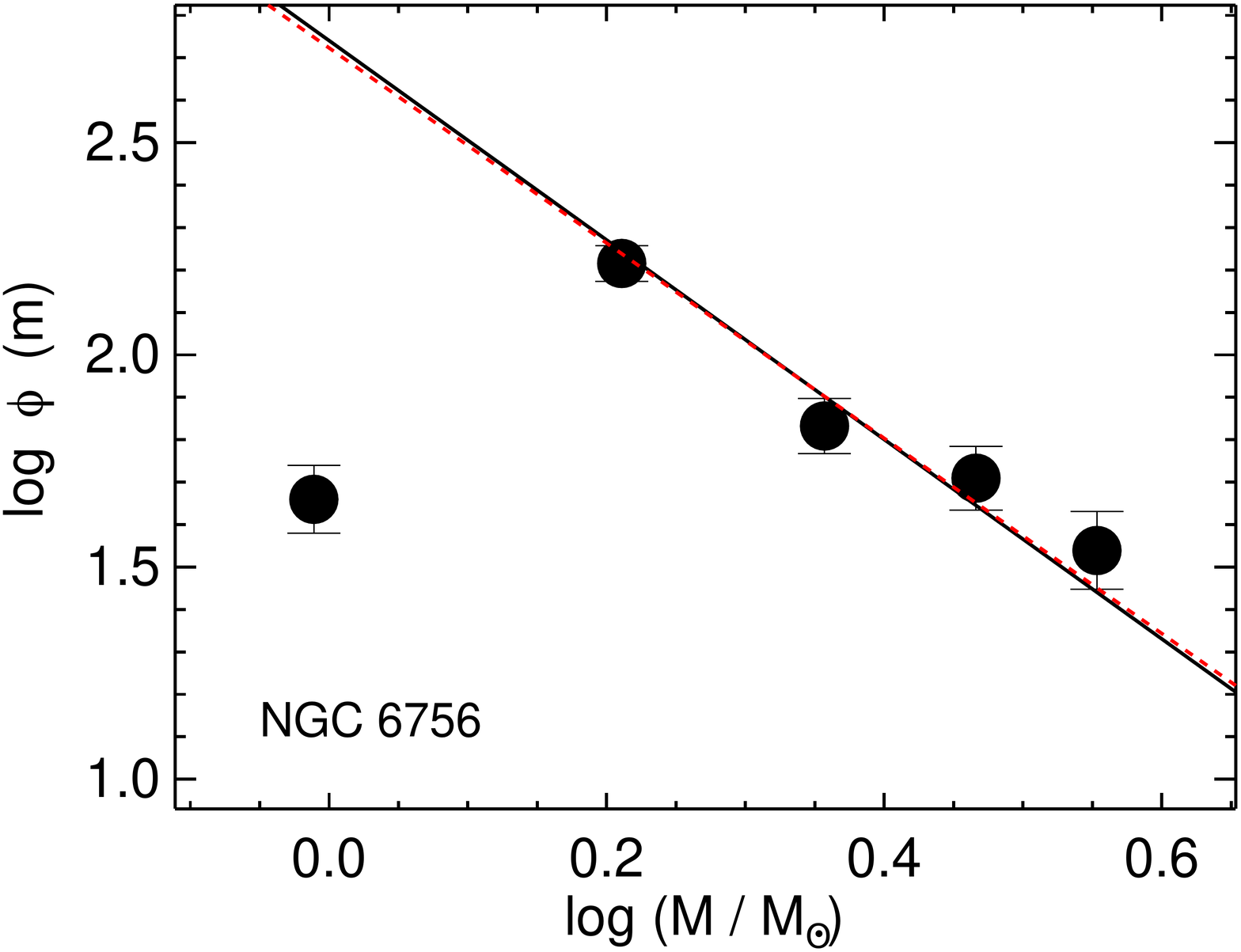}
    \includegraphics[width=0.333\textwidth]{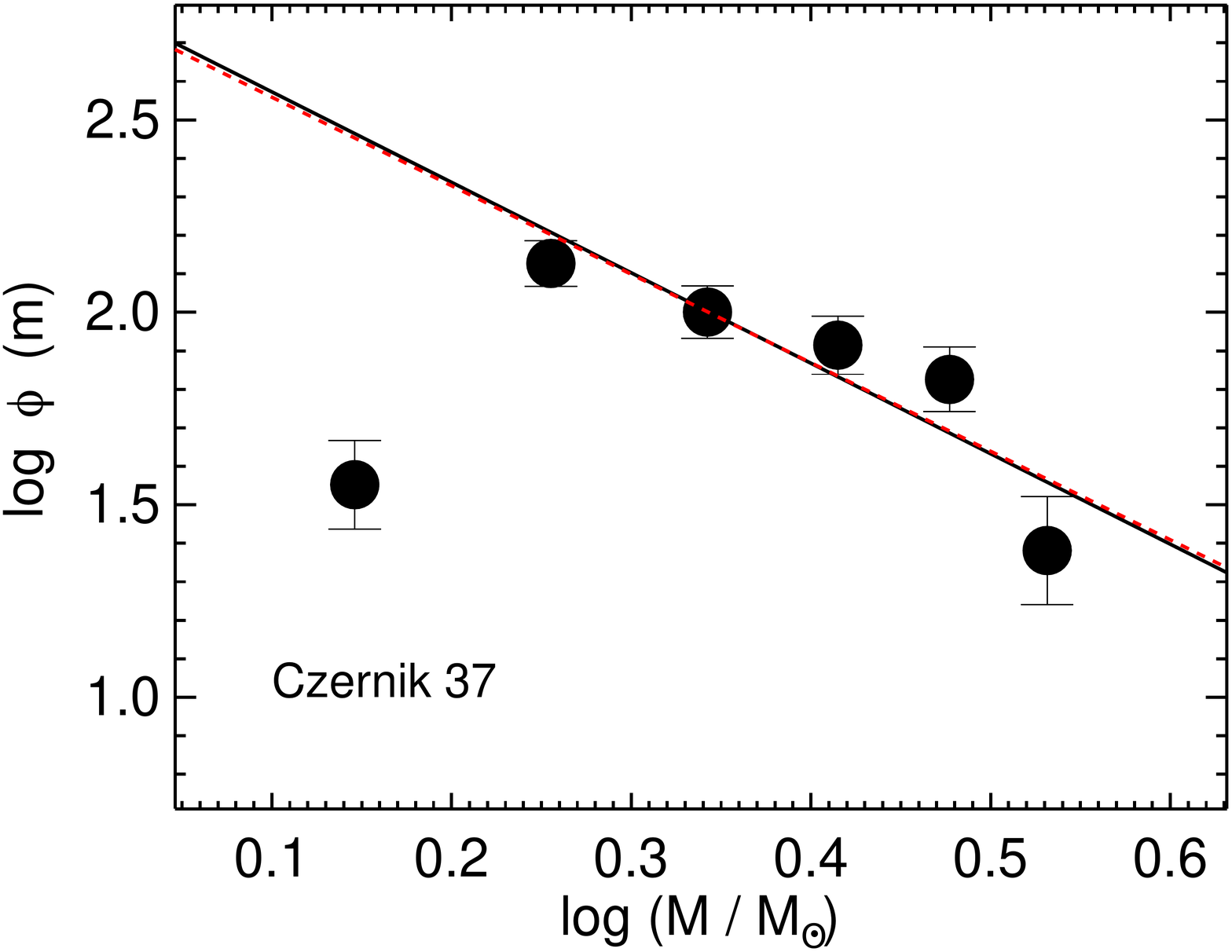}    
    \includegraphics[width=0.333\textwidth]{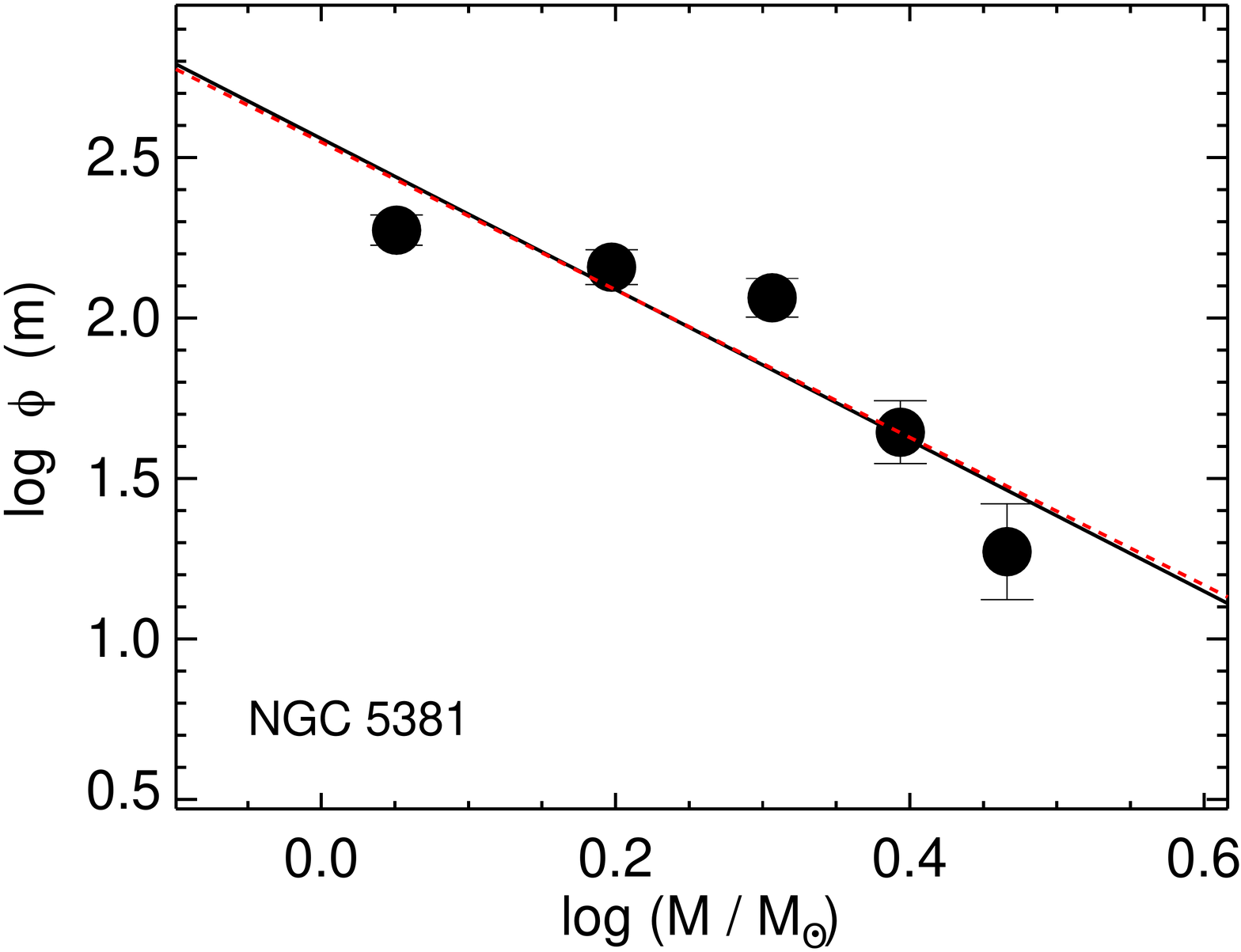}
    \includegraphics[width=0.333\textwidth]{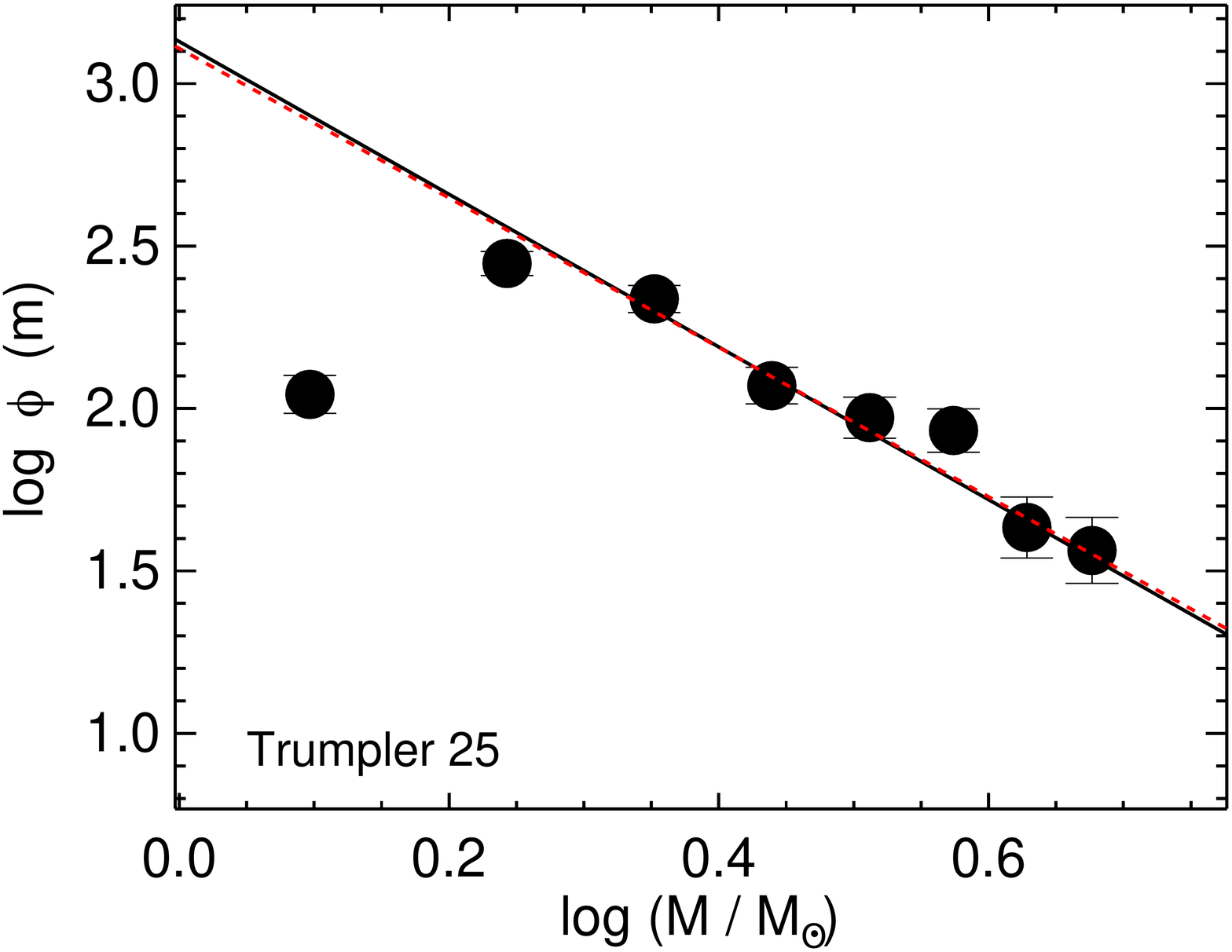}
    \includegraphics[width=0.333\textwidth]{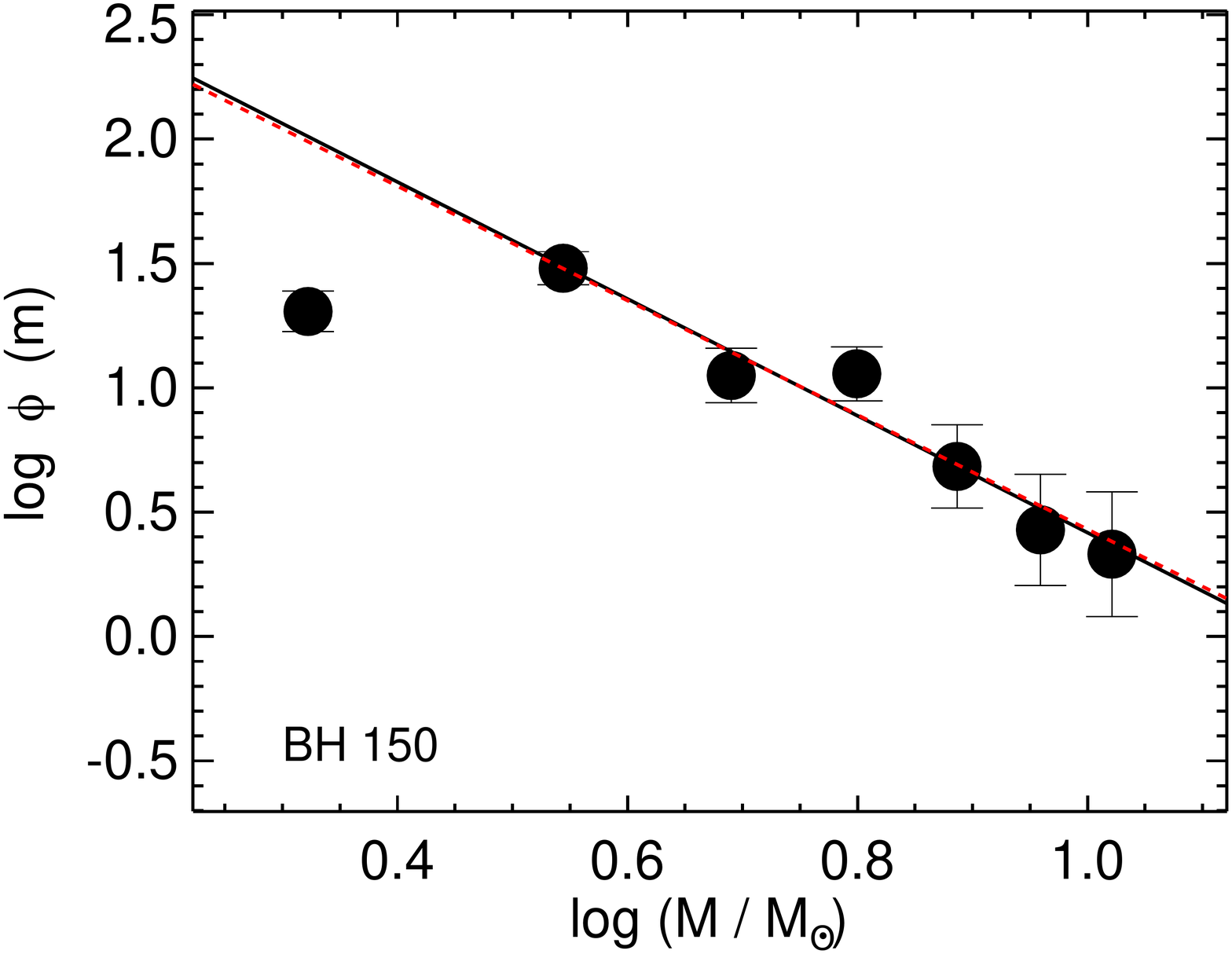}
    \includegraphics[width=0.333\textwidth]{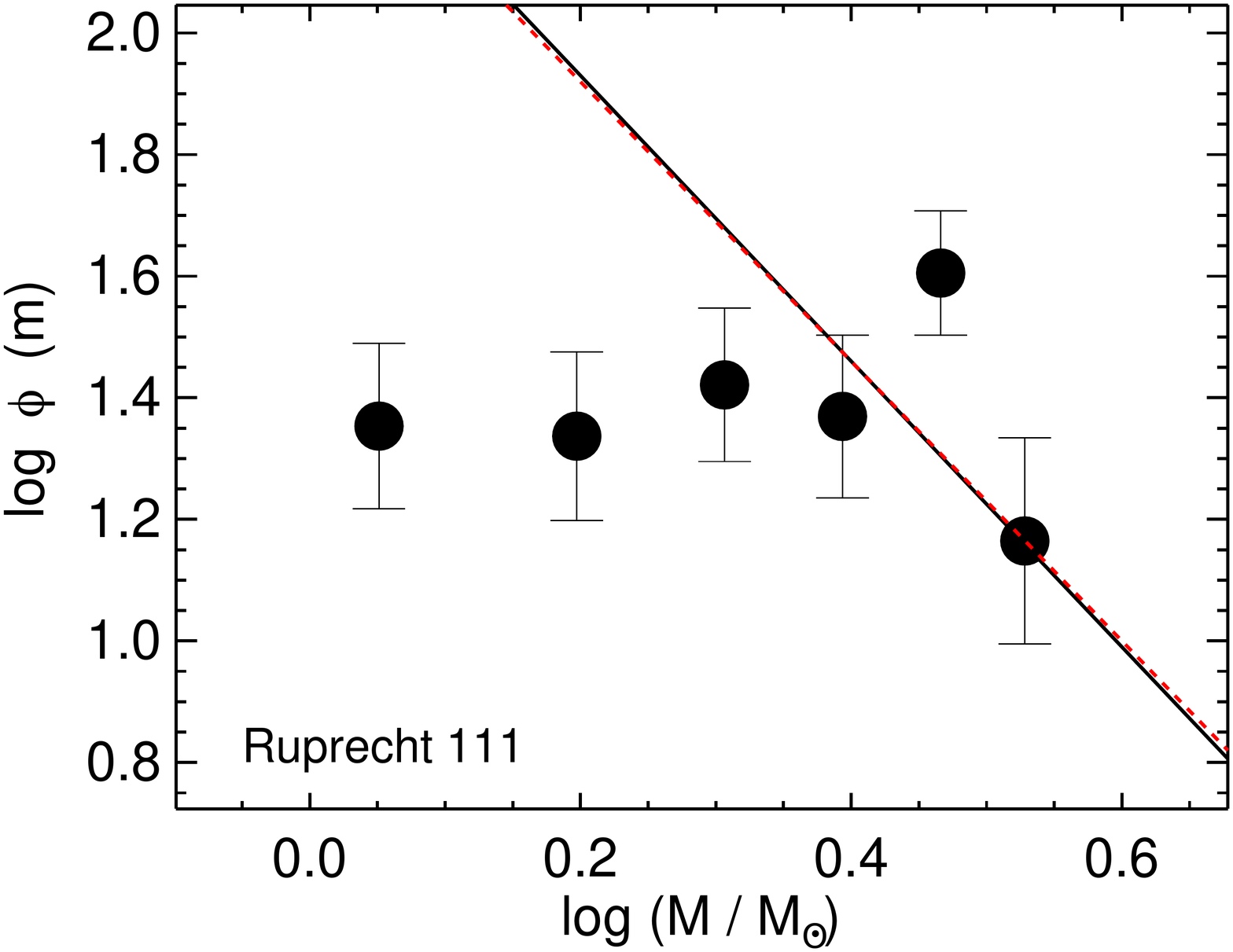}  
    \includegraphics[width=0.333\textwidth]{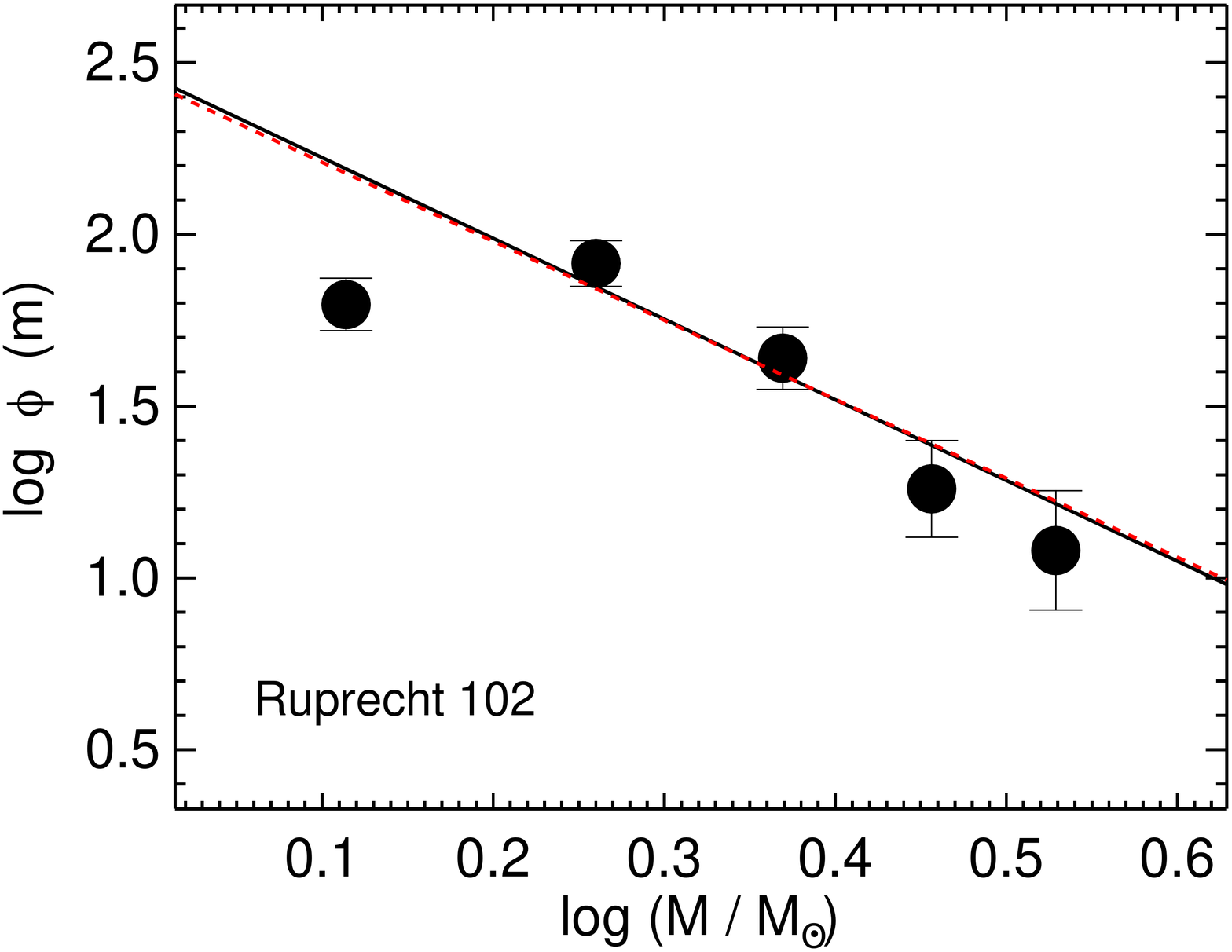} 
    \includegraphics[width=0.333\textwidth]{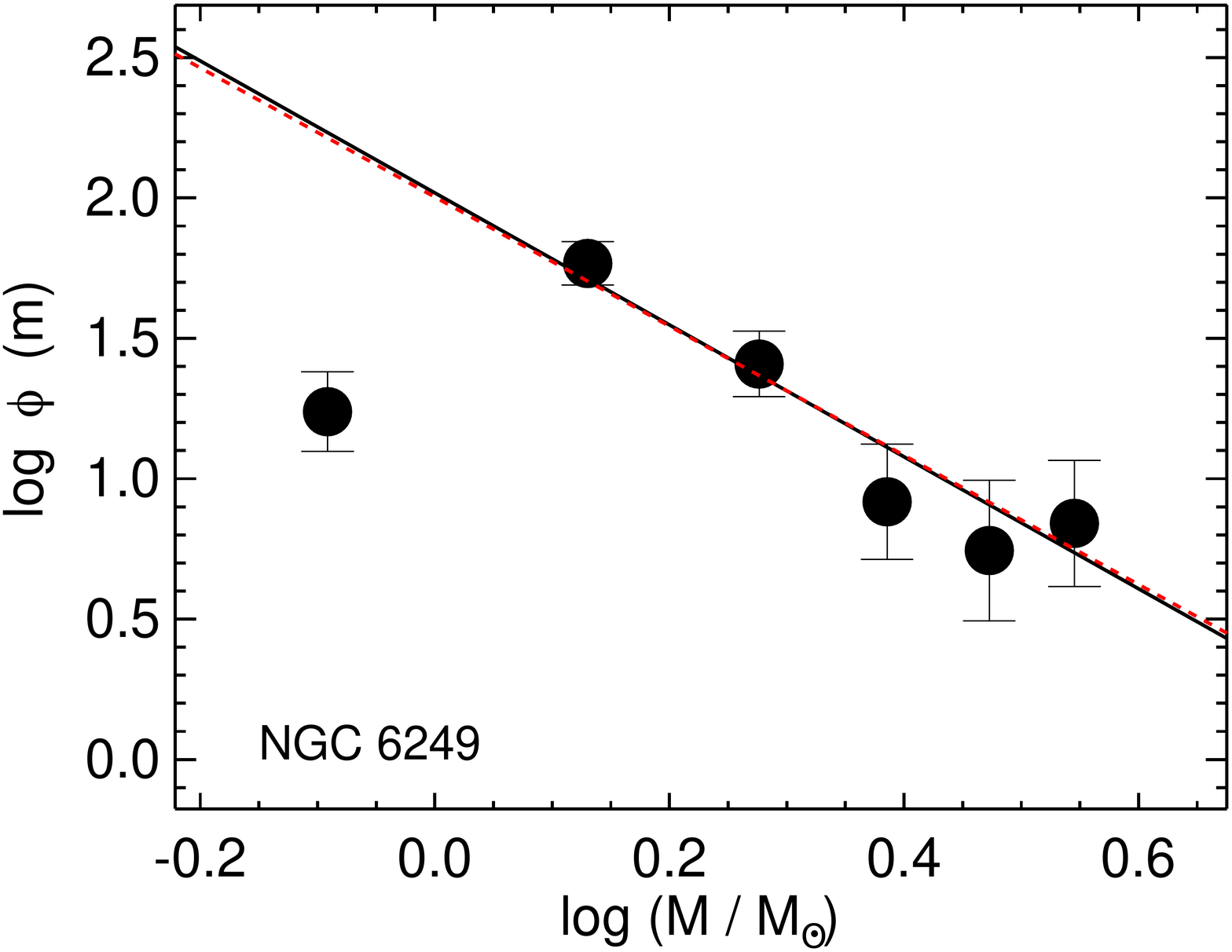}  
    \includegraphics[width=0.333\textwidth]{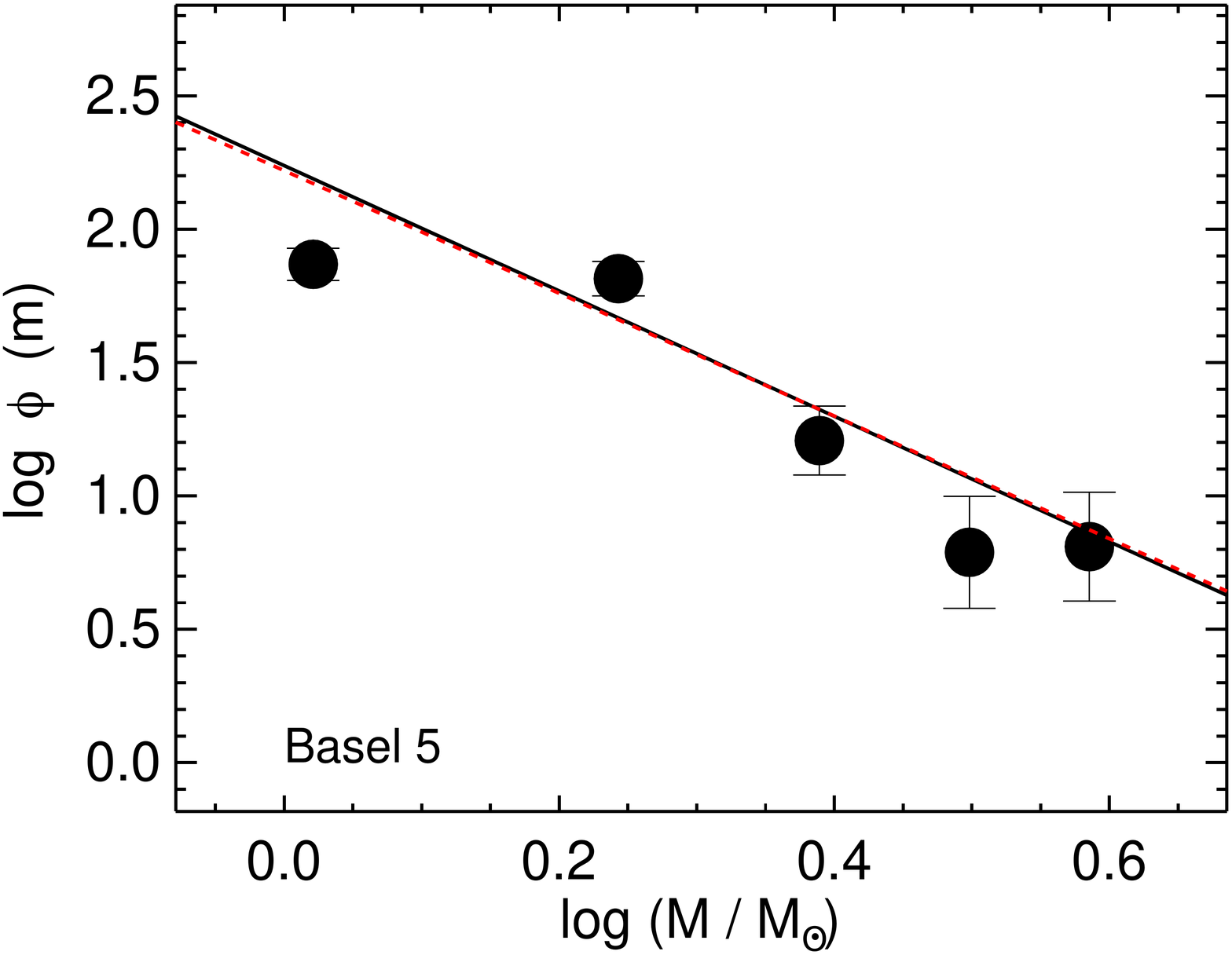}  
    \includegraphics[width=0.333\textwidth]{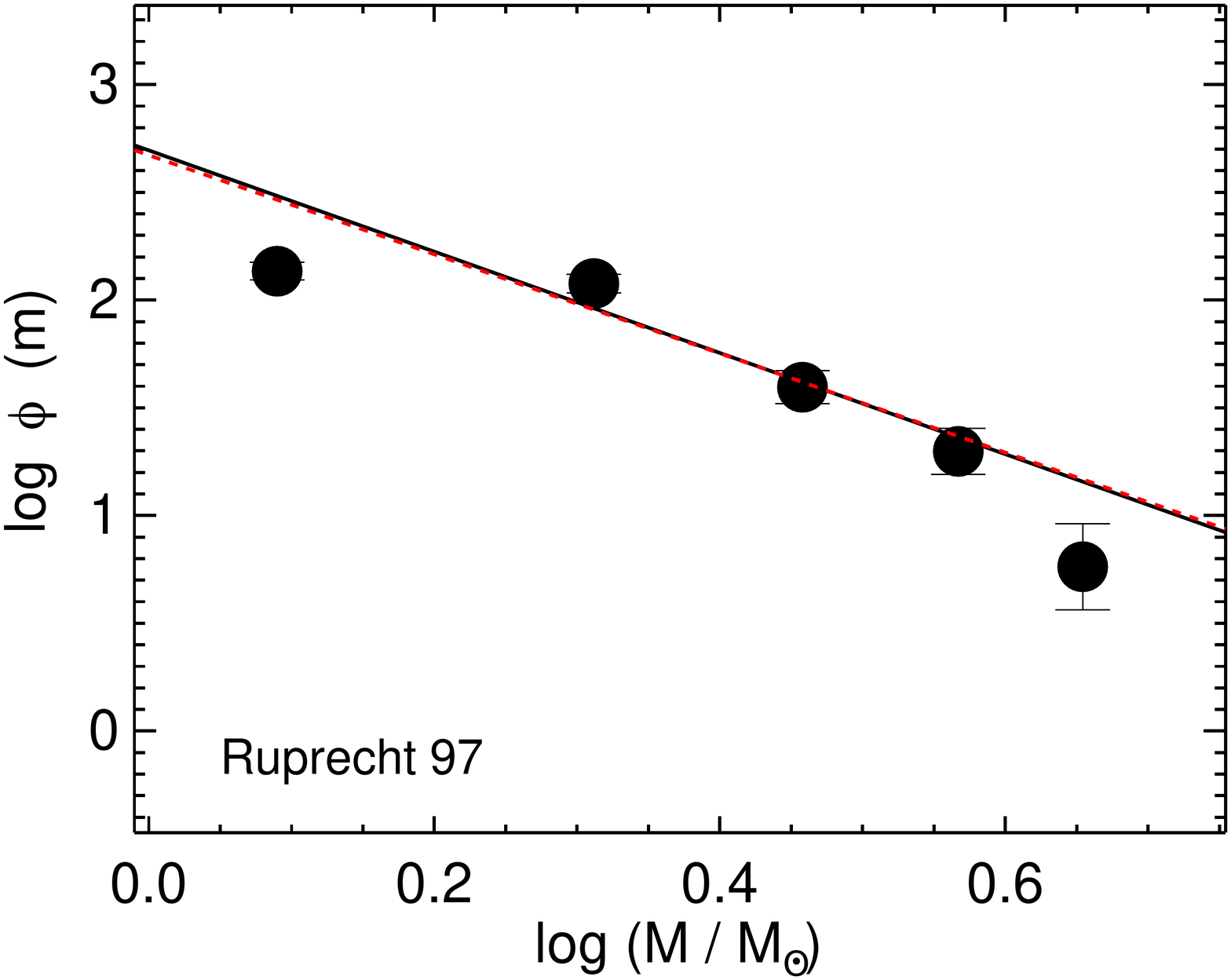}  
    \includegraphics[width=0.333\textwidth]{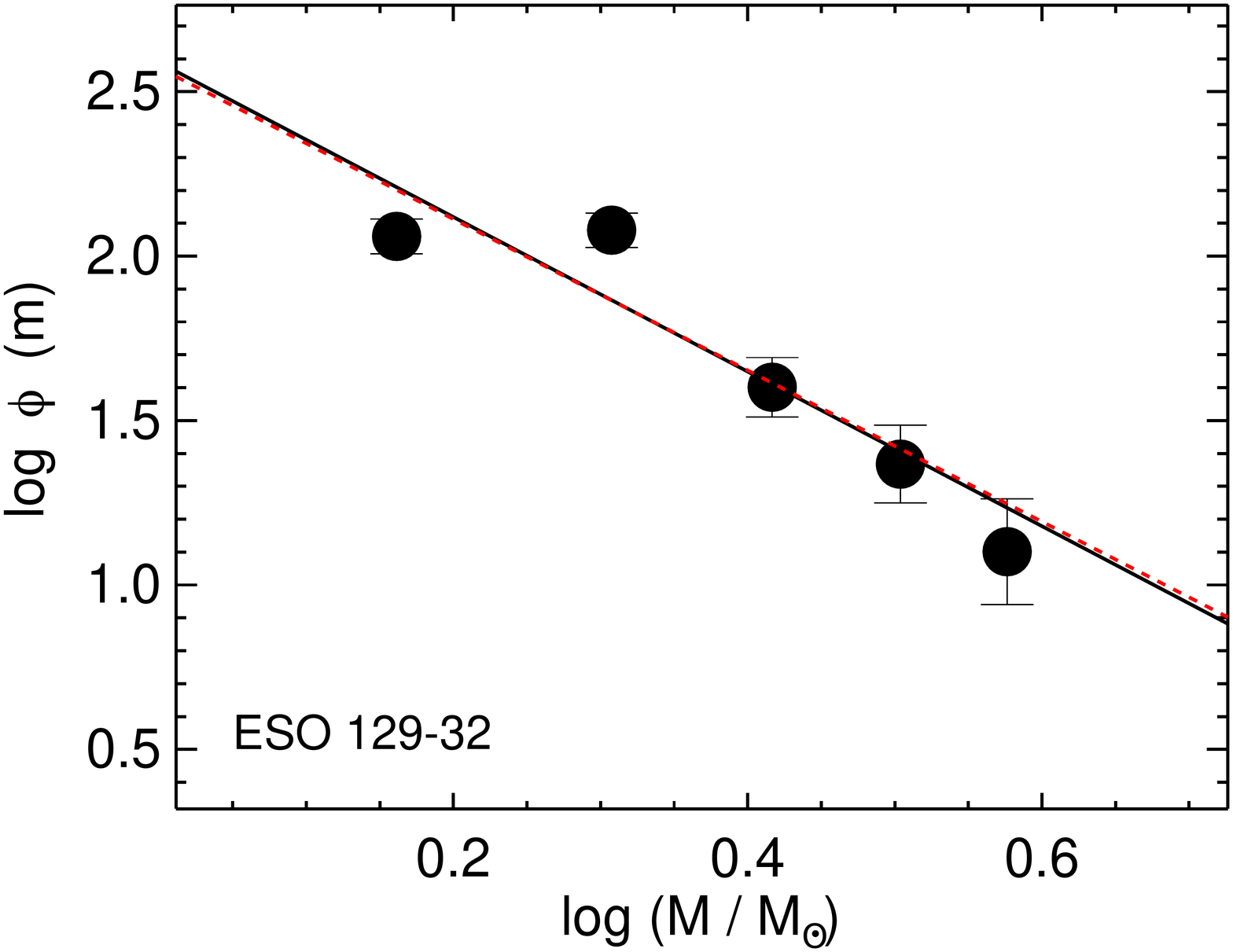} 
  }
\caption{ Mass functions for the studied OC sample. The scaled IMFs of Salpeter\,(1955) and Kroupa\,(2001) 
are overplotted  with black and red lines, respectively.  }

\label{mass_func_parte1}
\end{center}
\end{figure*}

Individual masses for member stars were estimated by interpolation from the observed $T_1$ magnitude along the best-fitted isochrone, properly shifted according to the OC reddening and distance modulus (Fig.~\ref{CMDs_parte1} and Table \ref{astroph_params}). The OCs' mass functions ($\phi(m)=dN/dm$) were then built by counting the number of stars in linear mass bins which were then converted to the logarithmic scale, as showed in Fig.~ \ref{mass_func_parte1}. Stars counts inside each mass bin were weighted by the membership likelihoods and properly corrected for photometric completeness. The total photometric OC masses were obtained from discrete sum of the mass bins, while the corresponding uncertainties were determined from Poisson statistics.

For comparison purposes, we overplotted in  Figs.~\ref{mass_func_parte1} the initial mass functions (IMF) of \citeauthor{Salpeter:1955}\,\,(\citeyear{Salpeter:1955}, continous black lines) and \citeauthor{Kroupa:2001}\,\,(\citeyear{Kroupa:2001}, red dashed lines), which were scaled according to the OC total photometric mass (Table \ref{astroph_params}). The signal of 
low-mass stars depletion is a consequence of preferential evaporation 
during the OCs dynamical evolution (see Section 5). Other OCs (e.g., ESO\,129-SC32, NGC\,5381) present less noticeable depletion in lower mass bins, since the observed mass function is more compatible with Kroupa's and Salpeter's IMF.

Finally, we estimated Jacobi radii from the expression:

\begin{equation}
   R_J = \left(\frac{M_{\textrm{clu}}}{3\,M_G}\right)^{1/3}\times R_G
\end{equation} 

\noindent
where $M_{\textrm{clu}}$ is the OC's photometric mass (Table \ref{astroph_params}). This formula assumes a circular orbit around a point mass galaxy ($M_{G}\sim1.0\times10^{11}\,M_{\odot}$; \citeauthor{Carraro:1994}\,\,\citeyear{Carraro:1994}; \citeauthor{Bonatto:2005}\,\,\citeyear{Bonatto:2005}; \citeauthor{Taylor:2016}\,\,\citeyear{Taylor:2016}). With the estimated OCs' masses, we also derived their half-light relaxation times (\citeauthor{Spitzer:1971}\,\,\citeyear{Spitzer:1971}):

\begin{equation}
   t_{\textrm{rh}}=8.9\times10^5\,\textrm{yr}\times\frac{M_{\textrm{clu}}^{1/2}\,r_{\textrm{h}}^{3/2}}{\langle m\rangle\,\textrm{log}_{10}(0.4M_{\textrm{clu}}/\langle m\rangle)}
\end{equation}

\noindent
where $\langle m\rangle$ is the mean mass of the OC stars.

\subsection{Comments on individual clusters}

The present OC sample was also investigated by \cite{Kharchenko:2013}, who employed proper motions from the PPMXL catalogue \citep{Roeser:2010} and photometric data from the near-infrared 2MASS catalogue \citep{Skrutskie:2006}. Their analysis was based on a dedicated data-processing pipeline to determine kinematic and photometric membership probabilities for stars in the cluster regions. In this section, we point out some similarities and differences between their and our results. Some other previous studies are also highlighted.

We found  a very good agreement with the results of \cite{Kharchenko:2013} for Collinder\,258, NGC\,6756, Czernik\,37, NGC\,5381, Trumpler\,25, Ruprecht\,111 and NGC\,6249, while some
differences arise for BH\,150, Ruprecht\,102, Basel\,5, Ruprecht\,97 and ESO\,129-SC32. 
For instance, distance moduli differ  from $\sim1\,$mag (Basel\,5) to $\sim2\,$mag 
(ESO\,129-SC32); differences in log($t$) vary  from $0.2\,$ (BH\,150) to $\sim1\,$ 
(Ruprecht\,102); $E(B-V)$ colour excesses are typically $\sim0.3\,$mag lower. 
We speculate that the main reason for these differences is related to the membership assignment procedures used. Kharchenko et al.'s method is based on lower quality proper motions data,
alongside the fact that parallaxes are not available in the PPMXL catalogue. Likewise, 
the photometric completeness of 2MASS data in regions projected close do the Galactic centre
may also play a role. Indeed, our CMDs (see Fig.~\ref{CMDs_parte1}) exhibit deeper main 
sequences, reaching nearly 3-4 mag below the clusters´ main sequence turnoffs.

\subsubsection{NGC\,6756}
NGC\,6756 was studied by \cite{Netopil:2013}, who employed publicy available Str\"omgren $uvby$ data in their analysis. Empirical relations between effective temperatures and reddening-free indices $[u-b]$ were previously calibrated. The cluster parameters were derived by means of the method proposed by \cite{Pohnl:2010}, which is based on the fit of differential evolutionary tracks (normalised to the zero age main-sequence, ZAMS) for a variety of metallicity/age combinations.

The derived metallicity ($[Fe/H]=0.10\pm0.14$), colour excess ($E(B-V)=1.03\pm0.05\,$mag) and age (log $t$/yr=8.10) are in agreement with our results (see Table \ref{astroph_params}), considering uncertainties. Their derived distance modulus ((m-M)$_0$=12.3\,mag), however, does not agree with our analysis. This discrepancy may be partially attributed to the different sets of isochrones employed in both studies and, more importantly, to the adopted criteria in the selection of member stars. In the analysis of NGC\,6756, \cite{Netopil:2013} employed only photometric data (see their table 2). Despite this, as stated above, the distance modulus obtained by \cite{Kharchenko:2013} ((m-M)$_0$ = 11.45\,mag) is in excellent agreement with our result.

Additionally, NGC\,6756 is present in the sample of \cite{Santos:2004}, who obtained equivalent widths from  integrated spectra and presented homogeneous scales of ages and metallicities for clusters younger than 10\,Gyr. They derived $[Fe/H]=0.0\pm0.2$ and log($t$/yr)=$8.48\pm0.15$, which are in fair agreement with our analysis (Table \ref{astroph_params}).

\subsubsection{Czernik\,37 and NGC\,5381}
Both OCs have also been investigated by \cite{Marcionni:2014}, who used the same
 public images as in Section \ref{data_collection_reduction}. They derived, for Czernik\,37,  $(m-M)_0=10.8\pm1.3\,$mag, log$(t/$yr)=$8.40\pm0.14$ and $E(B-V)=1.47\pm0.25\,$mag, and for NGC\,5381, $(m-M)_0=12.1\pm0.3\,$mag, 
log$(t/$yr)=$8.40\pm0.10$ and $E(B-V)=0.46\pm0.25\,$mag, respectively. 
Within the quoted uncertainties our values are in good agreement with theirs, except in the case of
NGC\,5381's colour excess. Nevertheless, we would like to point out that they did not employ 
GAIA DR2 data. Consequently, we provide with a larger and more reliable
list of member stars.     
 
\subsubsection{NGC\,6249}
\cite{Mermilliod:2008} determined the mean cluster radial velocity. Two stars ($\alpha=254^{\circ}\!\!.41346$, $\delta=-44^{\circ}\!\!.79994$ and $\alpha=254^{\circ}\!\!.42387$, $\delta=-44^{\circ}\!\!.78798$) observed by them are also present in our sample. They have been identified with numbers $\#1$ and $\#5$ in Fig.~\ref{CMDs_parte2}. Both stars could be considered members in a purely photometric analysis, but they were discarded as cluster members because they received very low membership likelihoods ($L<1$\%) due to very discrepant parallaxes ($\varpi_{\#1}=1.7587\,\pm\,$0.0435\,mas and $\varpi_{\#5}=1.4016\,\pm\,0.0524\,$mas) compared to the bulk of member stars (Fig.~\ref{T1mag_plx}).

\subsubsection{Ruprecht\,97}
\cite{Claria:2008} obtained Washington $CMT_1T_2$ photoeletric data for 6 red giant candidates of Ruprecht\,97. Their sample was defined based on the previous photometric study of \cite{Moffat:1975}. This 6 stars were also observed in the present study. Photometry from both studies proved to be consistent with each other, since the mean differences between our magnitudes and the literature ones resulted: $\langle C_{\textrm{our}}-C_{\textrm{lit}}\rangle=0.011\pm0.058\,$mag and $\langle T_{\textrm{1,our}}-T_{\textrm{1,lit}}\rangle=-0.062\pm0.019\,$mag, that is, a relatively small systematic offset was found for the $T_{1}$ magnitudes.  

By making use of Washington metallicity-sensitive indices \citep{Geisler:1991}, they identified stars \#4 and \#11 as members (their table 6), for which $[Fe/H]=-0.03\pm0.03$. Both stars were discarded from our membership analysis as cluster members, which could explain the difference in the derived cluster metallicity. Indeed, star \#4 is located at is located at $\sim6{\arcmin}$ from the cluster centre (Table \ref{struct_params}), beyond the cluster limiting radius and therefore in a region dominated by field stars (Fig.~\ref{RPDs_parte1}). After running our decontamination method, star \#11 received a membership likelihood of $\sim5$ per cent, because its parallax ($\varpi=0.5922\pm0.0298\,$mas) is incompatible with the bulk of cluster members (Figure \ref{T1mag_plx}). These results highlight the need of a proper characterization scheme to refine the selection of cluster members, thus leading to a more reliable determination of astrophysical parameters.

\subsubsection{BH\,150}
The physical nature of BH\,150 has been until now under debate. \cite{Kharchenko:2013} classified this object as ``dubious". On the other hand, \cite{Carraro:2005a} -- based only on the analysis of $V\times(B-V)$ and $V\times{(V-I)}$ CMDs -- could not derive any
astrophysical parameters, nor draw any conclusion on its existence as a genuine physical system
either. Due to the large field star contamination along the line of sight, their method did not allow them a reliable disentanglement between the object and the composite field population. Here, once we applied the decontamination procedure described in Section \ref{memberships}, we could identify a group of stars with compatible kinematics and parallaxes and to reveal clearer evolutionary sequences in its CMD (Fig.~\ref{CMDs_parte1}). Our results favour the hypothesis of a real OC.

%\subsubsection{Ruprecht\,97}
%The two brightest member stars of Ruprecht\,97 ($T_{1}<12\,$mag, see Fig.~\ref{CMDs_parte1}) received high membership likelihoods ($L\gtrsim0.9$) and they were also observed spectroscopically by \cite{Mermilliod:2008}, who provides mean cluster radial velocities ($V_\textrm{rad}$). 
%These $V_\textrm{rad}$ are $-10.27$\,$\pm$\,0.32 and $-1.74$\,$\pm\,$1.86 for the redder and
%bluer star, respectively.

%The brightest member star ($\alpha=179^{\circ}\!\!.40250$, $\delta=-62^{\circ}\!\!.73120$; $T_1=11.34\,$mag) and the second brightest one ($\alpha=179^{\circ}\!\!.36346$, $\delta=-62^{\circ}\!\!.70167$; $T_1=11.57\,$mag) have, respectively, the following values for $\mu_{\alpha}$\,(mas.yr$^{-1}$), $\mu_{\delta}$\,(mas.yr$^{-1}$), $\varpi$\,(mas) and $V_r\,$(km.s$^{-1}$): ($-5.854$\,$\pm$\,0.041, 0.399\,$\pm$\,0.042, 0.2256\,$\pm$\,0.0282, $-10.27$\,$\pm$\,0.32) and ($-6.373$\,$\pm$\,0.043, 0.443\,$\pm$\,0.040, 0.2469\,$\pm$\,0.0311, $-1.74$\,$\pm\,$1.86).

\section{Discussion}
\label{discussion}

\begin{figure*}
\begin{center}

  \includegraphics[width=1.0\textwidth]{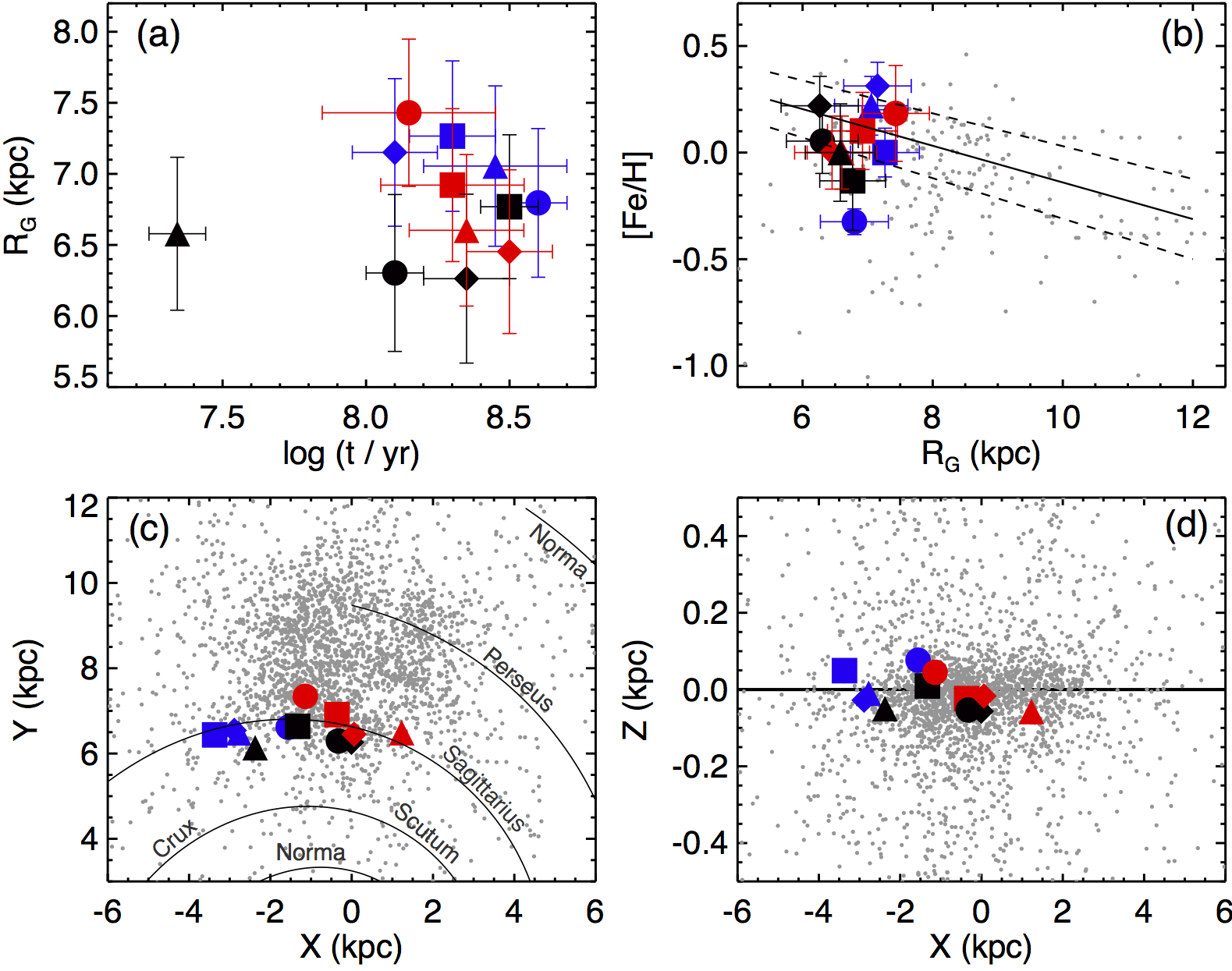}
%\caption{ Panel (a): Galactocentric distance $R_{\textrm{G}}\,$ versus log($t$/yr) for the studied OC sample. Symbols' colours were assigned according to the different $r_{h}/R_J$ 
%ranges described in the text. Red symbols: Collinder\,258 (\textcolor{red}{$\CIRCLE$}), NGC\,6756 (\textcolor{red}{$\blacktriangle$}), Czernik\,37 (\textcolor{red}{$\blacklozenge$}) and NGC\,6249 (\textcolor{red}{$\blacksquare$}); black symbols: Trumpler\,25 ($\CIRCLE$), BH\,150 ($\blacktriangle$), Ruprecht\,111 ($\blacksquare$), Basel\,5 ($\blacklozenge$); blue symbols: NGC\,5381 (\textcolor{blue}{$\CIRCLE$}), Ruprecht\,102 (\textcolor{blue}{$\blacktriangle$}), Ruprecht\,97 (\textcolor{blue}{$\blacklozenge$}) and ESO\,129-SC32 (\textcolor{blue}{$\blacksquare$}). Panel (b): Age versus metallicity $[Fe/H]$ plot. Panel (c): Distribution of the OCs projected on to the Galactic plane. The spiral pattern was taken from Vall\'ee\,(2008). Panel (d): Distribution of OCs perpendicular to the Galactic plane (horizontal line). The grey dots represent OCs taken from Khachenko et al. (2013).}

\caption{ Panel (a): Galactocentric distance $R_{\textrm{G}}\,$ versus log($t$/yr) for the studied OC sample. Symbols' colours were assigned according to the different $r_{h}/R_J$ 
ranges described in the text. Red symbols: Collinder\,258 (\textcolor{red}{$\CIRCLE$}), NGC\,6756 (\textcolor{red}{$\blacktriangle$}), Czernik\,37 (\textcolor{red}{$\blacklozenge$}) and NGC\,6249 (\textcolor{red}{$\blacksquare$}); black symbols: Trumpler\,25 ($\CIRCLE$), BH\,150 ($\blacktriangle$), Ruprecht\,111 ($\blacksquare$), Basel\,5 ($\blacklozenge$); blue symbols: NGC\,5381 (\textcolor{blue}{$\CIRCLE$}), Ruprecht\,102 (\textcolor{blue}{$\blacktriangle$}), Ruprecht\,97 (\textcolor{blue}{$\blacklozenge$}) and ESO\,129-SC32 (\textcolor{blue}{$\blacksquare$}). Panel (b): Radial metallicity $[Fe/H]$ distribution. The continuous line is the relationship derived by Netopil et al. (2016). The dashed lines represent its upper and lower limits. Panel (c): Distribution of the OCs projected on to the Galactic plane. The spiral pattern was taken from Vall\'ee\,(2008). Panel (d): Distribution of OCs perpendicular to the Galactic plane (horizontal line). The grey dots represent OCs taken from Kharchenko et al. (2013) and Dias et al. (2002).}

 \label{Rgal_versus_age}
\end{center}

\end{figure*}

The OC sample studied in this work present nearly similar ages and Galactocentric distances, with the sole exception of BH\,150 (see Fig.~\ref{Rgal_versus_age}, panel (a)). They are located close to the Galactic plane ($\vert Z\vert\leq75\,$pc) and are part of the Sagittarius arm (see panels (c) and (d)). Their colour-excess $E(B-V)$ vary from $\sim0.2$ up to $\sim1.6$ (Table \ref{astroph_params}), witnessing the different amount of dust and gas distributed along their line of sight, as expected. 

As shown in Fig.~\ref{Rgal_versus_age}, panel (b), they span metallicities from slightly sub-solar up to moderately more metal-rich values than the Sun, being most of them of solar metal content. In this panel we superimposed the $[Fe/H]$-R$_{G}$ relation (continuous line; slope -0.086\,dex/kpc) as derived in \citeauthor{Netopil:2016}\,\,(\citeyear{Netopil:2016}; their table 3) by fitting the radial metallicity distribution for a set of 88 OCs in the range $R_G<12\,$kpc. The upper and lower limits for this relation, as derived from the informed parameters uncertainties, are represented by dashed lines. OCs from the samples of \cite{Kharchenko:2013} and \cite{Dias:2002} are also shown (grey filled circles), for comparison purposes. Considering uncertainties, our cluster sample -- except possibly NGC 5381 -- agree well with the results of \cite{Netopil:2016}.

NGC\,5381 ($R_G=6.8\pm0.5\,$kpc) departs from the expected relation towards lower metallicity. Despite this, other clusters also show relative low-metallicity values. Indeed, the metallicity distribution derived by \cite{Netopil:2016} (their figure 10) shows azimuthal variations in $[Fe/H]$ from $\sim-0.2\,$ to $\sim+0.2$ for $R_G$ in the range $\sim6.3-7.3\,$kpc.

\begin{figure*}
\begin{center}

%\begin{minipage}{130mm}
%\parbox[c]{0.9\textwidth}
%  {   
    \includegraphics[width=1.0\textwidth]{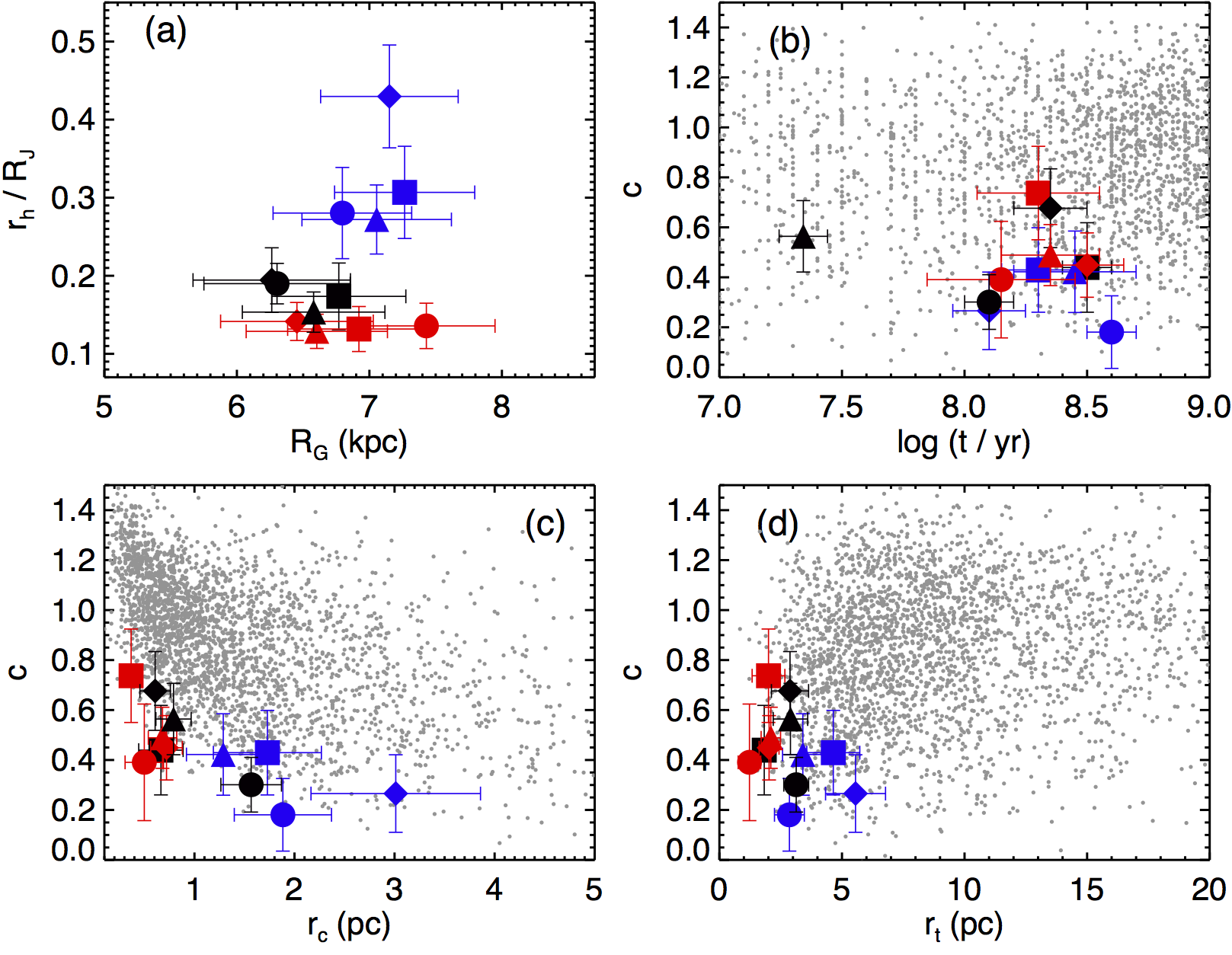}
%  }
\caption{ Relationships between different OC properties. Symbols´ colours are as
in Fig.~\ref{Rgal_versus_age}. Small dots represent OCs from the literature.}

\label{plots_radius_age_mphot_parte1}
%\end{minipage}

\end{center}
\end{figure*}

%$r_t/R_J$ versus $r_t$ plot. Same symbols convention as panel (a). Panel (c):   Panel (d): logarithmic half-light density versus logarithmic critical density (see equation \ref{rho_crit}). The dotted line represent the identity (log\,$\rho_{\textrm{h}}$ = log\,$\rho_{\textrm{crit}}$) locus. Part of Piskunov et al.'s sample was also represented, for comparison. }

\begin{figure*}
\begin{center}

%\parbox[c]{0.90\textwidth}
%  {
   
    \includegraphics[width=1.0\textwidth]{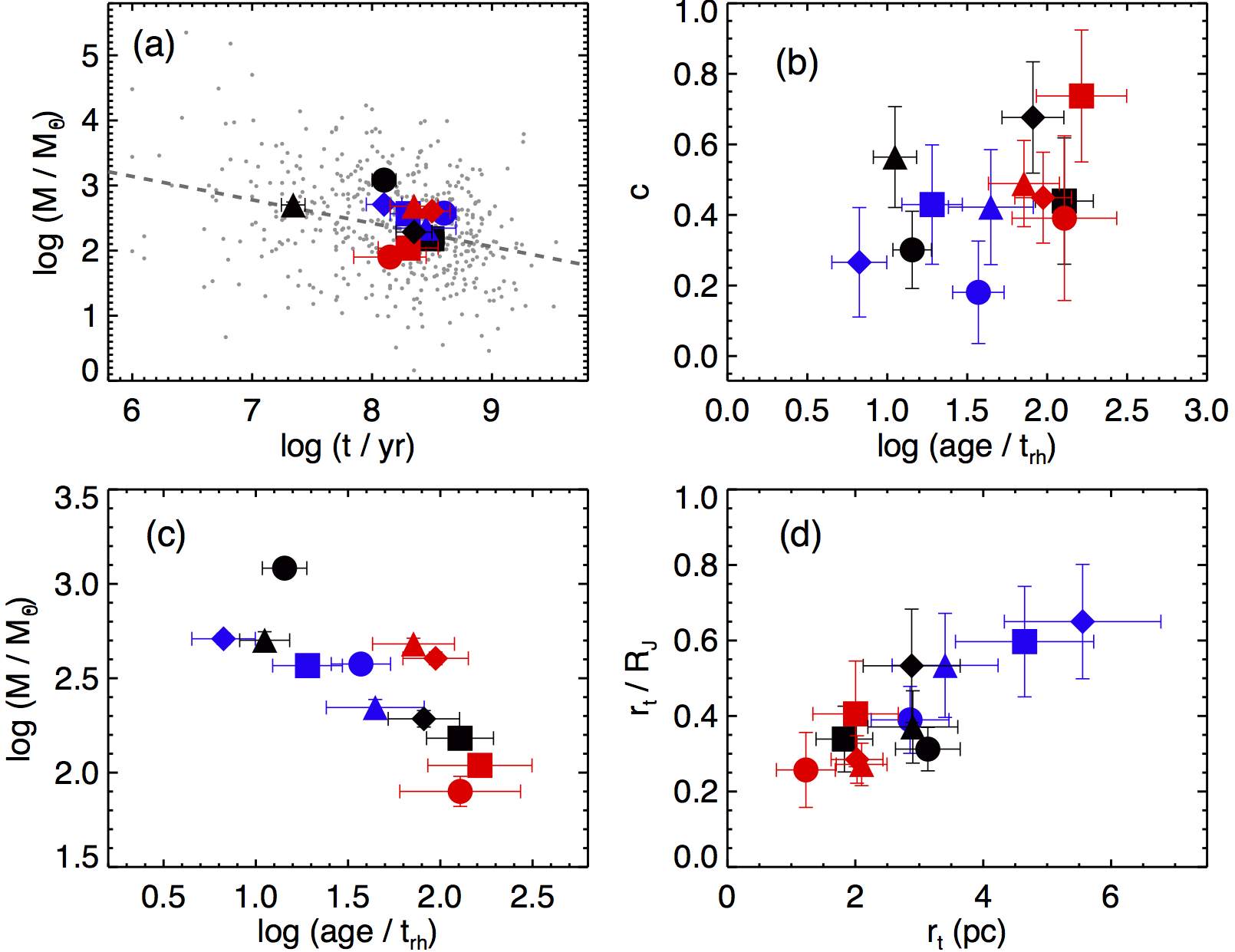}

%  }
\caption{ Relationships between different OC astrophysical parameters. 
Symbols and colours are as in Fig.~\ref{Rgal_versus_age}. Small dots represent 
Joshi et al.'s (2016) OC sample, while the dashed line represent their
derived relationship.}

\label{plots_radius_age_mphot_parte2}
\end{center}
\end{figure*}

In the subsequent analysis, we employ parameters associated to the dynamical evolution, namely mass, age, core, half-light, tidal and Jacobi radii, in order to characterise the dynamical stages of the investigated sample. Panel (a) of Fig.~\ref{plots_radius_age_mphot_parte1}  shows the $r_{h}/R_J$ versus $R_G$ plane, which gives us some hints on the dynamical evolution of the investigated OCs \citep{Baumgardt:2010}. Because of the similar $R_G$ values, the Galactic gravitational potential is
not expected to produce differential tidal effects \citep{Lamers:2005,Piatti:2018}. Therefore, we interpret any difference in the OC dynamical stages as being caused mainly by internal dynamical evolution. Consequently, the larger the $r_{h}/R_J$ ratio in panel (a), the relative less dynamically relaxed an OC. 

To the light of this correlation, we split our OC sample in three groups, distinguished by red ($r_h/R_J\lesssim0.15$), black ($0.15\lesssim r_h/R_J\lesssim0.21$) and blue symbols ($r_h/R_J\gtrsim0.21$), respectively, as also indicated in Fig.~\ref{Rgal_versus_age}. OCs in the blue group (NGC\,5381, Ruprecht\,102, Ruprecht\,97 and ESO\,129-SC32) are relatively less evolved; the black ones (Trumpler\,25, Ruprecht\,111 and Basel\,5; BH\,150 included) are at a relative intermediate stage of dynamical evolution, while the red ones (Collinder\,258, NGC\,6756, Czernik\,37 and NGC\,6249) are the most advanced in dynamical two-body relaxation. It is noticeable that all investigated OCs present $r_h/R_J<0.5$, which makes these stellar aggregates stable against rapid dissolution. Some studies (e.g., \citeauthor{Portegies-Zwart:2010}\,\,\citeyear{Portegies-Zwart:2010} and references therein) have extensively investigated the combined effects of mass loss by stellar evolution and dynamical evolution in the tidal field of the host galaxy and showed that, when clusters expand to a radius of $\sim$0.5\,$r_J$, they lose equilibrium and most of their stars overflow $r_J$.

In panel (b) of Fig.~\ref{plots_radius_age_mphot_parte1} we plot the concentration parameter $c$\,(=log($r_t/r_c$)) as a function of age for our OC sample and compare them with those in the literature. As can be seen, the OCs studied here are within those with smaller 
$c$ values for their ages. Likewise, 
there is a hint of relative different mass segregation (e.g., \citeauthor{de-La-Fuente-Marcos:1997}\,\,\citeyear{de-La-Fuente-Marcos:1997}; \citeauthor{Portegies-Zwart:2001}\,\,\citeyear{Portegies-Zwart:2001}), in the sense that the smaller the $c$ values, the less dynamically evolved an OC. Note that the selected OCs have $r_c$ that 
show a trend with $c$ (see panel (c)) following a much tighter relationship than that
observed for the vast majority of other known OCs. We can see that the less evolved OCs in our sample present less compact cores. This is an expected trend, since as internal relaxation transports energy from the (dynamically) warmer central core to the cooler outer regions, the core contracts as it loses energy \citep{Portegies-Zwart:2010}. Furthermore, the OCs in our sample are relatively small as judged by their tidal radii ($r_t$, see panel (d)) and present their stellar content well within their respective Jacobi radii (Figure \ref{plots_radius_age_mphot_parte2}, panel (d)). 

The classification scheme proposed in Fig.~\ref{plots_radius_age_mphot_parte1}, panel (a), is supported by the results presented in panel (b) of Fig.~\ref{plots_radius_age_mphot_parte2}, where $c$ is plotted as function of age/$t_{\textrm{rh}}$. As can be seen, those OCs with smaller $c$ values show smaller age/$t_{\textrm{rh}}$ ratios. In this sense, \cite{Vesperini:2009} provided an evolutionary picture from results of $N$-body simulations of star clusters in a tidal field. They showed that two-body relaxation causes star clusters to subsequently lose memory of their initial structure (e.g., initial density profile and concentration) and the concentration parameter $c$ increases steadily with time. This overall trend was also verified, although with considerable scatter, by \cite{Piatti:2016} and \cite{Angelo:2018} (figure 14 in both papers), who compared the concentration parameters of a set of investigated Galactic OCs with a sample of 236 OCs analysed homogeneously by \cite{Piskunov:2007}.
%The efficiency of heating on the visible pop- ulation of stars induced by dynamical interactions in the core of the system.

Indeed, the more evolved OCs (red symbols) present larger age/$t_{\textrm{rh}}$ compared to the less evolved ones (blue symbols). This is consistent with the overall evaporation scenario, in which the larger the age/$t_{\textrm{rh}}$ ratio, the more depleted the lower mass content. This can be seen in our clusters' mass functions (Fig.~\ref{mass_func_parte1}), where the ones from the red group show systematically higher depletion of their lowest mass bin when compared to those from the blue group. In all cases, the mass function steepnesses for the higher mass bins do not show noticeable deviations from linear trends as exhibited by Kroupa and Salpeter IMFs (log\,$\phi(m)$ $\propto$ -2.3\,log\,$m$, for $m>0.5\,$M$_{\odot}$). Since our 12 OCs are located at almost the same $R_G$, we expect no differential impact of the tidal field on the clusters' mass function depletion.

As stated by \cite{Bonatto:2004a}, 
OCs will leave at advanced evolutionary stages only a core, with most of the low-mass stars dispersed into the background. In this sense, it is noticeable that smaller OCs are those
relatively more evolved, so that they have had chance to lose low-mass stars, while the
more massive ones have concentrated toward the OC centres. 
Consequently, their $r_t$/$r_J$ ratios are 
smaller than those for relative less evolved OCs, which have mainly expanded within
the Jacobi volume.
(see panel (d) of  Fig.~\ref{plots_radius_age_mphot_parte2}).

Assuming $m_{\textrm{Kroupa}}$ (Table \ref{astroph_params}) as a rough estimate of the initial OC mass, the studied OCs may have lost more than $\sim60$\,percent of their initial mass. Note that Fig.~\ref{mass_func_parte1} shows that the OCs´ mass functions
depart from the linear trend at different stellar masses, which could be linked with
their relative different dynamical stages, in the sense that the more evolved a system
the more massive the lower mass end. On the other hand, the estimated OC masses are
within the range of those obtained by  \cite{Joshi:2016}, as shown in panel (a) of 
Fig.~\ref{plots_radius_age_mphot_parte2}. The dashed line is a linear fit performed by \citeauthor{Joshi:2016}\,\,(\citeyear{Joshi:2016}, their equation 8) to the mean masses within age bins of $\Delta$log($t$/yr)=0.5.

%Considering the whole sample (studied $+$ literature), we can see a general trend in which younger clusters are more massive, since they contain relatively larger number of stars of heavier masses. As they evolve, stellar clusters undergo through a significant amount of mass loss.} 

For completeness purposes we included
in panel (c) OC masses as a function of age/$t_{\textrm{rh}}$, which follows eq. (9). For all investigated clusters, the derived ages are larger than the corresponding $t_{rh}$ (7 $\lesssim$ age/$t_{\textrm{rh}}$$\lesssim$ 164), which means that they have had enough time to dynamically evolve. This statement is also true even if we consider $m_{\textrm{Kroupa}}$ (Table \ref{astroph_params}) to determine $t_{rh}$ (in which case we have $5\lesssim$ age/$t_{\textrm{rh}}$$\lesssim$ 140). Fig.~\ref{plots_radius_age_mphot_parte2} shows that the larger the age/$t_{\textrm{rh}}$, the smaller the cluster mass. This is an expected result, since clusters lose stars as its ages surpass many times its $t_{rh}$.

From the present analysis we advocate that the studied OCs are in different dynamical states.  Assuming that these objects have been subject to the same Galactic tidal effects -- they 
have similar ages (except  BH\,150) and Galactocentric distances --, the differences in their dynamical stages tell us about the wide family of OCs formed in the Sagittarius spiral
arm.

\section{Summary and concluding remarks}
\label{conclusions}

In the present paper we devised an analysis method to uniformly investigate a set of 12 OCs, located in the Sagittarius spiral arm. We took advantage of public Washington $CT_1$ photometric data combined with GAIA DR2 astrometry. The use of GAIA data proved to be an essential tool, since these objects are projected against dense stellar fields, which results in severe contamination in their CMDs.  

Once we estimated the structural parameters from King profile fitting to the OCs' RDPs, we searched for statistically significant concentrations of data in the 3D astrometric space, in order to assign membership likelihoods. PARSEC isochrones covering a wide range of astrophysical parameter values were automatically fitted to the photometric data of high-membership stars and the basic astrophysical parameters ($(m-M)_{0}$, $E(B-V)$, log $t$ and $[Fe/H]$) were derived. Then, OCs' mass functions were built and their total masses were estimated. 
%and presented evidence of low-mass stars depletion, which points out  a signature of dynamically evolved OCs. 

We confirmed BH\,150 as a genuine OC, as judged from the outcomes of our decontamination procedure showing real concentrations of stars in the astrometric space, which allowed the identification of clearer evolutive sequences in the object's CMD.  Its physical 
nature had been under debate, because of purely photometric analyses  or photometric data 
combined with lower quality astrometry could not unambiguously disentangle  the OC from field populations.

The studied OCs have similar Galactocentric distances, and hence are affected by nearly the
same Galactic gravitational potential. For this reason, we speculate that any
difference in the OC dynamical stages is caused by differences in the internal dynamical evolution. 
Based on this assumption, we split the OC sample in three groups according to their  $r_{h}/R_J$ ratios, which has resulted to be a good indicator of the relative OC dynamical evolution stages.
Particularly, we found that the larger the $r_{h}/R_J$ ratio, the less dynamically evolved 
an OC.

The set of investigated OCs are not in an advanced stage of dynamical evolution, since their 
concentration parameters span the lower part of the $c$ regime  ($c\lesssim0.75$). In general, the 
studied OCs present $c$ values which are within the smallest ones for OCs of similar core radii.
Their tidal radii reveal that they are relative small OCs as compared to the sizes of previously
studied OCs. We verified a general trend in which the higher the concentration parameter, the higher the age/$t_{\textrm{rh}}$. Those relative more dynamically evolved OCs have apparently 
experienced more important low-mass star loss.

\section{Acknowledgments}

This research has made use of the VizieR catalogue access tool, CDS, Strasbourg, France. This work has made use of data from the European Space Agency (ESA) mission Gaia (https://www.cosmos.esa.int/gaia), processed by the Gaia Data Processing and Analysis Consortium (DPAC, https://www.cosmos.esa.int/web/gaia/dpac/consortium). Funding for the DPAC has been provided by national institutions, in particular the institutions participating in the Gaia Multilateral Agreement.

%We thank the ...Brazilian financial agencies FAPEMIG (grant APQ-01858-12) and CNPq. We also thank the Gemini staff/resident astronomers for their support and service observing. This publication makes use of data products from the Two Micron All Sky Survey, which is a joint project of the University of Massachusetts and the Infrared Processing and Analysis Center/California Institute of Technology, funded by the National Aeronautics and Space Administration and the National Science Foundation. This research has made use of the WEBDA database, operated at the Institute for Astronomy of the University of Vienna, and of the SIMBAD database, operated at CDS, Strasbourg, France. This research has made use of Aladin and data from the UCAC4 and PPMXL catalogues.

%{\footnotesize
\bibliographystyle{mn2e}
\bibliography{referencias}

\begin{thebibliography}{}

\bibitem[\protect\citeauthoryear{{Angelo}, {Piatti}, {Dias} \& {Maia}}{{Angelo}
  et~al.}{2018}]{Angelo:2018}
{Angelo} M.~S.,  {Piatti} A.~E.,  {Dias} W.~S.,    {Maia} F.~F.~S.,  2018,
  \mnras, 477, 3600

\bibitem[\protect\citeauthoryear{{Angelo}, {Santos}, {Corradi} \&
  {Maia}}{{Angelo} et~al.}{2019}]{Angelo:2019a}
{Angelo} M.~S.,  {Santos} J.~F.~C.,  {Corradi} W.~J.~B.,    {Maia} F.~F.~S.,
  2019, \aap, 624, A8

\bibitem[\protect\citeauthoryear{{Baumgardt}, {Parmentier}, {Gieles} \&
  {Vesperini}}{{Baumgardt} et~al.}{2010}]{Baumgardt:2010}
{Baumgardt} H.,  {Parmentier} G.,  {Gieles} M.,    {Vesperini} E.,  2010,
  \mnras, 401, 1832

\bibitem[\protect\citeauthoryear{{Bica} \& {Bonatto}}{{Bica} \&
  {Bonatto}}{2011}]{Bica:2011}
{Bica} E.,  {Bonatto} C.,  2011, \aap, 530, A32

\bibitem[\protect\citeauthoryear{{Bica}, {Santiago}, {Dutra}, {Dottori}, {de
  Oliveira} \& {Pavani}}{{Bica} et~al.}{2001}]{Bica:2001}
{Bica} E.,  {Santiago} B.~X.,  {Dutra} C.~M.,  {Dottori} H.,  {de Oliveira}
  M.~R.,    {Pavani} D.,  2001, \aap, 366, 827

\bibitem[\protect\citeauthoryear{{Bonatto}, {Bica} \& {Pavani}}{{Bonatto}
  et~al.}{2004}]{Bonatto:2004a}
{Bonatto} C.,  {Bica} E.,    {Pavani} D.~B.,  2004, \aap, 427, 485

\bibitem[\protect\citeauthoryear{{Bonatto}, {Bica} \& {Santos} Jr.}{{Bonatto}
  et~al.}{2005}]{Bonatto:2005}
{Bonatto} C.,  {Bica} E.,    {Santos} Jr. J.~F.~C.,  2005, \aap, 433, 917

\bibitem[\protect\citeauthoryear{{Borissova}, {Ivanov}, {Lucas}, {Kurtev},
  {Alonso-Garcia}, {Ram{\'{\i}}rez Alegr{\'{\i}}a}, {Minniti}, {Froebrich},
  {Hempel}, {Medina}, {Chen{\'e}} \& {Kuhn}}{{Borissova}
  et~al.}{2018}]{Borissova:2018}
{Borissova} J.,  {Ivanov} V.~D.,  {Lucas} P.~W.,  {Kurtev} R.,  {Alonso-Garcia}
  J.,  {Ram{\'{\i}}rez Alegr{\'{\i}}a} S.,  {Minniti} D.,  {Froebrich} D.,
  {Hempel} M.,  {Medina} N.,  {Chen{\'e}} A.-N.,    {Kuhn} M.~A.,  2018,
  \mnras, 481, 3902

\bibitem[\protect\citeauthoryear{{Bressan}, {Marigo}, {Girardi}, {Salasnich},
  {Dal Cero}, {Rubele} \& {Nanni}}{{Bressan} et~al.}{2012}]{Bressan:2012}
{Bressan} A.,  {Marigo} P.,  {Girardi} L.,  {Salasnich} B.,  {Dal Cero} C.,
  {Rubele} S.,    {Nanni} A.,  2012, \mnras, 427, 127

\bibitem[\protect\citeauthoryear{{Caetano}, {Dias}, {L{\'e}pine}, {Monteiro},
  {Moitinho}, {Hickel} \& {Oliveira}}{{Caetano} et~al.}{2015}]{Caetano:2015}
{Caetano} T.~C.,  {Dias} W.~S.,  {L{\'e}pine} J.~R.~D.,  {Monteiro} H.~S.,
  {Moitinho} A.,  {Hickel} G.~R.,    {Oliveira} A.~F.,  2015, \na, 38, 31

\bibitem[\protect\citeauthoryear{{Cantat-Gaudin}, {Jordi}, {Vallenari},
  {Bragaglia}, {Balaguer-N{\'u}{\~n}ez}, {Soubiran}, {Bossini}, {Moitinho},
  {Castro-Ginard}, {Krone-Martins}, {Casamiquela}, {Sordo} \&
  {Carrera}}{{Cantat-Gaudin} et~al.}{2018}]{Cantat-Gaudin:2018a}
{Cantat-Gaudin} T.,  {Jordi} C.,  {Vallenari} A.,  {Bragaglia} A.,
  {Balaguer-N{\'u}{\~n}ez} L.,  {Soubiran} C.,  {Bossini} D.,  {Moitinho} A.,
  {Castro-Ginard} A.,  {Krone-Martins} A.,  {Casamiquela} L.,  {Sordo} R.,
  {Carrera} R.,  2018, \aap, 618, A93

\bibitem[\protect\citeauthoryear{{Carraro} \& {Chiosi}}{{Carraro} \&
  {Chiosi}}{1994}]{Carraro:1994}
{Carraro} G.,  {Chiosi} C.,  1994, \aap, 288, 751

\bibitem[\protect\citeauthoryear{{Carraro}, {Janes} \& {Eastman}}{{Carraro}
  et~al.}{2005}]{Carraro:2005a}
{Carraro} G.,  {Janes} K.~A.,    {Eastman} J.~D.,  2005, \mnras, 364, 179

\bibitem[\protect\citeauthoryear{{Clari{\'a}}, {Piatti}, {Mermilliod} \&
  {Palma}}{{Clari{\'a}} et~al.}{2008}]{Claria:2008}
{Clari{\'a}} J.~J.,  {Piatti} A.~E.,  {Mermilliod} J.~C.,    {Palma} T.,  2008,
  Astronomische Nachrichten, 329, 609

\bibitem[\protect\citeauthoryear{{de La Fuente Marcos}}{{de La Fuente
  Marcos}}{1997}]{de-La-Fuente-Marcos:1997}
{de La Fuente Marcos} R.,  1997, \aap, 322, 764

\bibitem[\protect\citeauthoryear{{Dias}, {Alessi}, {Moitinho} \&
  {L{\'e}pine}}{{Dias} et~al.}{2002}]{Dias:2002}
{Dias} W.~S.,  {Alessi} B.~S.,  {Moitinho} A.,    {L{\'e}pine} J.~R.~D.,  2002,
  \aap, 389, 871

\bibitem[\protect\citeauthoryear{{Dias}, {Monteiro}, {L{\'e}pine}, {Prates},
  {Gneiding} \& {Sacchi}}{{Dias} et~al.}{2018}]{Dias:2018a}
{Dias} W.~S.,  {Monteiro} H.,  {L{\'e}pine} J.~R.~D.,  {Prates} R.,  {Gneiding}
  C.~D.,    {Sacchi} M.,  2018, \mnras, 481, 3887

\bibitem[\protect\citeauthoryear{{Diolaiti}, {Bendinelli}, {Bonaccini},
  {Close}, {Currie} \& {Parmeggiani}}{{Diolaiti} et~al.}{2000}]{Diolaiti:2000}
{Diolaiti} E.,  {Bendinelli} O.,  {Bonaccini} D.,  {Close} L.,  {Currie} D.,
  {Parmeggiani} G.,  2000, \aaps, 147, 335

\bibitem[\protect\citeauthoryear{{Ferreira}, {Santos}, {Corradi}, {Maia} \&
  {Angelo}}{{Ferreira} et~al.}{2019}]{Ferreira:2019}
{Ferreira} F.~A.,  {Santos} J.~F.~C.,  {Corradi} W.~J.~B.,  {Maia} F.~F.~S.,
  {Angelo} M.~S.,  2019, \mnras, 483, 5508

\bibitem[\protect\citeauthoryear{{Gaia Collaboration}, {Brown}, {Vallenari},
  {Prusti}, {de Bruijne}, {Babusiaux} \& {Bailer-Jones}}{{Gaia Collaboration}
  et~al.}{2018}]{Gaia-Collaboration:2018}
{Gaia Collaboration} {Brown} A.~G.~A.,  {Vallenari} A.,  {Prusti} T.,  {de
  Bruijne} J.~H.~J.,  {Babusiaux} C.,    {Bailer-Jones} C.~A.~L.,  2018, ArXiv
  e-prints

\bibitem[\protect\citeauthoryear{{Geisler}}{{Geisler}}{1996}]{Geisler:1996}
{Geisler} D.,  1996, \aj, 111, 480

\bibitem[\protect\citeauthoryear{{Geisler}, {Claria} \& {Minniti}}{{Geisler}
  et~al.}{1991}]{Geisler:1991}
{Geisler} D.,  {Claria} J.~J.,    {Minniti} D.,  1991, \aj, 102, 1836

\bibitem[\protect\citeauthoryear{{Joshi}, {Dambis}, {Pandey} \&
  {Joshi}}{{Joshi} et~al.}{2016}]{Joshi:2016}
{Joshi} Y.~C.,  {Dambis} A.~K.,  {Pandey} A.~K.,    {Joshi} S.,  2016, \aap,
  593, A116

\bibitem[\protect\citeauthoryear{{Kharchenko}, {Piskunov}, {Schilbach},
  {R{\"o}ser} \& {Scholz}}{{Kharchenko} et~al.}{2013}]{Kharchenko:2013}
{Kharchenko} N.~V.,  {Piskunov} A.~E.,  {Schilbach} E.,  {R{\"o}ser} S.,
  {Scholz} R.-D.,  2013, \aap, 558, A53

\bibitem[\protect\citeauthoryear{{King}}{{King}}{1962}]{King:1962}
{King} I.,  1962, Astronomical Journal, 67, 471

\bibitem[\protect\citeauthoryear{{Kos}, {de Silva}, {Buder} \& {et al.}}{{Kos}
  et~al.}{2018}]{Kos:2018}
{Kos} J.,  {de Silva} G.,  {Buder} S.,    {et al.} 2018, \mnras, 480, 5242

\bibitem[\protect\citeauthoryear{{Kroupa}}{{Kroupa}}{2001}]{Kroupa:2001}
{Kroupa} P.,  2001, \mnras, 322, 231

\bibitem[\protect\citeauthoryear{{Lada} \& {Lada}}{{Lada} \&
  {Lada}}{2003}]{Lada:2003}
{Lada} C.~J.,  {Lada} E.~A.,  2003, \araa, 41, 57

\bibitem[\protect\citeauthoryear{{Lamers}, {Gieles}, {Bastian}, {Baumgardt},
  {Kharchenko} \& {Portegies Zwart}}{{Lamers} et~al.}{2005}]{Lamers:2005}
{Lamers} H.~J.~G.~L.~M.,  {Gieles} M.,  {Bastian} N.,  {Baumgardt} H.,
  {Kharchenko} N.~V.,    {Portegies Zwart} S.,  2005, \aap, 441, 117

\bibitem[\protect\citeauthoryear{{Landolt}}{{Landolt}}{1992}]{Landolt:1992}
{Landolt} A.~U.,  1992, \aj, 104, 340

\bibitem[\protect\citeauthoryear{{Maia}, {Corradi} \& {Santos} Jr.}{{Maia}
  et~al.}{2010}]{Maia:2010}
{Maia} F.~F.~S.,  {Corradi} W.~J.~B.,    {Santos} Jr. J.~F.~C.,  2010, \mnras,
  407, 1875

\bibitem[\protect\citeauthoryear{{Marcionni}, {Clari{\'a}}, {Parisi}, {Palma},
  {Oddone} \& {Ahumada}}{{Marcionni} et~al.}{2014}]{Marcionni:2014}
{Marcionni} N.,  {Clari{\'a}} J.~J.,  {Parisi} M.~C.,  {Palma} T.,  {Oddone}
  M.,    {Ahumada} A.~V.,  2014, \na, 33, 14

\bibitem[\protect\citeauthoryear{{Mermilliod}, {Mayor} \& {Udry}}{{Mermilliod}
  et~al.}{2008}]{Mermilliod:2008}
{Mermilliod} J.~C.,  {Mayor} M.,    {Udry} S.,  2008, \aap, 485, 303

\bibitem[\protect\citeauthoryear{{Moffat} \& {Vogt}}{{Moffat} \&
  {Vogt}}{1975}]{Moffat:1975}
{Moffat} A.~F.~J.,  {Vogt} N.,  1975, \aaps, 20, 125

\bibitem[\protect\citeauthoryear{{Moraux}}{{Moraux}}{2016}]{Moraux:2016}
{Moraux} E.,  2016, in {Moraux} E.,  {Lebreton} Y.,   {Charbonnel} C.,  eds,
  EAS Publications Series Vol.~80 of EAS Publications Series, {Open clusters
  and associations in the Gaia era}.
pp 73--114

\bibitem[\protect\citeauthoryear{{Netopil} \& {Paunzen}}{{Netopil} \&
  {Paunzen}}{2013}]{Netopil:2013}
{Netopil} M.,  {Paunzen} E.,  2013, \aap, 557, A10

\bibitem[\protect\citeauthoryear{{Netopil}, {Paunzen}, {Heiter} \&
  {Soubiran}}{{Netopil} et~al.}{2016}]{Netopil:2016}
{Netopil} M.,  {Paunzen} E.,  {Heiter} U.,    {Soubiran} C.,  2016, \aap, 585,
  A150

\bibitem[\protect\citeauthoryear{{Pavani} \& {Bica}}{{Pavani} \&
  {Bica}}{2007}]{Pavani:2007}
{Pavani} D.~B.,  {Bica} E.,  2007, \aap, 468, 139

\bibitem[\protect\citeauthoryear{{Perren}, {V{\'a}zquez} \& {Piatti}}{{Perren}
  et~al.}{2015}]{Perren:2015}
{Perren} G.~I.,  {V{\'a}zquez} R.~A.,    {Piatti} A.~E.,  2015, \aap, 576, A6

\bibitem[\protect\citeauthoryear{{Piatti}}{{Piatti}}{2016}]{Piatti:2016}
{Piatti} A.~E.,  2016, \mnras, 463, 3476

\bibitem[\protect\citeauthoryear{{Piatti}, {Dias} \& {Sampedro}}{{Piatti}
  et~al.}{2017}]{Piatti:2017a}
{Piatti} A.~E.,  {Dias} W.~S.,    {Sampedro} L.~M.,  2017, \mnras, 466, 392

\bibitem[\protect\citeauthoryear{{Piatti} \& {Mackey}}{{Piatti} \&
  {Mackey}}{2018}]{Piatti:2018}
{Piatti} A.~E.,  {Mackey} A.~D.,  2018, \mnras, 478, 2164

\bibitem[\protect\citeauthoryear{{Piskunov}, {Schilbach}, {Kharchenko},
  {R{\"o}ser} \& {Scholz}}{{Piskunov} et~al.}{2007}]{Piskunov:2007}
{Piskunov} A.~E.,  {Schilbach} E.,  {Kharchenko} N.~V.,  {R{\"o}ser} S.,
  {Scholz} R.-D.,  2007, \aap, 468, 151

\bibitem[\protect\citeauthoryear{{Plummer}}{{Plummer}}{1911}]{Plummer:1911}
{Plummer} H.~C.,  1911, \mnras, 71, 460

\bibitem[\protect\citeauthoryear{{P{\"o}hnl} \& {Paunzen}}{{P{\"o}hnl} \&
  {Paunzen}}{2010}]{Pohnl:2010}
{P{\"o}hnl} H.,  {Paunzen} E.,  2010, \aap, 514, A81

\bibitem[\protect\citeauthoryear{{Portegies Zwart}, {McMillan} \&
  {Gieles}}{{Portegies Zwart} et~al.}{2010}]{Portegies-Zwart:2010}
{Portegies Zwart} S.~F.,  {McMillan} S.~L.~W.,    {Gieles} M.,  2010, \araa,
  48, 431

\bibitem[\protect\citeauthoryear{{Portegies Zwart}, {McMillan}, {Hut} \&
  {Makino}}{{Portegies Zwart} et~al.}{2001}]{Portegies-Zwart:2001}
{Portegies Zwart} S.~F.,  {McMillan} S.~L.~W.,  {Hut} P.,    {Makino} J.,
  2001, \mnras, 321, 199

\bibitem[\protect\citeauthoryear{{Reid}}{{Reid}}{1993}]{Reid:1993a}
{Reid} M.~J.,  1993, \araa, 31, 345

\bibitem[\protect\citeauthoryear{{Roeser}, {Demleitner} \&
  {Schilbach}}{{Roeser} et~al.}{2010}]{Roeser:2010}
{Roeser} S.,  {Demleitner} M.,    {Schilbach} E.,  2010, \aj, 139, 2440

\bibitem[\protect\citeauthoryear{{Salpeter}}{{Salpeter}}{1955}]{Salpeter:1955}
{Salpeter} E.~E.,  1955, \apj, 121, 161

\bibitem[\protect\citeauthoryear{{Santos} J.~F.~C. \& {Piatti}}{{Santos} \&
  {Piatti}}{2004}]{Santos:2004}
{Santos} J.~F.~C. J.,  {Piatti} A.~E.,  2004, \aap, 428, 79

\bibitem[\protect\citeauthoryear{{Skrutskie}, {Cutri}, {Stiening}, {Weinberg}
  \& {et al.}}{{Skrutskie} et~al.}{2006}]{Skrutskie:2006}
{Skrutskie} M.~F.,  {Cutri} R.~M.,  {Stiening} R.,  {Weinberg} M.~D.,    {et
  al.} 2006, \aj, 131, 1163

\bibitem[\protect\citeauthoryear{{Spitzer} Jr. \& {Hart}}{{Spitzer} \&
  {Hart}}{1971}]{Spitzer:1971}
{Spitzer} Jr. L.,  {Hart} M.~H.,  1971, \apj, 164, 399

\bibitem[\protect\citeauthoryear{{Taylor}, {Boylan-Kolchin}, {Torrey},
  {Vogelsberger} \& {Hernquist}}{{Taylor} et~al.}{2016}]{Taylor:2016}
{Taylor} C.,  {Boylan-Kolchin} M.,  {Torrey} P.,  {Vogelsberger} M.,
  {Hernquist} L.,  2016, \mnras, 461, 3483

\bibitem[\protect\citeauthoryear{{Vesperini}, {McMillan} \& {Portegies
  Zwart}}{{Vesperini} et~al.}{2009}]{Vesperini:2009}
{Vesperini} E.,  {McMillan} S.~L.~W.,    {Portegies Zwart} S.,  2009, \apj,
  698, 615

\end{thebibliography}
%}

%%%%%%%%%%Original do macro%%%%%%%%%%%%
%\begin{thebibliography}{99}
%\bibitem[\protect\citeauthoryear{Baird}{1981}]{b1} Baird S.R., 1981,
%ApJ, 245, 208
%\bibitem[\protect\citeauthoryear{Beichman et al.}{1985a}]{b2} Beichman
%C.A., Neugebauer G., Habing H.J., Clegg P.E., Chester T.J., 1985a,
%{\it IRAS\/} Point Source Catalog. Jet Propulsion Laboratory,
%Pasadena
%\bibitem[\protect\citeauthoryear{Beichman et al.}{1985b}]{b3} Beichman
%C.A., Neugebauer G., Habing H.J., Clegg P.E., Chester T.J., 1985b,
%{\it IRAS\/} Explanatory Supplement. Jet Propulsion Laboratory,
%Pasadena
%\end{thebibliography}
%%%%%%%%%%%%%%%%%%%%%%%%%%%%%%%%

\newpage

%%%%%%%%%%%%%%%%%%%%%%%%%%%%%%%%%%%%%%%%%%%%%%%

%\begin{figure*}

%\parbox[c]{1.0\textwidth}
%{

%   \includegraphics[width=0.5\textwidth]{mctime_ESO518-3.pdf}
%   \includegraphics[width=0.5\linewidth]{mctime_Ruprecht121.pdf}
%  \includegraphics[width=0.5\linewidth]{mctime_ESO134-12.pdf}
%   \includegraphics[width=0.5\linewidth]{mctime_NGC6573.pdf}
%   \includegraphics[width=0.5\linewidth]{mctime_ESO260-7.pdf}
%   \includegraphics[width=0.5\linewidth]{mctime_ESO065-7.pdf}
%}

%\caption{Same of Figure \ref{fig:mctime}, but for the OCs ESO\,518-3 (top left panel), Ruprecht\,121 (top right panel), ESO\,134-12 (middle left panel), NGC\,6573 (middle right panel), ESO\,260-7 (bottom left panel) and ESO\,065-7 (bottom right panel).}
%\label{fig:mctime_all_OCs}
%\end{figure*}

%%%%%%%%%%%%%%%%%%%%%%%%%%%%%%%%%%%%%%%%%%%%%%%

\bsp

\label{lastpage}

\end{document}